\documentclass[aps,pra,floatfix,10pt,twocolumn,superscriptaddress]{revtex4-1}
\usepackage{graphicx}
\usepackage{ulem}
\usepackage{soul}
\usepackage{amsmath}
\usepackage{amsfonts}
\usepackage{amssymb}
\usepackage{color}
\usepackage{float}
\usepackage{verbatim}
\usepackage{stackrel}
\usepackage[capitalize]{cleveref}
\usepackage{tikz}
\usetikzlibrary{shapes.geometric, arrows}
\usetikzlibrary{decorations.markings}
\usetikzlibrary{decorations.pathmorphing}
\usetikzlibrary{decorations.pathreplacing}

\makeatletter
\allowdisplaybreaks
\makeatother

\newcommand{\ket}[1]{$\left|#1\right\rangle$}
\newcommand{\Om}[1]{\small $\omega_{#1}$}

\begin{document}
\title{Time resolved multi-photon effects in the fluorescence spectra \\ of two-level systems at rest and in motion.}
\date{\today}
\author{Emil Vi\~nas Bostr\"om}
\affiliation{Division of Mathematical Physics and ETSF, Lund University, PO Box 118, 221 00 Lund, Sweden}
\author{Andrea D'Andrea}
\affiliation{Istituto di Struttura della Materia, CNR, Area della Ricerca di Roma 1 (Montelibretti) Italy}

\author{Michele Cini}
\affiliation{ Dipartimento di Fisica, Universita` di Roma Tor Vergata, Via della Ricerca Scientifica 1, I-00133 Rome, Italy}

\author{Claudio Verdozzi}
\affiliation{Division of Mathematical Physics and ETSF, Lund University, PO Box 118, 221 00 Lund, Sweden}

%%%%%%%%%%%%%%%%%%%%%%%%%%%%%%%%%%%%%%%%%%%%%%%%%%%%%%%%%%%%
%%%%%%%%%%%%%%%%%%%%%%%%%%%%%%%%%%%%%%%%%%%%%%%%%%%%%%%%%%%%

\begin{abstract}
We study the time-resolved fluorescence spectrum in two-level systems interacting with an incident coherent field, both in the weak and intermediate coupling regimes. For a single two-level system in the intermediate coupling case, as time flows, the spectrum develops distinct features, that are not captured by a semi-classical treatment of the incident field. Specifically, for a field on resonance with the atomic transition energy, the usual Mollow spectrum is replaced by a four peak structure, and for a frequency that is half of the atomic transition energy, the time-dependent spectrum develops a second harmonic peak with a superimposed Mollow triplet. In the long-time limit, our description recovers results previously found in the literature.  
After analyzing why a different behavior is observed in the quantum and classical dynamics, the reason for the occurrence of a second harmonic signal in a two-level system is explained via a symmetry analysis of the total (electron and photon) system, and in terms of a three level system operating in limiting regimes. We find an increased second harmonic signal in an array of two-level systems, suggesting a superradiance-like enhancement for multiple two-level systems in cavity setups. Finally, initial explorative results are presented for two-level model atoms entering and exiting a cavity, which hint at an interesting interplay between cavity-photon screening and atomic dynamics effects.
\end{abstract}
\maketitle

\section{Introduction}
Fluorescence, a type of luminescence  \cite{Einstein,Dirac,WignerWeisskopf}, is a hallmark of quantum mechanics at work: A system that 
has absorbed electromagnetic radiation re-emits it at a later time, while the spins of the electrons
involved in the de-excitation process conform to specific selection rules. 

In addition to being an operational mechanism in several biological systems \cite{Biopaper}, fluorescence serves
in many technologies of different complexity, ranging from simple indoor lighting to in-depth spectroscopic 
characterization at the atomic scale. 
Investigations of fluorescence started before the advent 
of quantum mechanics, but it would be the latter that provided the conceptual framework for a 
microscopic description \cite{Einstein,Dirac,WignerWeisskopf}.

Spectroscopic methods have a rich history as a means of investigating the internal structure of matter. In particular, with the development of ever more sophisticated laser systems, the electronic dynamics in atomic and solid state systems can now be mapped out in real time while maintaining a high frequency resolution~\cite{Neppl15,Chew15}. 
This allows to study in a precise manner the basic processes of light-matter interaction, and even to characterize properties of light itself, as for example emitted via fluorescence.

A minimal-complexity model to study fluorescence emission and fluorescence spectroscopy is a 
two-level system \cite{AllenEberly,Shorebook,Schleich} (for example, a spinless electron that can be in either of two nondegenerate quantum states,  or in any linear superposition thereof) interacting with a single radiation mode via dipolar coupling. 
If the two levels are thought to be selected from an atom, the model is also referred to as a
``two-level atom'' (with the additional option that the center of mass of the atom can be either at rest
or in motion). 

Two-level systems came into prominence with Rabi et {\it al.}'s work for a magnetic moment exposed to a classical 
circularly polarized field \cite{Rabi}. In a subsequent study by Bloch and Siegert the linearly polarized case was then considered \cite{Bloch} (in this situation, the solution is more complicated compared to Rabi's original case \cite{Stenholm}). 
The next important development took place when the radiation mode was also treated quantum-mechanically,
and the so-called rotating wave approximation (RWA) was introduced \cite{JaynCum,CarmaRWA,JaynCumReview,K.Fujii,QXie2017}. 
Designed for weak-coupling and near-resonance regimes, the RWA permits an explicit treatment of the 
time dependence \cite{Eberly1400}, and provides a convenient route to the so-called  dressed-level (or -atom) approach 
\cite{Schleich,Guerin,Ultrastrong}, where the levels of the system split and are renormalized (shifted)
by the radiation field. In turn, within this approach a clear picture emerges \cite{BooKAdvances} of the
Mollow spectrum \cite{usualMollow}, a three-peak structure due to the fluorescent response
of a two-level system to a resonant or quasi resonant radiation mode.

Nevertheless, a number of interesting physical situations are outside the reach of RWA, as e.g. 
the intense pulsed regime, where field monochromaticity is absent and off-resonant coupling
cannot be avoided. The need under some circumstances to go beyond
the RWA has in fact been recognized in several contexts (also by comparing exact and RWA solutions \cite{Swain1,AdAnoRWA}). For example, when discussing modifications of the shape of the three peaks in Mollow spectra \cite{Keitel,BrowneKeitel,PerfettoMollow}, spontaneous
emission in three-level systems \cite{Q.Xu}, or when center of mass dynamics is
included \cite{ScullyCapasso} (for a recent review, see e.g. \cite{OBrien}).

As these few, incomplete remarks suggest, two-level systems coupled to radiation in different ``flavors'' remain of capital relevance to this day to probe and redefine the knowledge boundaries in (quantum) optics \cite{IOP2017,CavityReview,Shore_honor}.
This can occur via generalization of the basic model(s) together with deeper mathematical analysis (see e.g. \cite{Kollar,Kollar1,Eckle}), to address
unexplored coupling regimes \cite{FreeRydberg,Sanchez19}, or novel areas of applications. For example, 
cavity quantum optics and the Unruh effect \cite{ScullyCapasso}, quantum mechanical interference \cite{Narducci}, two-photon relaxation \cite{Malekakhlagh19}, 
quantum phase transition \cite{phasetrans}, interaction of a photons matter qubits \cite{Leuchs}
and Mollow spectra in ultracold atoms
\cite{MollowCold}.

The quantum nature of light manifests in a clear way at low photon number and for large light-matter coupling. These two aspects contribute distinctly. This is different from the strong-field regime, where a semiclassical treatment becomes appropriate and where the effective coupling parameter is the product of field strength and coupling strength \cite{Stenholm}.
Concerning the few-photon limit, this can be e.g. reached in high quality-factor cavities
~\cite{Guerlin07,Gleyzes07,Murch13}. On the other hand, to attain the strong coupling (also denoted polaritonic) regime, a possibility is offered by the insertion of a quantum well into a distributed Bragg reflector cavity~\cite{Pilozzi07}, i.e. by coupling the photon field and an optical inter-band transition (a Wannier exciton). 

{\it Scope of this work.-}
In the present study we consider two-level, one-electron systems interacting with two optical modes (the coherent-pump and a de-excitation field). Specifically,  we 
address the (so far largely unexplored) multi-photon effects in fluorescence spectra, that depend separately on field intensity and light-matter coupling strength. This will be done in situations of progressive complexity: a single two-level atom at rest, 
an array of two-level atoms at rest, and finally a single two-level atom moving through an optical cavity.  
To this end, it is necessary to employ a theoretical framework suitable for both non-linear effects (in relation to certain experimental setups~\cite{Joshi92,Joshi98,Dorfman18}) and an explicitly time-dependent light-matter coupling. 

Several years ago, three of the present authors introduced an exact solution method~\cite{Cini93,Cini95} for a large class of multi-photon spectroscopy models.  Their method is based on a recursion technique (see e.g. Ref.~\cite{Cini88}) in the frequency domain, i.e. for the stationary limit of fluorescence. They also pointed out the difference between the exact solutions and the RWA.
The aim was to address systems where interactions with the environment are as weak as possible. That is, 
where both inhomogeneous (such as due to static and dynamical disorder, mode leakage, etc. ) and homogeneous 
(atomic collisions, but also non-radiative decay, etc.) decoherence factors plays a minor role. Put differently, the focus was on a regime where both energy-dependent broadening (not considered in the rest of this paper, because assumed to be controllable), and energy-independent one  (denoted by $\Gamma$ and retained in the paper) are as small as possible. 
As discussed above, current experimental capabilities provide practical and close-to-ideal realisations of these premises
with optical cavity setups, and make time-resolved studies possible. In this way, it is possible to investigate the actual development of the fluorescence signal before the steady state signal sets in.

\begin{figure}
 \begin{tikzpicture}[scale=0.5, level/.style={thick}, virtual/.style={thick,densely dashed,color=orange}, cohe/.style={thick,<->,>=stealth,color=blue!70!green!40!white,
 decorate,decoration={snake, amplitude=1.0, segment length=8, pre length=4, post length=3}}, emit/.style={thick,->,>=stealth,color=orange!80!white,decorate,decoration={snake, amplitude=1.0, segment length=8, post length=2}}, inci/.style={thick,->,>=stealth,color=blue!70!green!40!white,
 decorate,decoration={snake, amplitude=1.0, segment length=8, pre length=4, post length=3}}, dbl/.style={<->,shorten >=2pt,shorten <=2pt,>=stealth,color=black}]
    
\draw[decorate,decoration={brace,amplitude=3pt}] 
    (4.5,4.) node(t_k_unten){} -- 
    (10.5,4.) node(t_k_opt_unten){}; 
\node at (7.5,4.5){$N$};    
    
    \draw[level] (1., 1.2) -- (-1.1, 1.2) node[left] {\normalsize \ket{2}};
    \draw[level] (1.,-1.2) -- (-1.1,-1.2) node[left] {\normalsize \ket{1}};
    \draw[<->,>=stealth,black,thick] (-0.05,-1.1) -- (-0.05,1.1) node[right, midway] {\normalsize $\epsilon$};
    \draw[inci] (-2.0,0) node[black, left] {\normalsize $\omega_a, g_a$} -- (-0.3,0);
    \draw[emit] (0.6,0) -- (2.1,0) node[black, right] {\normalsize $\omega_b, g_b$};
    \node at (-2.6,2.2) {\normalsize $(a)$};
 
    \draw[inci] (7.6,3.1) node[black, left] {\normalsize $$} -- (9.6,3.1);
    \draw[emit] (7.6,2.7) -- (9.6,2.7) node[black, right] {\normalsize $$};    
    \draw[very thick,gray] (3.5,4.5) arc (140:220:2.5);
    \draw[very thick,gray] (11.5,4.5) arc (40:-40:2.5);
    \fill[color=green!35!blue!60!white] (3.75,1.75) rectangle (5.25,3.75);
    \draw[level] (4,3.5) -- (5,3.5);
    \draw[level] (4,2.0) -- (5,2.0);
    \fill[color=green!35!blue!60!white] (5.75,1.75) rectangle (7.25,3.75);
    \draw[level] (6,3.5) -- (7,3.5);
    \draw[level] (6,2.0) -- (7,2.0);
    \node at (8.5,2.25) {\large $\cdots$};
    \fill[color=green!35!blue!60!white] (9.75,1.75) rectangle (11.25,3.75);
    \draw[level] (10,3.5) -- (11,3.5);
    \draw[level] (10,2.0) -- (11,2.0);
    \node at (2.5,4.3) {\normalsize $(b)$};
    \draw[dbl] (4.5, -4.2) -- (10.5, -4.2) node[midway, yshift=0.25cm] {$L$};

    \draw[very thick,gray] (4.8,0) arc (140:220:2.5);
    \draw[very thick,gray] (10.2,0) arc (40:-40:2.5);
    \draw[cohe] (5,-1.1) -- (10,-1.1);
    \draw[cohe] (5,-2.0) -- (10,-2.0);
    \fill[color=green!35!blue!60!white] (7.5,-1.55) circle (1.2);
    \draw[level] (7,-0.8) -- (8,-0.8);
    \draw[level] (7,-2.3) -- (8,-2.3);
    \draw[->,>=stealth,black,thick] (7.5,-1.55) -- (9.5,-1.55) node[right] {$p$};
    \draw[emit] (7.0,-0.2) -- (6.4,0.8);
    \draw[emit] (8.0,-0.2) -- (8.6,0.8);
    \draw[emit] (7.0,-2.9) -- (6.4,-3.9);
    \draw[emit] (8.0,-2.9) -- (8.6,-3.9);
    \node at (11.1,0.0) {\normalsize $(c)$};
 \end{tikzpicture}
 \caption{Schematic of the systems considered in the paper. In panel $(a)$, a single two-level system of transition energy $\epsilon$ interacting with a coherent field of frequency $\omega_a$ (with coupling strength $g_a$) and a fluorescent field of frequency $\omega_b$ (with coupling strength $g_b$). In panel $(b)$, an array of $N$ two-level systems interacting with a coherent field and fluorescent field. In panel $(c)$, a two-level atom of momentum $p$ passing through a cavity of length $L$ where it interacts with a coherent cavity field and emits fluorescent photons.}
 \label{fig:schematic}
\end{figure}
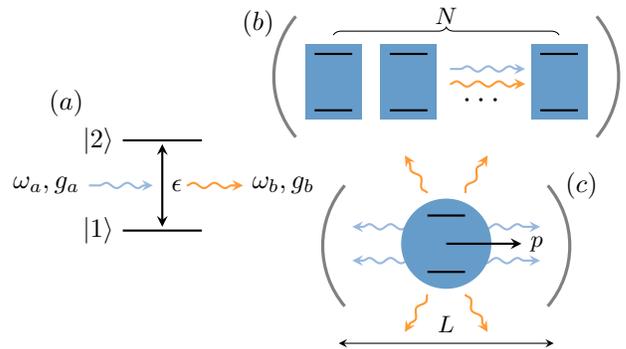

With these considerations in mind, here we take a different methodological route from that in Refs.~\cite{Cini93,Cini95}, by working in a real-time (and again free of RWA) picture, and computing the 
exact time-dependent fluorescence response. To establish the effect of multi-photon contributions and counter-rotating terms, we specialize to the Mollow regime (on-resonance situation)  and to second-harmonic generation (SHG) (off-resonance situation). In the literature, Mollow spectra are discussed in terms of two level systems \cite{usualMollow}, and thus our work conforms to previous treatments. In contrast, SHG is commonly discussed in terms of three level systems \cite{3SHG}; however, we will show that a genuine two-level system admits SHG. Overall, our study here can thus be summarized as
an exploration of multi-photon effects in Mollow and SHG fluorescence spectra across three different two-level-system setups (a single system at rest, many systems at rest, a single system in motion). 

A remark about the units used in this work: 
unless otherwise stated at specific points in the paper, the energy unit is $\epsilon=\epsilon_2-\epsilon_1$, the distance between the two levels in the system, and the time unit is $\hbar/\epsilon$ \cite{PARAMS}.

{\it Organisation of the paper.-}
The rest of this article is organized in three parts: In Sec.~\ref{sec:single_system} we study the time-dependent fluorescence spectrum of an isolated two-level system (see Fig.~\ref{fig:schematic}a). In Sec.~\ref{sec:array}, we consider an array of two-level systems interacting with a common coherent field (see Fig.~\ref{fig:schematic}b). In this case, the fluorescence signal shows an enhancement compared to the single two-level system case, consistently with a superradiance-like mechanism. Finally, in Sec.~\ref{sec:atomic_motion} we consider a two-level model atom passing through an optical cavity (see Fig.~\ref{fig:schematic}c). We explicitly treat the quantum motion of the atomic center of mass, and show that this results in a fluorescence spectrum which differs from that obtained from a semi-classical treatment of the atomic motion. Conclusive remarks and outlook are in Sec. \ref{conclude}.

%
%
%%%%%%%%%%%%%%%%%%%%%%%%%%%%%%%%%%%%%%%%%%%%%%%%%%%%%%%%%%%%
%%%%%%%%%%%%%%%%%%%%%%%%%%%%%%%%%%%%%%%%%%%%%%%%%%%%%%%%%%%%
%
%

\section{A single two-level system}\label{sec:single_system}
We start with the simplest of the three situations, i.e. 
a single two-level system interacting with two optical modes. After presenting model and method of solution, we investigate the time evolution and the long time limit of the system's fluorescence spectrum in the Mollow regime. We consider the case when the Rabi frequency $g$ becomes a moderate fraction of the level spacing $\epsilon$ of the two-level system (about $10$ \%). The results show that for a field in resonance with the atomic transition energy, the usual three peak Mollow spectrum is replaced by a four peak structure. Instead, when the frequency is half of the atomic transition energy, we obtain an SHG spectrum with a superimposed Mollow splitting. The emergence of an SHG signal in a two-level system is
anticipated by the analysis of the spectrum of a three-level system, and further validated by 
a symmetry analysis of the coupled electron-photon states. 
\subsection{Model and method} \label{model_method}
Our two-level system interacting with an incident and a fluorescent light-field~\cite{Cini93,Cini95} is
described by the following Hamiltonian:
\begin{align}\label{Heq1}
 \hat{H}(t) = \hat{H}_a + \hat{H}_r + \hat{H}_i(t).
\end{align} 
We assume that the atom is occupied by a single spinless electron, so that
\begin{align}\label{Heq2}
 \hat{H}_a = \epsilon_1 \hat{c}_1^\dagger \hat{c}_1 + \epsilon_2 \hat{c}_2^\dagger \hat{c}_2 = \epsilon\hat{\sigma}_z.
\end{align} 
Here $\hat{c}_i$ destroys an electron in the orbital $|i\rangle$ with energy $\epsilon_i$, $\hat{\sigma}_z$ is the $z$-component Pauli operator, and $\epsilon = \epsilon_2 - \epsilon_1$. The free radiation modes are described by the Hamiltonian
\begin{align}\label{Heq3}
 \hat{H}_r = \omega_a \hat{a}^\dagger \hat{a} + \omega_b \hat{b}^\dagger \hat{b},
\end{align} 
where $\hat{a}$ annihilates a photon of the incident field with frequency $\omega_a$, which we assume to be in a coherent state defined by $\hat{a}|\alpha\rangle = \alpha|\alpha\rangle$. Similarly $\hat{b}$ annihilates a photon of the fluorescent field with frequency $\omega_b$. The light-matter interaction Hamiltonian is given by
\begin{align}\label{eq:ham_i}
 \hat{H}_i(t) &= \left[g_a(t)(\hat{a}^\dagger + \hat{a}) + g_b(t)(\hat{b}^\dagger + \hat{b})\right](\hat{c}_1^\dagger \hat{c}_2 + \hat{c}_2^\dagger \hat{c}_1) \nonumber \\
        &= \left[g_a(t)(\hat{a}^\dagger + \hat{a}) + g_b(t)(\hat{b}^\dagger + \hat{b})\right]\hat{\sigma}_x,
\end{align}
where $g_a(t)$ and $g_b(t)$ are the (time-dependent) couplings of the electron to the incident and fluorescent fields respectively, and can have any time-dependence. Also, $\hat{\sigma}_x$ is the $x$-component Pauli operator. In the following we consider the case $g_b(t) = g_b e^{-\Gamma t}$, which introduces a frequency independent, phenomenological damping of rate $\Gamma$. The use of $\Gamma$ takes into account in a qualitative way effects left out, e.g. non-radiative transitions and/or mode leakages in a cavity geometry (see also Appendix~\ref{App_A0_Lind}).

Concerning the role of spontaneous decay in our description, 
we note that in the stationary regime
(e.g. due to a steady photon pump) it is often legitimate 
to overlook this type of decay with respect to the stimulated one. 
Away from stationarity, other factors
come into play, depending on the situation: i) in the single-atom case
(and with Einstein's description of radiation-matter interaction as conceptual reference),
spontaneous decay does 
not induce a thermal bath; here, the coherence
of the overall optical response is not altered by neglecting such
decay (see e.g. \cite{KnightAllen}) ii) in the the many-atom case
(e.g the Dicke's regime as discussed in Sect.~\ref{sec:array}), 
thermal-bath effects and the significance of spontaneous decay 
are hindered by the collective effect of super-radiance \cite{Since}.

To describe the dynamics according to Eq.~(\ref{Heq1}), we use the exact configuration interaction method. In this way, the full wavefunction of the coupled atom-light system is represented in the basis \ket{i,n,m} $\equiv$ \ket{i}\ket{n}\ket{m}, with \ket{i} the state of the atom (where \ket{1} is the ground state and \ket{2} the excited state), \ket{n} a number state of the incident field and \ket{m} a number state of the fluorescent field. We start from the initial state $|\psi_0\rangle = |1,\alpha,0\rangle$
(here, the number state \ket{n} has been replaced by the coherent state \ket{\alpha} of the pump field),
and time-evolve it with the full Hamiltonian $H$ using the short iterated Lanczos technique \cite{ParkLight,Bostrom16} (see also Appendix \ref{App_A0_Lanc}).

We note here that the recursion method originally employed to
 study fluorescence in the stationary limit \cite{Cini93,Cini95} can be used for more general setups,
 e.g. to generate exact solutions with several electronic levels and several bosonic modes \cite{Cini88}. However,
its applicability is not immediate for genuinely
time-dependent Hamiltonians, and to address the transient response of a system. Hence, 
the need to proceed here with a real-time approach.

It has been shown \cite{Eberly} that to make contact with experimental time-resolved light spectra, the transition probability needs to be convolved with the resolution function of a Fabry-Perot spectrometer. Also, depending on the experiment performed on a quantum
system, a detection of $N$-photon correlations can be used \cite{DelValle}. 

However, here we employ a definition of the spectrum different from the one considered in \cite{Eberly,DelValle}. Namely, we consider the probability $\mathcal{P}$ that at least one photon (at a given frequency) is emitted. Such a spectrum would not be obtained by an interferometer, Fabry-Perot or otherwise, but rather by an apparatus including e.g. a prism and a photomultiplier. In other words, this we would correspond to measure particle-like photons, rather than waves, and probing simultaneously the atomic state.

Accordingly, our observable of main interest is the probability to find $m$ photons in the fluorescent field of frequency $\omega_b$ at time $t$, given by
\begin{align}\label{eq:prob_n}
\!\!\mathcal{P}_m(t,\omega_b) = \sum_{ni} |\langle i,n,m|\mathcal{T}\left[e^{-i\int_0^t \hat{H}(t')dt'}\right]|1,\alpha,0 \rangle|^2,
\end{align}
where the dependence on $\omega_b$ in RHS is implicitly contained in $\hat{H}(t)$. Since $g_b(t)$ has an exponential decay, $\mathcal{P}$ will be independent of $t$ in the long time limit. In the following we will focus on the quantity $\mathcal{P} = \sum_{m>0}\mathcal{P}_m$, giving the probability that at least one fluorescent photon has been emitted. 

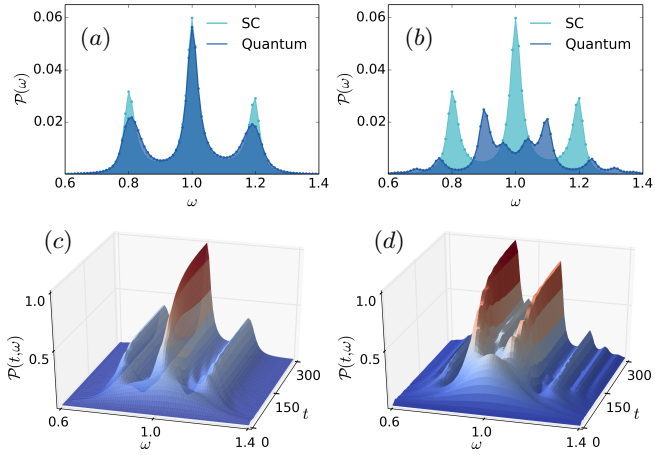
\begin{figure}
  \begin{tikzpicture}
  \node at (-2.3, 2) {\includegraphics[width=0.5\columnwidth]{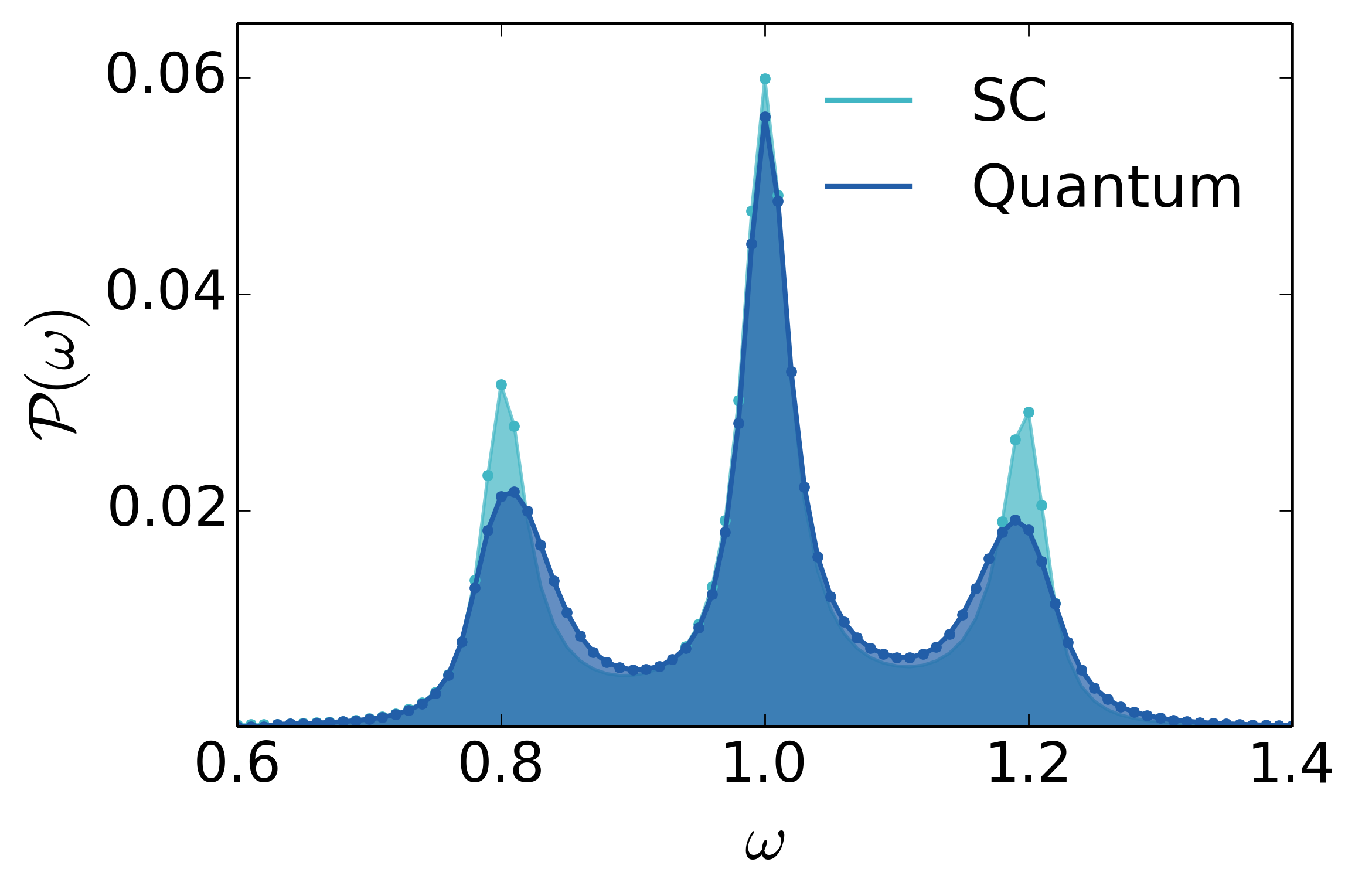}};
  \node at ( 2.0, 2) {\includegraphics[width=0.5\columnwidth]{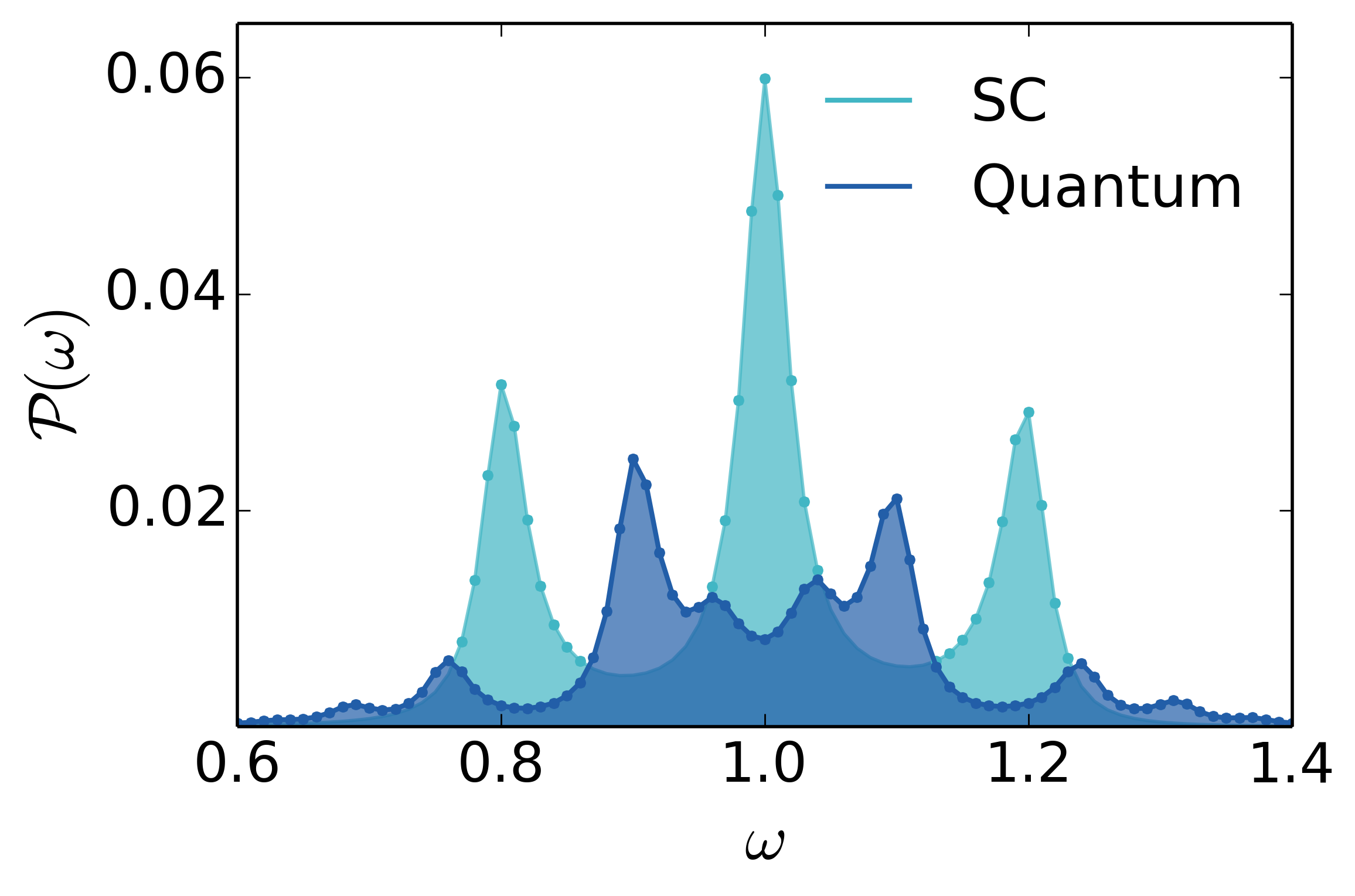}};
  \node at (-2.4,-1.1) {\includegraphics[width=0.48\columnwidth]{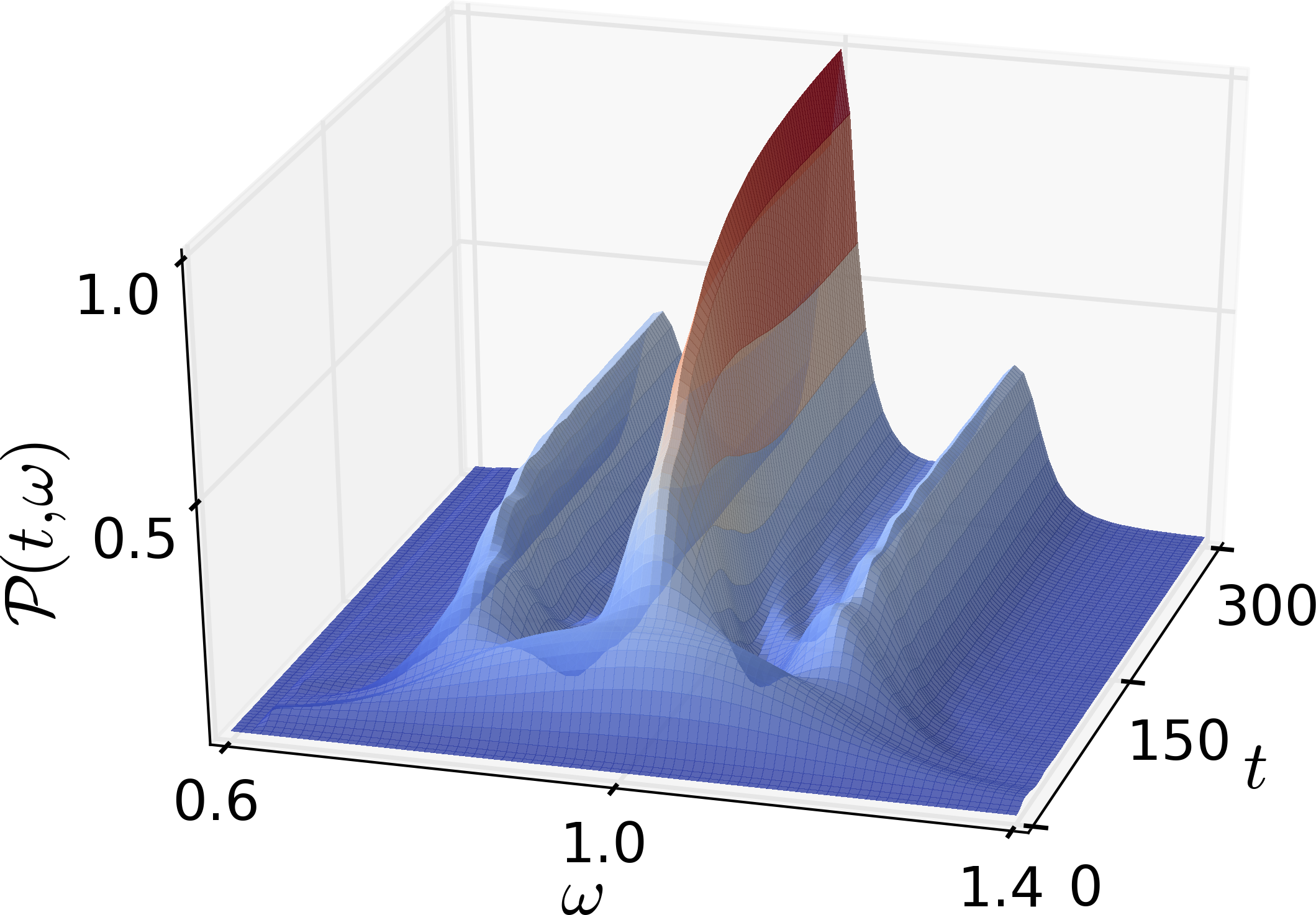}};
  \node at ( 2.0,-1.1) {\includegraphics[width=0.48\columnwidth]{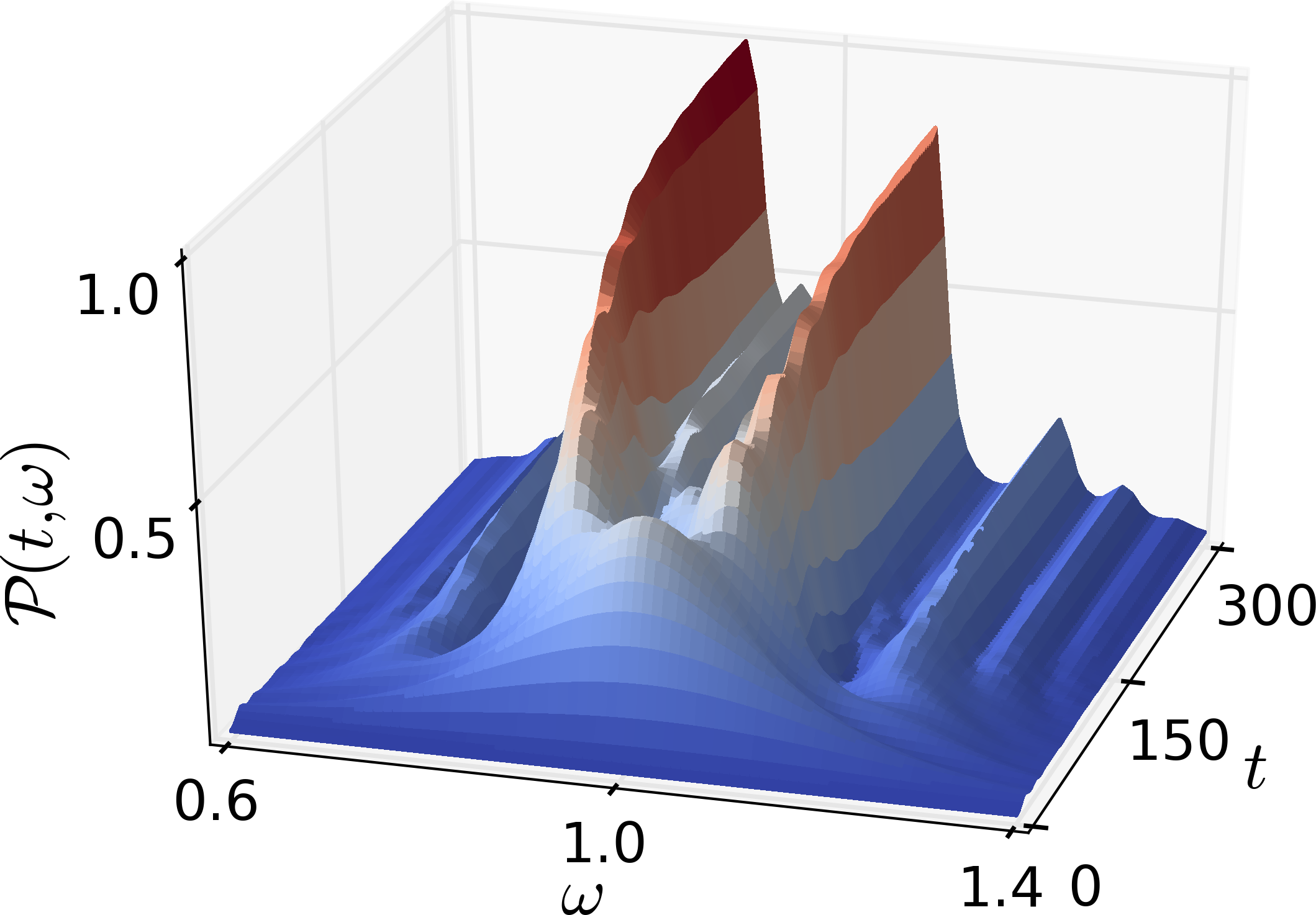}};  
  \node[black] at (-3.3,2.9) {$(a)$};
  \node[black] at ( 1.0,2.9) {$(b)$};
  \node[black] at (-3.8,0.2) {$(c)$};
  \node[black] at ( 0.6,0.2) {$(d)$};
 \end{tikzpicture}
 \caption{Fluorescence spectra of a two-level system interacting with a coherent field with frequency $\omega_a = \epsilon$ and an average number of photons $\alpha^2 = 25$ [panels $(a)$ and $(c)$] and $\alpha^2 = 1$ [panels $(b)$ and $(d)$]. The light-matter coupling $g_a$ is chosen so that $g_a\alpha = 0.1$ in both cases, giving $g_a = 0.02$ and $g_a = 0.1$ respectively. The coupling to the fluorescent field is $g_b = 0.01$. Panels $(a)$ and $(b)$ show the asymptotic spectrum $\mathcal{P}(\omega) = \mathcal{P}(t\to\infty,\omega)$ obtained using either a quantized coherent field or taking the semi-classical limit of Eq.~(\ref{eq:semi}). Panels $(c)$ and $(d)$ show the corresponding time-dependent spectrum using a quantized coherent field. The units of energy and time are respectively given by $\epsilon = \epsilon_2 - \epsilon_1$ and $\hbar/\epsilon$.}
 \label{fig:mollow}
\end{figure}

The semi-classical limit of our model is obtained by taking $\alpha \to \infty$ and $g \to 0$, while keeping $g\alpha$ constant. In this limit $\hat{H}_r = \omega \hat{b}^\dagger \hat{b} \nonumber$ and the interaction Hamiltonian is 
\begin{align}\label{eq:semi}
\!\!\!\!\!\!\hat{H}_i^{sc}(t) = \left[2g_a\alpha\cos(\omega_a t) + g_b(t)(\hat{b}^\dagger + \hat{b})\right](\hat{c}_1^\dagger \hat{c}_2 + \hat{c}_2^\dagger \hat{c}_1).
\end{align}
The $t \to \infty$ limit of the model has been studied in previous works~\cite{Cini93,Cini95}. In that case, the Hamiltonian was split as $\hat{H}(t) = \hat{H}'_0(t) + \hat{H}'(t)$ with $\hat{H}'(t) = g_b(t)(\hat{c}_1^\dagger \hat{c}_2 + \hat{c}_2^\dagger \hat{c}_1)(\hat{b}^\dagger + \hat{b})$, and the probability $\mathcal{P}_1$ was evaluated under the assumption that $\hat{H}'(t)$ acts only once during the time evolution. Physically this corresponds to a first-order treatment of the fluorescent field, in contrast to the exact numerical solution of the present work, which retains 
 the effects of the interaction at all orders.

%%%%%%%%%%%%%%%%%%%%%%%%%%%%%%%%%%%%%%%%%%%%%%%%%%%%%%%%%%%%
%%%%%%%%%%%%%%%%%%%%%%%%%%%%%%%%%%%%%%%%%%%%%%%%%%%%%%%%%%%%

\subsection{Two-level system and Mollow spectra} \label{MollowSec}
In the following we let the transition energy $\epsilon = 1$ define our unit of energy, and further fix the parameters $g_b = 0.01$ and $\Gamma = 0.02$. We start by studying the fluorescence spectrum for $\omega_a = 1$ as a function of $\alpha$ and $g_a$, keeping the product $g_a\alpha$ fixed. The results are displayed in Fig.~\ref{fig:mollow}, and show both the asymptotic spectrum (as $t \to \infty$) and the explicit time-evolution. For $\alpha = 5$ and $g_a = 0.02$ we see the well-known Mollow spectrum, that is qualitatively reproduced by the semi-classical approximation. Keeping the product $g_a\alpha$ fixed and taking $\alpha = 1$ and $g_a = 0.1$ the semi-classical result is unchanged, while the full quantum treatment gives a fluorescence spectrum with four peaks and additional substructure.

To understand the spectra we consider the states \ket{i,n} with the atom in state $i$ and $n$ photons in the incident field. For $\omega_a = \epsilon$ the levels \ket{1,n} and \ket{2,n-1} are degenerate, and mixed by the light-matter interaction. Diagonalizing the Hamiltonian in this subspace (i.e. neglecting the counter-rotating terms), we find the energies
\begin{align}
\epsilon_{\pm,n} = \frac{\epsilon_1 + \epsilon_2}{2} + (n-1/2)\omega_a  \pm g_a\sqrt{n}.
\end{align}
For a coherent state with large $\alpha$ the Poisson distribution is sharply peaked around $\alpha^2$, so that $n \sim \alpha^2$ and the energies differences between successive $n$ are
\begin{align}
\epsilon_{\pm,n+1} - \epsilon_{\mp,n} &= \omega_a \pm g_a\sqrt{n+1} \pm g_a\sqrt{n} \xrightarrow{n\rightarrow |\alpha|^2} \pm 2g_a\alpha \nonumber \\
\epsilon_{\pm,n+1} - \epsilon_{\pm,n} &= \omega_a \pm g_a\sqrt{n+1} \mp g_a\sqrt{n} \xrightarrow{n\rightarrow |\alpha|^2} 0.
\end{align}
In this limit the energy splittings are independent of $n$, and transitions between successive levels give a three-peaked structure. For small $\alpha$, this is no longer the case, since the splitting $\epsilon_{+,n} - \epsilon_{-,n} = 2g_a\sqrt{n}$. The non-uniformity in the level spacing is a clear sign of the quantum nature of the light (the low photon number limit), and is e.g. responsible for photon blockade effect~\cite{Yang05,Ridolfo12}. In the present context, it gives rise to the additional features observed in the fluorescence spectrum of Fig.~\ref{fig:mollow}.

%%%%%%%%%%%%%%%%%%%%%%%%%%%%%%%%%%%%%%%%%%%%%%%%%%%%%%%%%%%%
%%%%%%%%%%%%%%%%%%%%%%%%%%%%%%%%%%%%%%%%%%%%%%%%%%%%%%%%%%%%
%%%%%%%%%%%%%%%%%%%%%%%%%%%%%%%%%%%%%%%%%%%%%%%%%%%%%%%%%%%%
%%%%%%%%%%%%%%%%%%%%%%%%%%%%%%%%%%%%%%%%%%%%%%%%%%%%%%%%%%%%

\subsection{Prelude to SHG in a two level system: frequency doubling in an ordinary three-level system}\label{3levs}

\begin{figure}
  \begin{tikzpicture}[
      scale=0.6,
      level/.style={thick},
      virtual/.style={thick,densely dashed,color=orange},
      trans/.style={thick,->,shorten >=2pt,shorten <=2pt,>=stealth},
      emit/.style={thick,<-,shorten >=2pt,shorten <=2pt,>=stealth,color=blue!60!white},
      away/.style={thick,->,shorten >=2pt,shorten <=2pt,>=stealth,color=orange},
    ]
    \draw[level] (2.5cm,-2em) -- (0cm,-2em) node[left] {\ket{1}};
    \draw[level] (2.5cm,12em) -- (0cm,12em) node[left] {\ket{3}};
    \draw[virtual] (2.5cm,5em) -- (0cm,5em) node[left] {\ket{2}};
    \draw[away] (2.5cm,5em) .. controls +(right:0.5cm) .. node[above,sloped] {} (3cm,14em);
    \draw[trans] (0.5cm,-2em) -- (0.5cm,5em) node[midway,left] {};
    \draw[trans] (0.5cm,5em) -- (0.5cm,12em) node[midway,left] {\Om{a}};
    \draw[emit] (2cm,-2em) -- (2cm,12em) node[midway,right,yshift=0.55cm] {\Om{b}};
    
    \node at (8.4, 1.8) {\includegraphics[width=0.62\columnwidth]{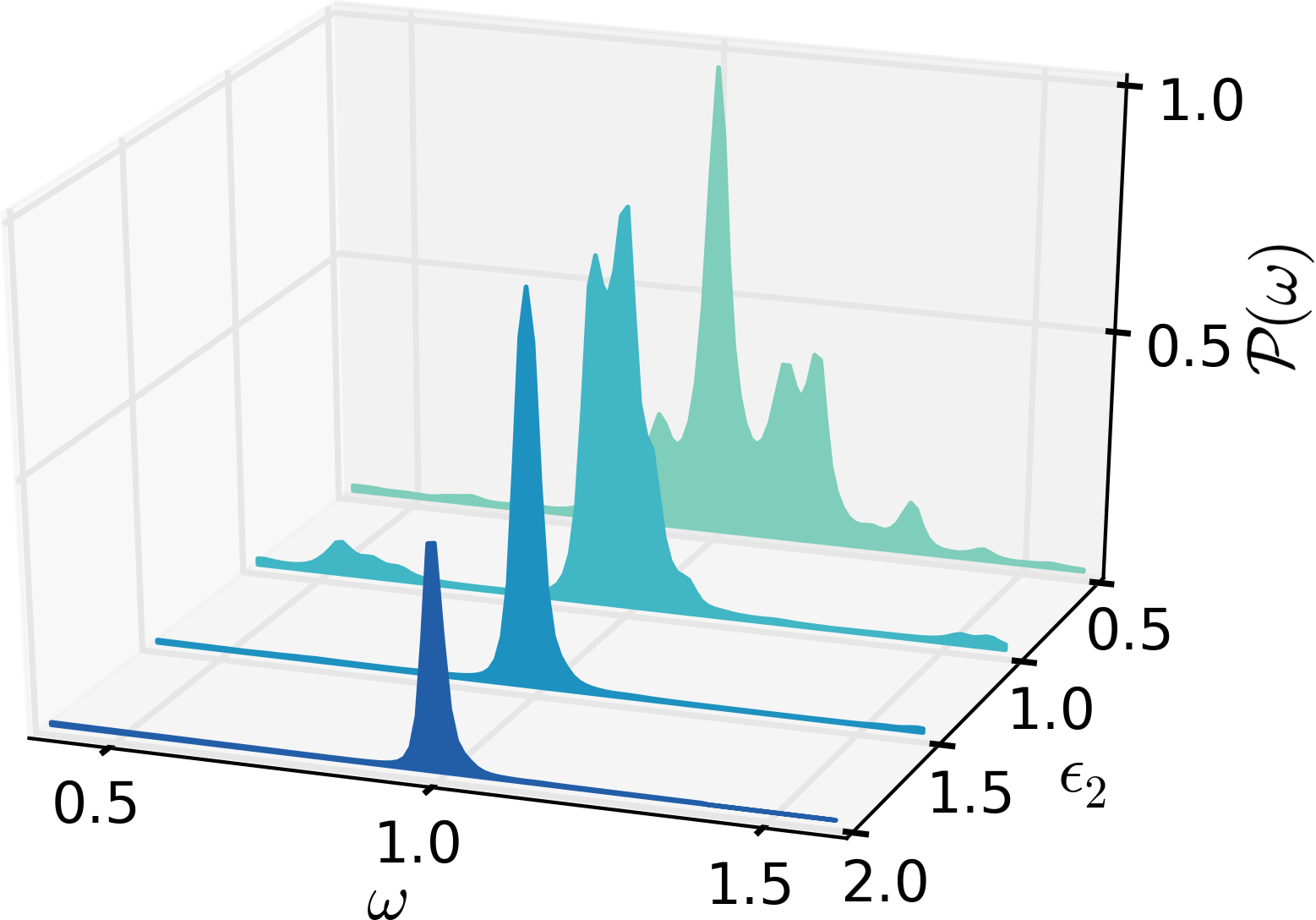}};
    \node[black] at (4.4,4.8) {$(a)$};

  \end{tikzpicture}

  \begin{tikzpicture}[
      scale=0.6,
      level/.style={thick},
      virtual/.style={thick,densely dashed,color=orange},
      trans/.style={thick,->,shorten >=2pt,shorten <=2pt,>=stealth},
      emit/.style={thick,<-,shorten >=2pt,shorten <=2pt,>=stealth,color=blue!60!white},
      away/.style={thick,->,shorten >=2pt,shorten <=2pt,>=stealth,color=orange},
    ]
    \draw[level] (2.5cm,-2em) -- (0cm,-2em) node[left] {\ket{1}};
    \draw[level] (2.5cm,12em) -- (0cm,12em) node[left] {\ket{3}};
    \draw[virtual] (2.5cm,5em) -- (0cm,5em) node[left] {\ket{2}};
    \draw[away] (2.5cm,5em) .. controls +(right:0.5cm) .. node[above,sloped] {} (3cm,14em);
    \draw[trans] (0.5cm,-2em) -- (0.5cm,5em) node[midway,left] {\Om{a}};
    \draw[trans] (0.5cm,5em) -- (0.5cm,12em) node[midway,left] {};
    \draw[trans] (1cm,-2em) -- (1cm,12em) node[midway,xshift=0.5cm] {};
    \draw[emit] (2cm,-2em) -- (2cm,12em) node[midway,right,yshift=0.55cm] {\Om{b}};
    
    \node at (8.4, 1.8) {\includegraphics[width=0.62\columnwidth]{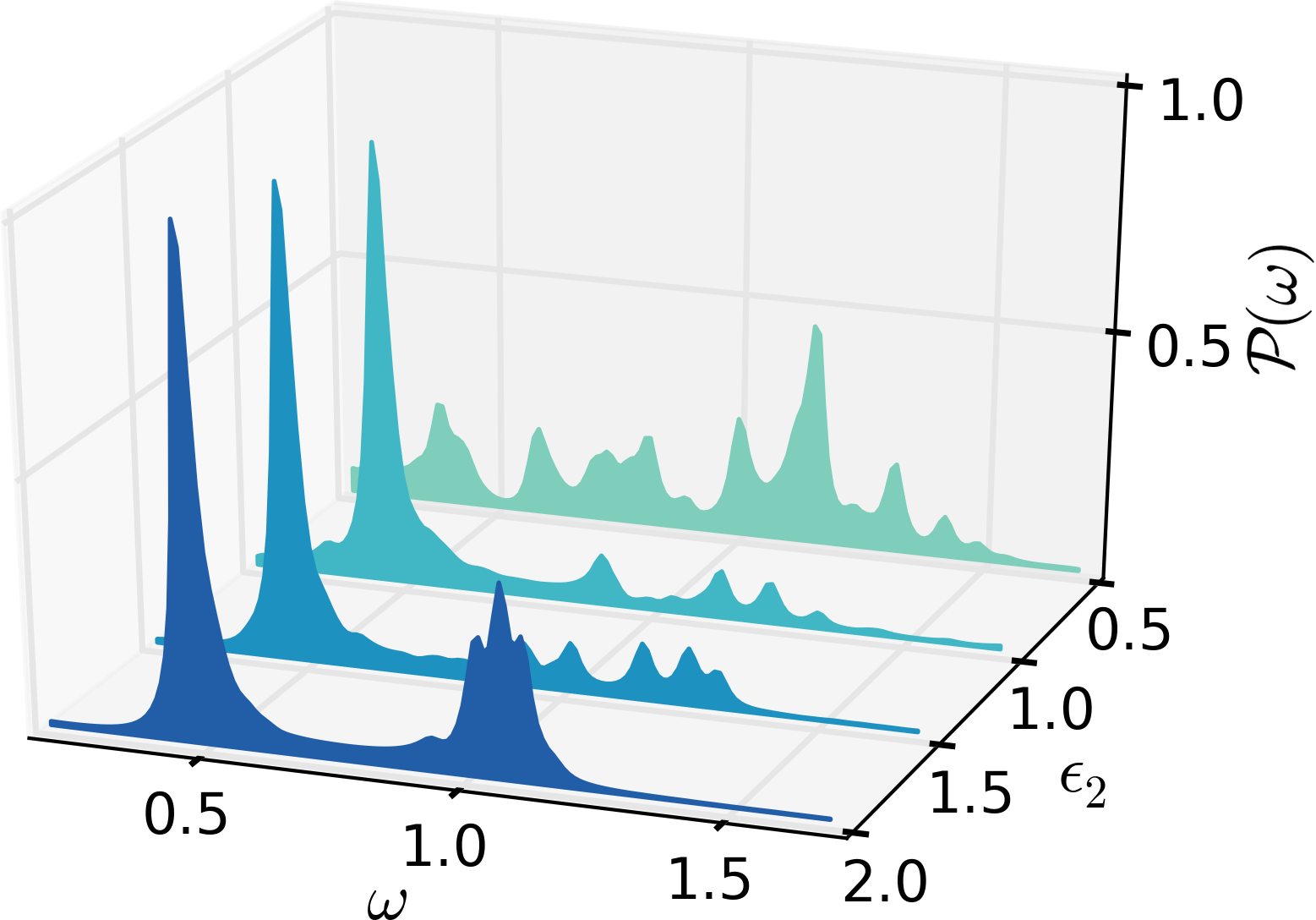}};
    \node[black] at (4.4,4.8) {$(b)$};
  \end{tikzpicture}
 \caption{Fluorescence spectra for a three-level system interacting with a coherent field with a frequency $\omega_a = 1$ and an average number of photons $\alpha^2 = 1$. The coupling between the system and the coherent field is given by $f=g_a = 0.1$, while the coupling to the fluorescent field is $g_b = 0.01$. The energy levels of the atom are $\epsilon_1 = 0$ and $\epsilon_3 = 1$, with $\epsilon_2$ taking the values $0.5$, $1$, $1.5$ and $2$. In panel $(a)$ the coherent field is coupled to the transitions $1 \leftrightarrow 2$ and $2 \leftrightarrow 3$, while in panel $(b)$ the coherent field in addition couples to transition $1 \leftrightarrow 3$. The couplings in the system are indicated schematically in the panels to the left. For ease of reading, the $z$-axis has been multiplied by 100. The unit of energy is given by $\epsilon = \epsilon_3 - \epsilon_1$.}
 \label{fig:three_level}
\end{figure}
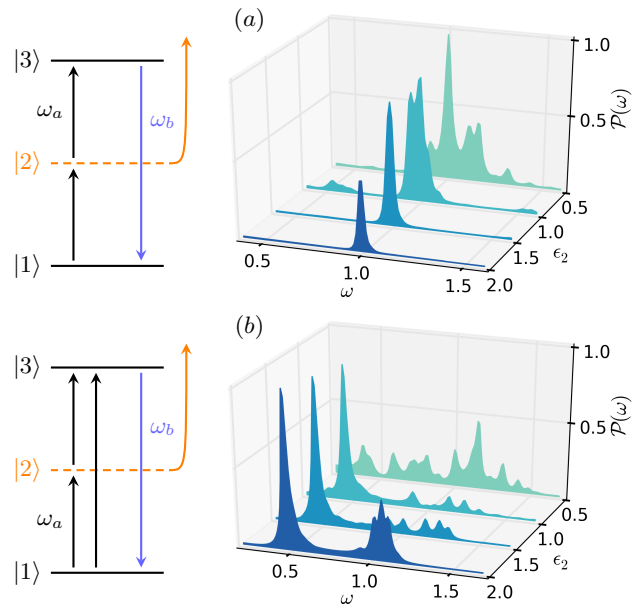

Before addressing SHG in a two-level system, we make a detour into the more familiar theory of SHG in three-level systems. Let the Hamiltonian be $\hat{H}(t) = \hat{H}_e + \hat{H}_r + \hat{H}_i(t)$, with $\hat{H}_r$ as in previous sections, and the electronic Hamiltonian be
\begin{align}\label{Hs3}
\hat{H}_e = \epsilon_1 \hat{c}_1^\dagger \hat{c}_1 + \epsilon_2 \hat{c}_2^\dagger \hat{c}_2 + \epsilon_3 \hat{c}_3^\dagger \hat{c}_3.
\end{align}
For the light-matter interaction we consider two scenarios, illustrated in Fig.~\ref{fig:three_level}. In the first the incident field couples to the transitions $|1\rangle \leftrightarrow |2\rangle$ and $|2\rangle \leftrightarrow |3\rangle$, and the fluorescent field to the transition $|1\rangle \leftrightarrow |3\rangle$. The interaction Hamiltonian is then
\begin{align}\label{Hi_a}
\hat{H}_i^{(1)}(t) &= f(t)(\hat{a}^\dagger + \hat{a})(\hat{c}_1^\dagger \hat{c}_2 + \hat{c}_2^\dagger \hat{c}_3 + H.c.) \nonumber \\
& + g_b(t)(\hat{b}^\dagger + \hat{b})(\hat{c}_1^\dagger \hat{c}_3 + \hat{c}_3^\dagger \hat{c}_1).
\end{align}
In the second case the incident field also couples to the transition $|1\rangle \leftrightarrow |3\rangle$, with a strength $g(t)$, and the Hamiltonian is
\begin{align} \label{Hi_b}
\hat{H}_i^{(2)}(t) &= \hat{H}_i^{(1)}(t) + g_a(t)(\hat{a}^\dagger + \hat{a})(\hat{c}_1^\dagger \hat{c}_3 + \hat{c}_3^\dagger \hat{c}_1).
\end{align}

If the levels \ket{1} and \ket{3} are of definite and different symmetry, while the level \ket{2} is assumed to be of mixed symmetry, both these models allow SHG using a perturbative treatment. However, in the case of a parity-invariant electronic Hamiltonian, where all electronic states have definite parity, both models forbid SHG in a perturbative approximation. The reason is that there is no way to arrange the parities such that both the excitation and emission steps are allowed: the parities $\pi_1$ and $\pi_3$ of levels $|1\rangle$ and $|3\rangle$ need to be different (for the fluorescent transition to be allowed), while the parity $\pi_2$ of level $|2\rangle$ needs to be different both from $\pi_1$ and $\pi_3$ (in order for the exciting transitions to be allowed).

We now study the fluorescence spectrum in the non-perturbative limit. We take $\epsilon_2 - \epsilon_1 = 0.5$ and $\epsilon_3 - \epsilon_1 = 1$, and let the incident field be resonant with the transition energies ($\omega_a = 0.5$). We choose the couplings as $f(t) = g_a(t) = 0.1$, take $\alpha = 1$, and as before let $g_b = 0.01$ and $\Gamma = 0.02$. The results are reported in Fig.~\ref{fig:three_level}. For the Hamiltonian $\hat{H}_i^{(1)}$ we see a broadened SHG peak centered around $\omega_b = 1$, while for $\hat{H}_i^{(2)}$ the spectrum has two contributions corresponding to Rayleigh scattering and SHG centered at $\omega_b = 0.5$ and $\omega_b = 1$ respectively. As expected, both models predict a non-zero SHG signal.

A connection with a simpler two-level system can be made by taking the limit $\epsilon_2 \to \infty$, illustrated here by considering $\epsilon_2 - \epsilon_1 = 1$, $1.5$ and $2$ (see Fig.~\ref{fig:three_level}). With the interaction $\hat{H}_i^{(1)}$ the fluorescence signal narrows around $\omega = 1$ as $\epsilon_2 - \epsilon_1$ is increased, since the coupling between degenerate states causing the broadening decreases as the levels are energetically separated (cf. the discussion of the Mollow spectrum). For even larger values of $\epsilon_2 - \epsilon_1$ (not shown), the peak tends to zero since excitation of the atom becomes increasingly unlikely. With the interaction $\hat{H}_i^{(2)}$ both the Rayleigh and SHG peaks narrow as $\epsilon_2 - \epsilon_1$ is made large. However, due to the presence of the coupling between the incident field and the transition $|1\rangle \leftrightarrow |3\rangle$, a finite fluorescence signal remains even for $\epsilon_2 -\epsilon_1 \to \infty$, i.e 
when effectively the Hamiltonian $\hat{H}_i^{(2)}$ collapses onto that of the two-level system. The specific shape
of the SHG profile is discussed in the next Section.

As a final consideration, we note that in quantum optics it is sometimes useful to perform an adiabatic elimination of intermediate levels, to end up with a reduced-space effective Hamiltonian~\cite{Alsing87,Gou89}. In appendix \ref{App_A}, such a reduction is performed to map the three-level system of
Fig.~\ref{fig:three_level} onto a two-level one~\cite{Brion07}, and to see if SHG can be {\it exactly} described in a two-level system in the limit $\epsilon_2 \to \infty$ of a three level system. The results in the appendix show that this is not the case. In fact, going beyond the RWA and neglecting off-resonant (but dipole-allowed) 
transitions in the system are two elements which play distinct roles. Specifically, the occurrence of the
Mollow structure is not only due to the removal of the RWA, but also due to the
explicit inclusion of the virtual state in the total Hamiltonian. Preventing transitions to this state
removes the Mollow structure, also without RWA and for strong light-matter coupling. 

\subsection{Two-levels again: Second harmonic generation}\label{SHGagain}

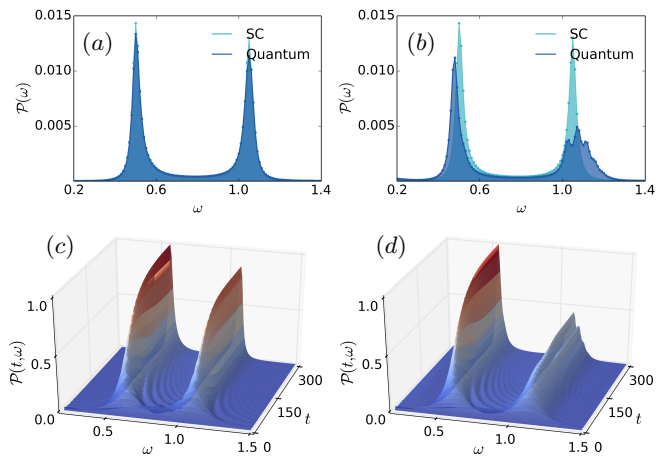
\begin{figure}
  \begin{tikzpicture}
  \node at (-2.3, 2) {\includegraphics[width=0.5\columnwidth]{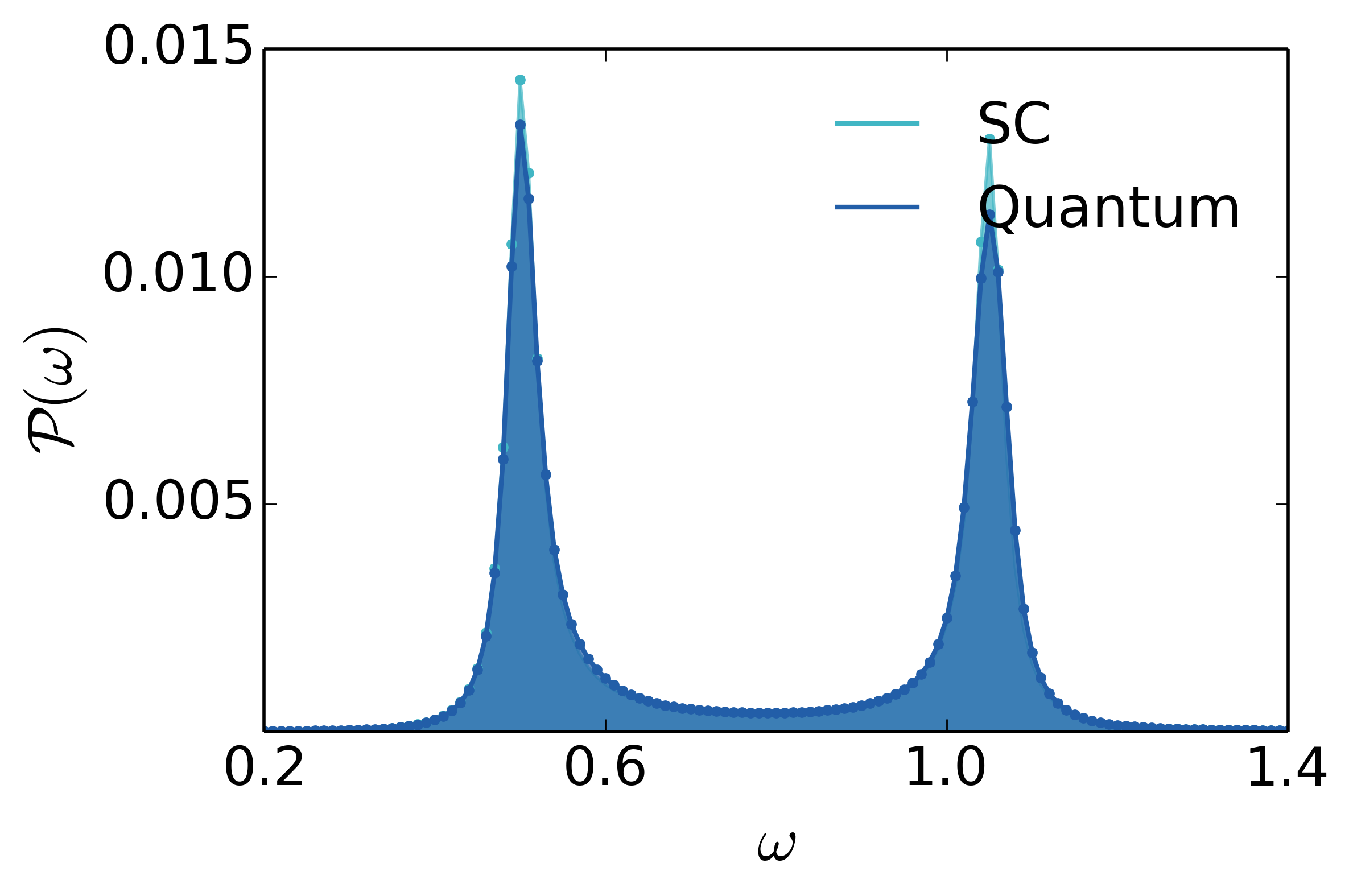}};
  \node at ( 2.0, 2) {\includegraphics[width=0.5\columnwidth]{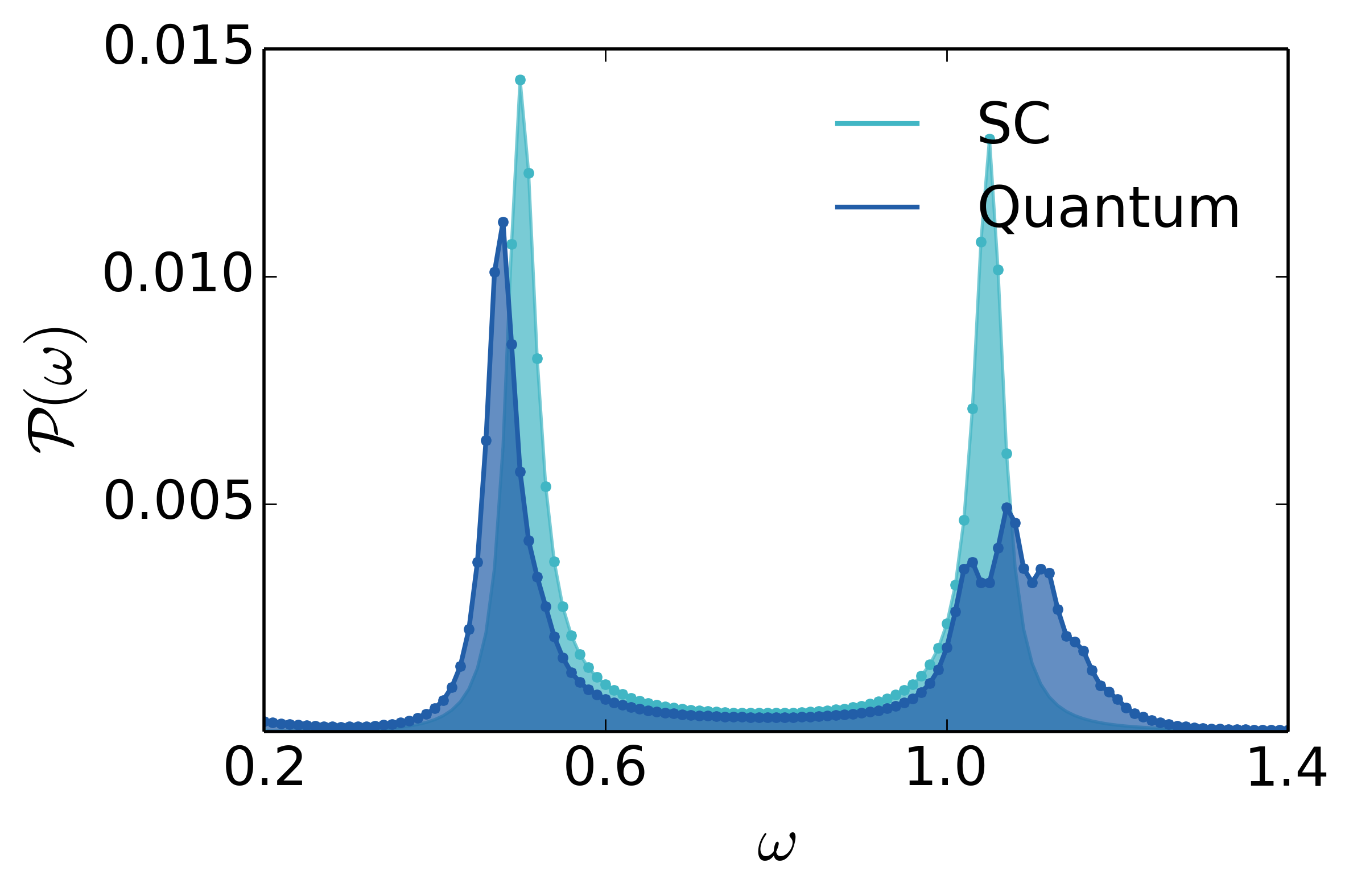}};
  \node at (-2.4,-1.1) {\includegraphics[width=0.48\columnwidth]{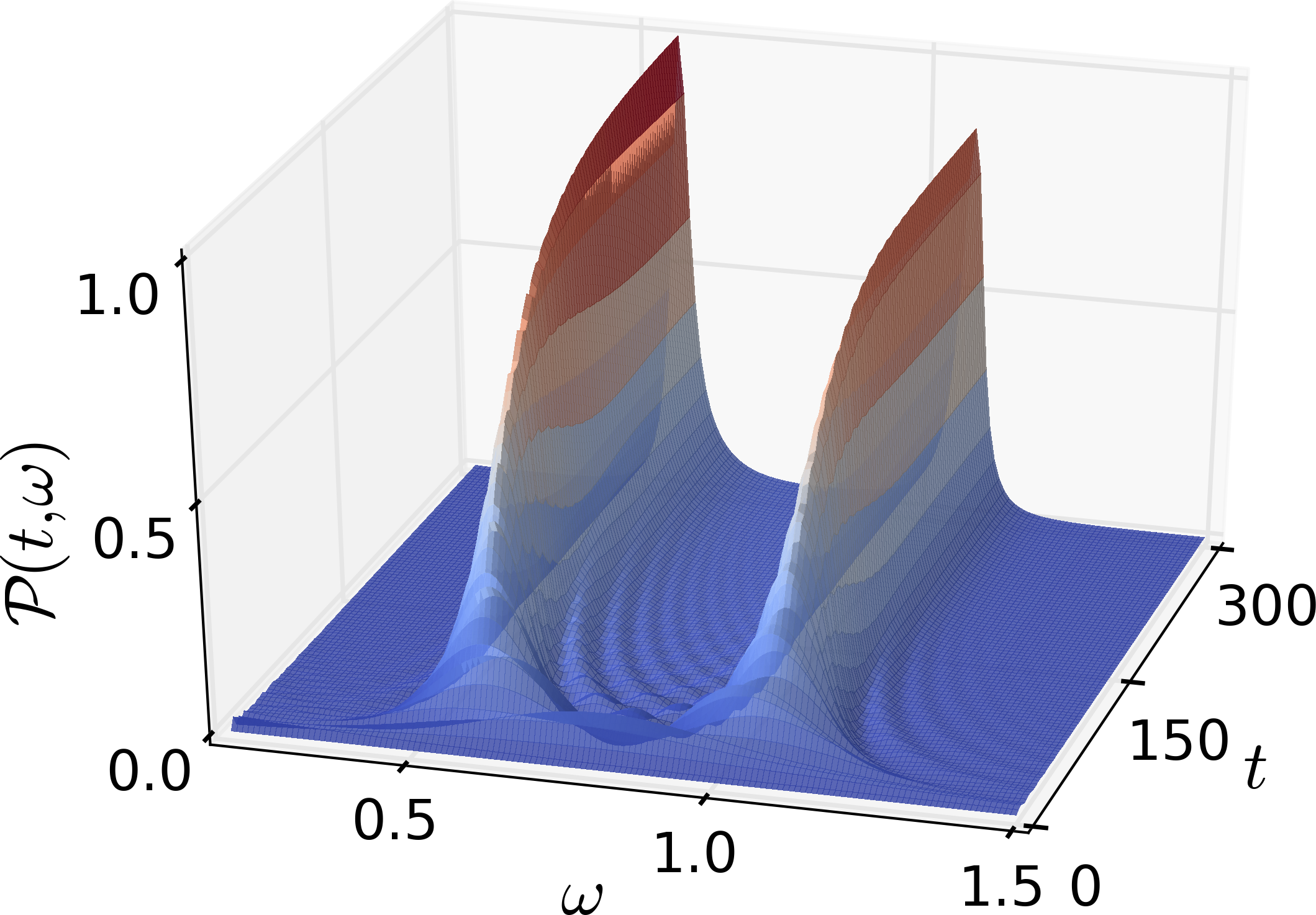}};
  \node at ( 2.0,-1.1) {\includegraphics[width=0.48\columnwidth]{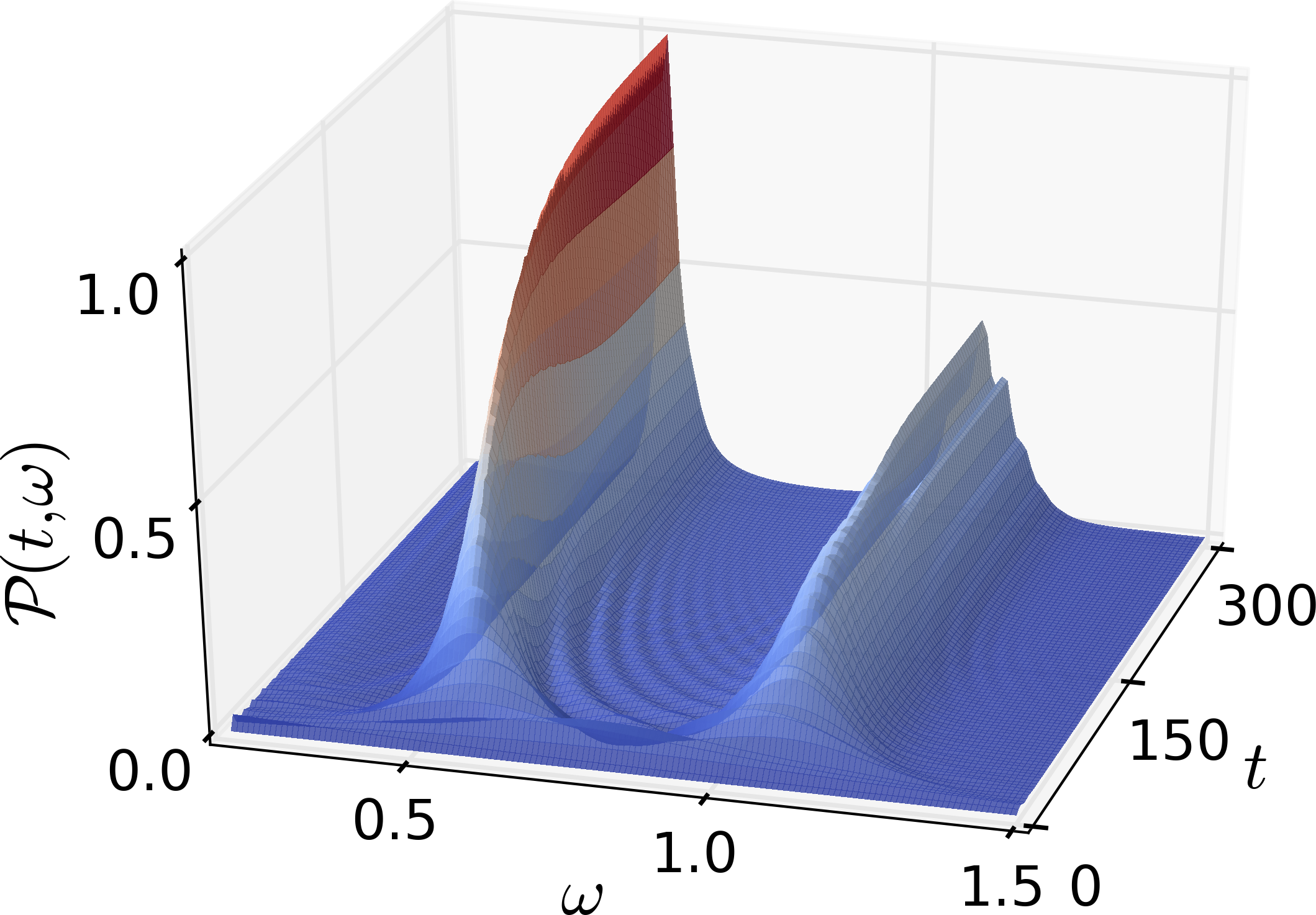}};  
  \node[black] at (-3.3,2.9) {$(a)$};
  \node[black] at ( 1.0,2.9) {$(b)$};
  \node[black] at (-3.8,0.2) {$(c)$};
  \node[black] at ( 0.6,0.2) {$(d)$};
 \end{tikzpicture}
 \caption{Fluorescence spectra of a two-level system interacting with a coherent field with a frequency $\omega_a = \epsilon/2$ and an average number of photons $\alpha^2 = 25$ [panels $(a)$ and $(c)$] and $\alpha^2 = 1$ [panels $(b)$ and $(d)$]. The light-matter coupling $g_a$ is chosen so that $g_a\alpha = 0.1$ in both cases, giving $g_a = 0.02$ and $g_a = 0.1$ respectively. The coupling to the fluorescent field is $g_b = 0.01$. Panels $(a)$ and $(b)$ show the asymptotic spectrum $\mathcal{P}(\omega) = \mathcal{P}(t\to\infty,\omega)$ obtained using either a quantized coherent field or taking the semi-classical limit of Eq.~(\ref{eq:semi}). Panels $(c)$ and $(d)$ show the corresponding time-dependent spectrum using a quantized coherent field. The units of energy and time are respectively given by $\epsilon = \epsilon_2 - \epsilon_1$ and $\hbar/\epsilon$.}
 \label{fig:shg}
\end{figure}

%%%%%%%%%%%%%%%%%%%%%%%%%%%%%%%%%%%%%%%%%%%%%%%%%%%%%%%%%%%%
%%%%%%%%%%%%%%%%%%%%%%%%%%%%%%%%%%%%%%%%%%%%%%%%%%%%%%%%%%%%

With the information gained so far, we finally move to study the fluorescence spectrum
in a two level system as described in Eqs.~(\ref{Heq1}-\ref{eq:ham_i}) \cite{confuse}. Since a common theme of this work is fluorescence in the SHG regime, we consider the case of $\omega_a = \epsilon/2$, with all other parameters as in Sections.~\ref{model_method}, \ref{MollowSec}. 

\begin{figure}
  \begin{tikzpicture}
  \node at (0,0) {\includegraphics[width=\columnwidth]{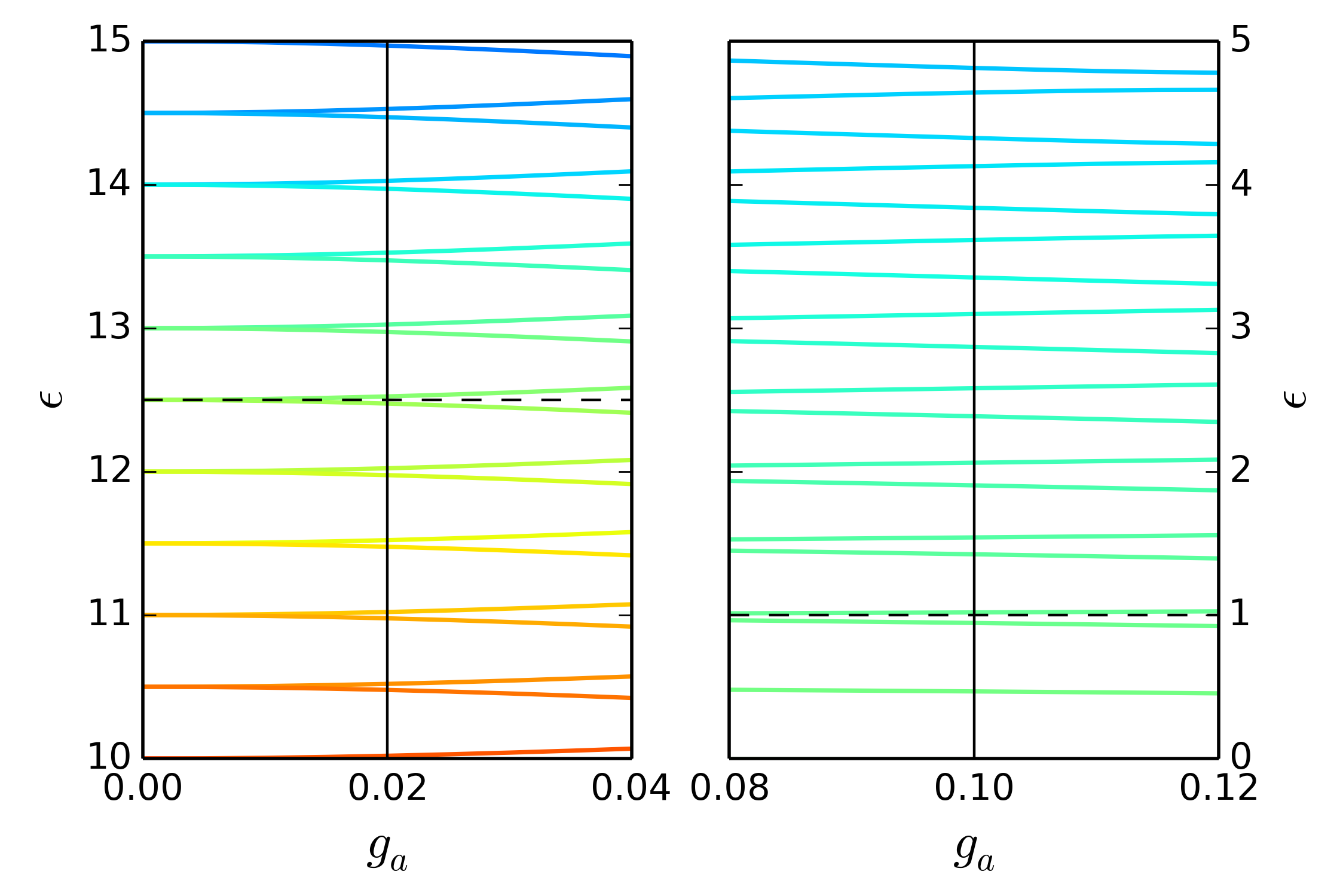}}; 
  \node[black] at (-3.0,2.2) {$(a)$};
  \node[black] at ( 0.8,2.2) {$(b)$};
 \end{tikzpicture}
 \caption{Energy levels as a function of coupling strength for a two-level system interacting with a single light field. In both panels, the solid black line indicates the coupling $g_a$ and the dashed line the average energy $\epsilon_{av} = \omega_a \alpha^2$ of the setups described in the main text. Panel $(a)$ shows the energy levels in a range around the coupling $g_a = 0.02$ and energy $\epsilon_{av}$ for $\omega_a = 0.5$ and $\alpha = 25$, while panel $(b)$ shows the energy levels in a range around the coupling $g_a = 0.1$ and energy $\epsilon_{av}$ for $\omega_a = 0.5$ and $\alpha = 1$. The unit of energy is given by $\epsilon = \epsilon_2 - \epsilon_1$.}
 \label{fig:levels}
\end{figure}

In Fig.~\ref{fig:shg} we show results for $\alpha = 1$ and $\alpha = 5$, while keeping $g_a\alpha = 0.1$ fixed. For a small light-matter coupling the semi-classical calculation is seen to be in good agreement with the quantum treatment, while for large coupling the results differ by the presence of a superimposed Mollow spectrum on the second harmonic peak. This additional feature can be understood similarly to the Mollow spectrum of Fig.~\ref{fig:mollow}, as due to an energy level splitting depending on the coupling strength $g_a$ and photon number $\alpha^2$. The energy levels are shown in Fig.~\ref{fig:levels}, and we see that for small $g_a$ and large $\alpha$ the splittings are small and uniform (between successive levels), while for large $g_a$ and small $\alpha$ the splittings are large and non-uniform. We note that for small $g_a$, in contrast to the Mollow spectrum discussed above, the level splitting is expected to be $\sim g_a^2$ to leading order, since the coupling between the levels is of second order in the interaction.
To summarize, i) the ability of the quantum treatment to discriminate between photon number and coupling strength (they always appear via their product in the semiclassical treatment) and ii) the photon fluctuations at low photon number, are likely the reasons for the Mollow structure in the quantum SHG signal. This is in line with the evidence provided by the three-level system results in Fig.~\ref{fig:three_level} where, depending on the closure of the ``three-level triangle'' via the pump field, the Mollow structure in the SHG peak is observed or not.\

For the results just presented the photon energy $\hbar\omega_a$ of the coherent field is commensurate with the atomic transition energy, i.e. $\hbar\omega_a = \epsilon/n$. Although our approach is completely general and we can in principle study any frequency, a systematic scan of the parameters is outside the scope of the of the present work. 
However, we mention that according to additional calculations (not shown) the spectra for frequencies $\hbar\omega_a/\epsilon \approx 0.3 - 0.8$  resemble the second harmonic spectrum above, although with the position of the Rayleigh peak displaced to $\hbar\omega_b \approx \hbar\omega_a$. Instead, for frequencies $\hbar\omega_a/\epsilon \approx 0.9 - 1.1$ the Rayleigh and harmonic peaks blend together to form the four peak Mollow structure discussed above.

{\it Parity conservation.-}
Although borne by the exact numerical results of Sect.~\ref{SHGagain} (and also supported by the connection
to three-level physics as discussed in Sect.~\ref{3levs}), the occurrence of a SHG in a two-level system might  remain at some extent counter-intuitive. For second harmonic generation (SHG) to occur, two photons are needed to excite the atom. This requires atomic levels of equal parity, since in the low intensity limit where the absorption process is well described by perturbation theory, the two-photon absorption probability otherwise vanishes. However, the emission of a double frequency photon requires the atomic levels to have opposite parity. This apparent paradox however vanishes at stronger light-matter coupling, where a more appropriate description of the system is in terms of dressed atomic levels. In this regime the electronic states are mixed by the coupling to the light field, and no longer have definite parity.

To understand how SHG can happen in a two-level system, we look at the parity of the eigenstates of $\hat{H}$. A parity operator can be defined through~\cite{He12}
\begin{align}
 \hat{\Pi} = (\hat{n}_1 - \hat{n}_2)e^{i\pi \hat{n}_a}e^{i\pi \hat{n}_b},
\end{align}
where $\hat{n}_a$ and $\hat{n}_b$ are the photon number operators of the incident and fluorescent fields respectively. It is straightforward to show that $[\hat{H},\hat{\Pi}] = 0$, from which it follows that the eigenstates of $\hat{H}$ can be classified according to the eigenvalues of $\hat{\Pi}$. The general structure of the eigenstates with the electron and the incident field coupled is~\cite{He12}
\begin{align}
 |\Psi_e^k\rangle &= \sum_{n} c_{2n}^k|1,2n\rangle + \sum_{n} c_{2n+1}^k|2,2n+1\rangle \\
 |\Psi_o^k\rangle &= \sum_{n} c_{2n+1}^k|1,2n+1\rangle + \sum_{n} c_{2n}^k|2,2n\rangle,
\end{align}
where $e$ and $o$ denote even- and odd-parity states respectively. We see that an eigenstate with a well-defined parity in the coupled system has in general an undefined parity in the electronic and photonic subspaces. Therefore an argument against SHG in the two-level system, which relies on the conservation of only electronic parity is not applicable.

Consider now the evolution of a system starting from the initial state $|\Psi\rangle = |1,\alpha\rangle$. Since the coherent state is a superposition of number states, it explicitly breaks parity symmetry. We therefore expect the time-evolution to induce transitions between the initial state and all eigenstates allowed by energy conservation, which in particular means that an electron may be excited to the upper atomic level. However, even for an initial state of definite parity (as for example $|\Psi\rangle = |1,n\rangle$) the time-evolution will mix eigenstates, but now in a definite parity sector. 

Thus, in stark contrast with the predictions of perturbation theory, 
an SHG signal also occurs for parity conserving dynamics. In addition,
and differently from a three-level system, the closure of the multi-photonic
triangle (which seems needed for the appearance of the three-peaked
Mollow structure), always occurs in a two level system, as an emerging 
symmetry mixing behavior. Furthermore, Mollow-like overtones 
can also be present in higher-order harmonics \cite{Cini93,Cini95},
a feature that could become relevant in the time domain (e.g. in the ultrafast regime).  
These summary notions provide rigorous foundation and motivation to investigate SHG  
in more complicated two-level system setups and out of the stationary limit, 
i.e. when the effects discussed so far will appear within characteristic timescales
(for example, when investigating an atom moving through a cavity of finite length).
Accordingly, time-resolved multi-photon fluorescence is the main theme of the rest of 
the paper.

%%%%%%%%%%%%%%%%%%%%%%%%%%%%%%%%%%%%%%%%%%%%%%%%%%%%%%%%%%%%
%%%%%%%%%%%%%%%%%%%%%%%%%%%%%%%%%%%%%%%%%%%%%%%%%%%%%%%%%%%%
%%%%%%%%%%%%%%%%%%%%%%%%%%%%%%%%%%%%%%%%%%%%%%%%%%%%%%%%%%%%
%%%%%%%%%%%%%%%%%%%%%%%%%%%%%%%%%%%%%%%%%%%%%%%%%%%%%%%%%%%%

\section{An array of two-level systems}\label{sec:array}
We now consider $N$ two-level systems interacting with one incident and one fluorescent light mode.
This brings in the possibility that the different two-level
systems interact cooperatively with the radiation, with an
enhancement of the radiation field that goes under the name of superradiance \cite{Dicke54}
(more precisely, depending on the nature of the initial state, one can speak of superradiance or
supercoherence \cite{Bonifacio}).

The standard Dicke model \cite{Dicke54}, which describes $N$  two-level systems
interacting with a single optical mode, plays an archetypal role in the study of superradiance and 
has been the focus of extensive investigations (see e.g. \cite{Bonifacio71a,revDicke1,revDicke2,Schwendimann}). 
An exact solution within the RWA of the Dicke model at resonance has been known for quite some time
(for this reason, the model is also referred to as the Tavis-Cummings model \cite{Tavis68})
but some aspects related to this model are still being debated \cite{reviewDicke}. A notable case concerns the existence of a no-go theorem for superradiance, also in connection to the role of
dipolar couplings between electrons and photons and the interactions among the
different two-level systems. Furthermore, consideration is needed for the superradiant behavior when $N\rightarrow \infty$, i.e. the periodic array limit. Mathematically, the intensity diverges in that situation, but in fact a continuous electronic band structure emerges in the thermodynamical limit,
and the bulk polaritonic regime applies, with a finite optical response \cite{polaritonic_response}. 
 
These different aspects are not addressed  here.
Rather, our simple analysis is aimed to gain qualitative insight into how the behavior
of the single two-level system discussed above changes in a more complex setup. 
As before, we position ourselves in the fully-resonant and the SHG regimes for the incident field. We
will consider the problem from two different but complementary perspectives: (i) We first study the equilibrium properties of the system in the large-$N$ limit using a generalization of the Dicke model. (ii) Secondly, we investigate the non-equilibrium properties of the system for finite $N$ (similar to what done earlier for $N=1$) using an exact numerical treatment. We find that, compared to the single two-level system case, the fluorescence signal shows an enhancement compatible with a superradiance-like mechanism, both in the resonant and SHG regimes.

As an obvious extension of the $N=1$ case considered earlier, the total Hamiltonian becomes
\begin{align}
 H=\hat{H}_a+\hat{H}_f+\hat{H}_i(t),
\end{align} 
with 
\begin{align}\label{eq:ham_ia}
&\hat{H}_a= \sum_{i=1}^{2N} \epsilon_i \hat{c}_i^\dagger \hat{c}_i,\\
&\hat{H}_f= \omega_a \hat{a}^\dagger \hat{a} + \omega_b \hat{b}^\dagger \hat{b}\\
&\hat{H}_i(t)= g_a(t)(\hat{a}^\dagger + \hat{a})\sum_{i=1}^N (\hat{c}_{2i-1}^\dagger \hat{c}_{2i} + \hat{c}_{2i}^\dagger \hat{c}_{2i-1}) \nonumber \\
 &\;\;\;\;\;\;\;\;\;+ g_b(t)(\hat{b}^\dagger + \hat{b})\sum_{i=1}^N (\hat{c}_{2i-1}^\dagger \hat{c}_{2i} + \hat{c}_{2i}^\dagger \hat{c}_{2i-1}).
\end{align}
Here $\hat{c}_{2i-1}$ refers to the ground state and $\hat{c}_{2i}$ to the excited state of two-level system $i$. Due to the specific way the fields couple to the two-level systems, we can rewrite the Hamiltonian in a more compact form as
\begin{align}\label{eq:ham_array}
\hat{H}&= \sum_{i=1}^{N} \omega_i \hat{\sigma}_{z,i} + \omega_0 \hat{a}^\dagger \hat{a} + \omega \hat{b}^\dagger \hat{b} \nonumber \\ 
 &+ \left[g_a(t)(\hat{a}^\dagger + \hat{a}) + g_b(t)(\hat{b}^\dagger + \hat{b})\right] \sum_{i=1}^N \hat{\sigma}_{x,i},
\end{align}
where $\omega_i = \epsilon_{2i} - \epsilon_{2i-1}$. When all excitation energies $\omega_i = \omega_s$, the fields only couple to the total spin operators $\hat{S}_z = \sum \hat{\sigma}_{z,i}$ and $\hat{S}_x = \sum \hat{\sigma}_{x,i}$. The Hamiltonian then closely resembles the Dicke Hamiltonian \cite{Dicke54}, and is written
\begin{align}
\hat{H}&= \omega_s \hat{S}_z + \omega_a \hat{a}^\dagger \hat{a} + \omega_b \hat{b}^\dagger \hat{b} \nonumber \\ 
 &+ \left[g_a(t)(\hat{a}^\dagger + \hat{a}) + g_b(t)(\hat{b}^\dagger + \hat{b})\right] \hat{S}_x.
\end{align}

\subsection{Ground state}
We start by discussing the ground state properties. For an initial state with all two-level systems in their ground state, we can further simplify the Hamiltonian using the Holstein-Primakoff transformation~\cite{HolPri40}. Writing the spin operators in terms of bosonic creation and annihilation operators, $\hat{S}_z = -(N/2) + \hat{s}^\dagger \hat{s}$ and $\hat{S}_+ = s^\dagger\sqrt{N - \hat{s}^\dagger \hat{s}}$, it is straightforward to check that the commutation relations defining the spin algebra are preserved. Taking the limit where $N \gg \langle \hat{s}^\dagger \hat{s}\rangle$, the $\hat{S}_+$ operator can be expanded as $\hat{S}_+ \approx s^\dagger N + \mathcal{O}(N^{-1})$, and the Hamiltonian becomes 
\begin{align}\label{eq:ham_hp}
\hat{H}&= \omega_s \hat{s}^\dagger \hat{s} + \omega_a \hat{a}^\dagger \hat{a} + \omega_b \hat{b}^\dagger \hat{b} \nonumber \\
 &+ N\left[g_a(t)(\hat{a}^\dagger + \hat{a}) + g_b(t)(\hat{b}^\dagger + \hat{b})\right](\hat{s}^\dagger + \hat{s}),
\end{align}
and the system maps into three coupled oscillators. 

\begin{figure}
 \begin{tikzpicture}
  \node at (0,0) {\includegraphics[width=\columnwidth]{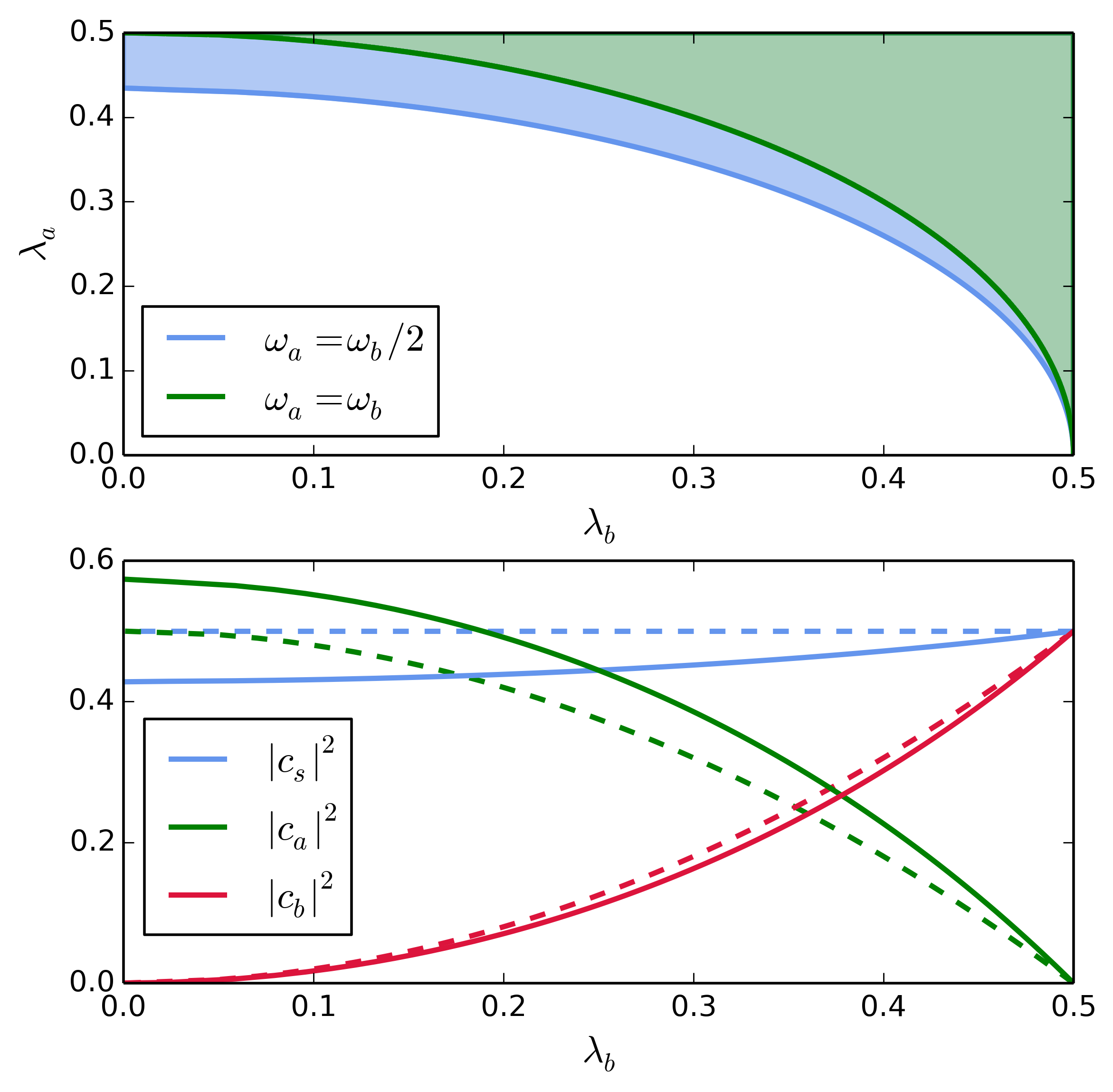}}; 
  \node[black] at (3.6, 3.5) {$(a)$};
  \node[black] at (3.6,-2.0) {$(b)$};
 \end{tikzpicture}
 \caption{Panel $(a)$: the critical coupling strength for the ground-state superradiant transition as a function of $\lambda_a = Ng_a$ and $\lambda_b = Ng_b$. The green line refers to the resonant case $\omega_s = \omega_a = \omega_b = 1$, while the blue line shows the non-resonant case $\omega_s = \omega_b = 1$ and $\omega_a = 1/2$. Panel $(b)$: expansion coefficients $|c_i|^2$ of the lowest eigenstate of Eq.~(\ref{eq:ham_osc}) in the basis of oscillators from Eq.~(\ref{eq:ham_hp}). The dashed lines refer the resonant case $\omega_s = \omega_a = \omega_b = 1$, while the solid lines show the non-resonant case $\omega_s = \omega_b = 1$ and $\omega_a = 1/2$. The unit of energy is given by $\omega_s$.}
 \label{fig:superrad}
\end{figure}

For $g_b = 0$, the Hamiltonian of Eq.~(\ref{eq:ham_hp}) reduces to the usual Dicke Hamiltonian, where a transition to a superradiant state takes place at a critical coupling $g_a \approx (2N)^{-1}$, with the field intensity of the radiation becoming proportional to $N^2$ \cite{reviewDicke}. To see this, we follow closely and reproduce here the discussion given in Ref. \cite{reviewDicke}, starting by writing the Hamiltonian in terms of canonical coordinates:
\begin{align}
\hat{H}&= \frac{1}{2}(\hat{p}_s^2 + \hat{p}_a^2) + \frac{1}{2} \begin{pmatrix} \hat{x}_s & \hat{x}_a \end{pmatrix} \begin{pmatrix} \omega_s & 2\lambda \\ 2\lambda & \omega_a \end{pmatrix} \begin{pmatrix} \hat{x}_s \\ \hat{x}_a \end{pmatrix},
\end{align}
where, to keep the notation light, we have introduced the coupling parameter $\lambda = Ng_a$. We also assume that the field is resonant with the two-level systems, i.e. $\omega_a = \omega_s = 1$.
To identify the superradiant transition, we look for the point $\lambda_c$ where the number of photons in the lowest normal mode of $\hat{H}$ diverges. The lowest mode is identified by moving to normal canonical coordinates and momenta of $\hat{H}$, given by $\hat{x}_\pm = (\hat{x}_s \pm \hat{x}_a)/\sqrt{2}$ and $\hat{p}_\pm = (\hat{p}_s \pm \hat{p}_a)/\sqrt{2}$, respectively, and with normal frequencies $\Omega^2_\pm = 1 \pm 2\lambda$.
In the ground state, the expectation value of  $\hat{x}^2_\pm$ is $\langle \hat{x}^2_\pm \rangle = (2\Omega_\pm)^{-1}$, which
diverges at coupling $\lambda_c$, whilst $\langle p^2_\pm \rangle$ remains well behaved. We then go back to
the original coordinates $\hat{x}_{a/b},\hat{p}_{a/b}$, expressing them in terms of $\hat{x}_\pm,\hat{p}_\pm$, and note that $n_a = \langle \hat{a}^\dagger \hat{a}\rangle =  (\langle \hat{p}^2_a \rangle + \langle \hat{x}^2_a \rangle-1)/2$.
Then, when $\lambda \rightarrow \lambda_c$, the diverging contribution comes from $\langle \hat{x}^2_\pm\rangle $, i.e.
\begin{align}
n_a  \stackrel[\lambda \rightarrow \lambda_c]{}{\approx} \frac{1}{4 \sqrt{1+2\lambda}} + \frac{1}{4 \sqrt{1-2\lambda}}
\end{align}
which diverges for $|\lambda| = \lambda_c=1/2$. Thus, for $\lambda >0$ we can identify the ground state superradiant transition as the point where the lowest eigenvalue $\Omega_-$ vanishes.

To extend the above discussion from Ref.~\cite{reviewDicke} to the original Hamiltonian Eq.~(\ref{eq:ham_hp}) with the fluorescent field included, we  write
\begin{align}\label{eq:ham_osc}
\hat{H}&= \frac{1}{2}(\hat{p}_s^2 + \hat{p}_a^2 + \hat{p}_b^2) \nonumber \\ 
   &+ \frac{1}{2} \begin{pmatrix} \hat{x}_s &  \hat{x}_a &  \hat{x}_b \end{pmatrix} \begin{pmatrix} \omega_s & 2\lambda_a & 2\lambda_b \\ 2\lambda_a & \omega_a & 0 \\ 2\lambda_b & 0 & \omega_b \end{pmatrix} \begin{pmatrix}  \hat{x}_s \\  \hat{x}_a \\  \hat{x}_b \end{pmatrix},
\end{align}
where again $\lambda_a = Ng_a$ and $\lambda_b = Ng_b$. At this point,
if we take $\omega_s = \omega_a = \omega_b = 1$, the lowest normal mode eigenvalue $\Omega_0$ of $H$  is given by
\begin{align}
 \Omega_0(\lambda_a,\lambda_b) &= 1 - 2\sqrt{\lambda_a^2 + \lambda_b^2},
\end{align}
which is the generalization to the case of two fields both in resonance with the $N$ two-level systems. If instead,
to address the SHG regime, we assume that $\omega_s = \omega_b = 1$ and $\omega_a = 1/2$, we obtain 
\begin{equation}\label{eq:superrad}
 \Omega_0(\lambda_a,\lambda_b) = \frac{1}{12}\left[11 - 4\beta\left(\frac{2}{A}\right)^{1/3} 
- 4\left(\frac{A}{2}\right)^{1/3}\right],
\end{equation}
where $A= \alpha - \sqrt{\alpha^2 -4\beta^3}$, 
$\alpha = 1/32 + 9\lambda_a^2 - 18\lambda_b^2$, and $\beta = 1/16 + 12\lambda_a^2 + 12\lambda_b^2$.
As before, we search for signatures of a superradiant transition by looking at the points where $\Omega_0$ vanishes. In the resonant case this is easily done, and one obtains the semi-circle solution set $\lambda_a = \sqrt{1/4-\lambda_b^2}$. We note that for either $\lambda_a = 0$ or $\lambda_b = 0$, we recover the critical coupling of the standard Dicke model discussed above. In the non-resonant case, we solve instead Eq.~(\ref{eq:superrad}) numerically to find the result in Fig.~\ref{fig:superrad}a. Interestingly, we find that the signature of a superradiant transition occurs for smaller values of the coupling when one of the fields is non-resonant with the atomic transitions.

The above argument indicates that the lowest eigenstate of the Hamiltonian undergoes a superradiant transition. However, we have not yet determined how this state is related to the original photon fields, and therefore at this point it is not clear how these fields behave at the transition. In analogy with the Dicke model, the average number of photons in each of the original light fields (in the ground state) is proportional to $|c_{a/b}|^2(2\Omega_0)^{-1}$, where $|c_i|^2 = |\langle i|\Omega_0\rangle|^2$ are the projections of the normal mode $|\Omega_0\rangle$ onto the original oscillators. The fields should therefore undergo a transition to a superradiant state for values of $\lambda_a$ and $\lambda_b$ such that (i) $\Omega_0 = 0$, and (ii) $|c_{a/b}|^2$ are finite. The coefficients are found by numerical diagonalization of the Hamiltonian in Eq.~(\ref{eq:ham_osc}), and shown in Fig.~\ref{fig:superrad}b for values of $\lambda_a$ and $\lambda_b$ at the superradiant transition. It is apparent that for $\lambda_{a/b} > 0$ we always have $|c_{a/b}|^2 > 0$, so that condition (ii) above is always satisfied. We thus find that at the transition points found above, both the incident and the coherent fields behave as if a superradiant state is attained.

\subsection{Real-time simulations for finite $N$}
Having discussed some ground state features of a two-level system array coupled to radiation, we now return to explore the time evolution of the system defined by Eq.~(\ref{eq:ham_array}). We start from an initial state with all atoms in their ground state, and therefore we do not expect to see a superradiant emission burst. However, we are interested in exploring how the fluorescence spectrum changes as we approach coupling strengths close to the (equilibrium) superradiant transition. We consider the parameters $\omega_s = \omega_b = 1$, $\omega_a = 0.5$, and choose $|\alpha|^2 = M=9$ for the average number of photons in the cavity. With $M$ of the same of order as the  number of atoms $N$, the energy of the field should be enough to simultaneously excite all the atoms.
This means that for a two-level array with bare couplings given by e.g. $g_a = 0.03$ and $g_b = 0.01$ (to be used below in the actual simulations),
the minimal value of $N$ needed to get the critical values of $\lambda_a$ and $\lambda_b$ indicated in Fig.~\ref{fig:superrad} is given by $N \approx 14$.  This is done by checking for which value of $N$ that $\lambda_a = Ng_a$ and $\lambda_b = Ng_b$ cross the blue line in Fig.~\ref{fig:superrad}a.  

Since the arguments of the previous section are only strictly valid in the ground state and for large $N$ (due to the use of the Holstein-Primakoff transformation), there is no guarantee that they would hold in real time. Thus, our estimate for $N$ just provides a hint of the order magnitude of $N$ where we can expect superradiant effects to appear. In addition, the non-equilibrium signatures of superradiance are typically expressed through the scaling of the duration and intensity of the superradiant burst with the number of two-level systems $N$, given respectively by $1/N$ and $N^2$. We therefore consider below the fluorescence spectrum for $N \in \{1, 2, \ldots, 10 \}$, and look for signatures consistent with these scaling laws.

\begin{figure}
 \begin{tikzpicture}
  \node at (0.0, 2.2) {\includegraphics[width=0.915\columnwidth]{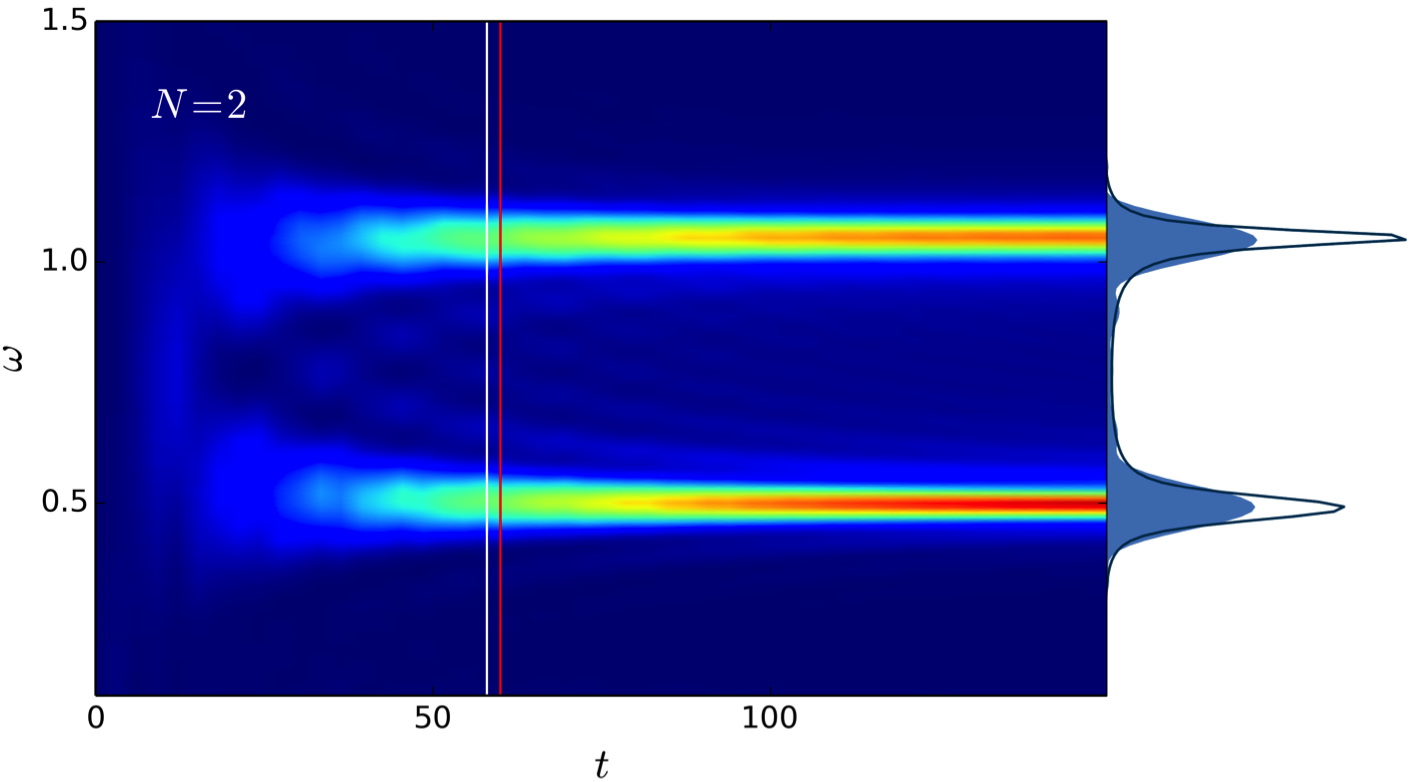}};
  \node at (4.5, 2.4) {\includegraphics[width=0.08\columnwidth]{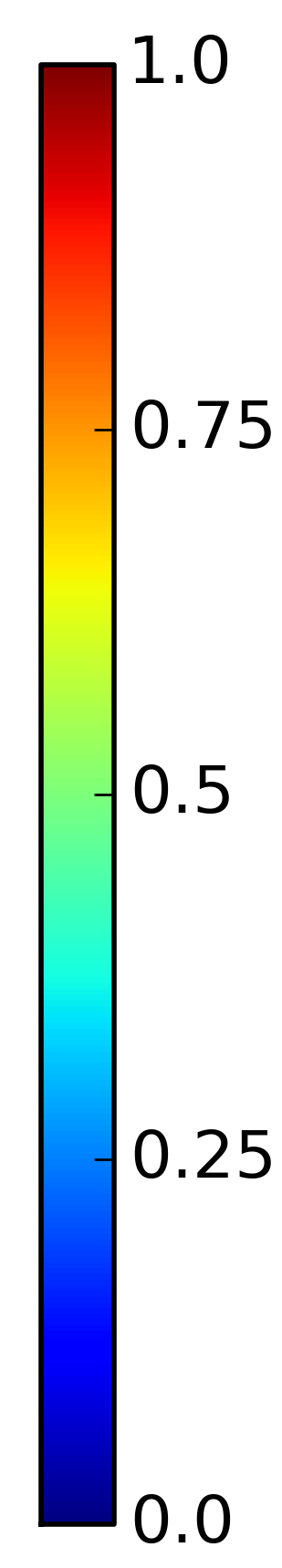}};
  \node at (0.0,-2.3) {\includegraphics[width=0.915\columnwidth]{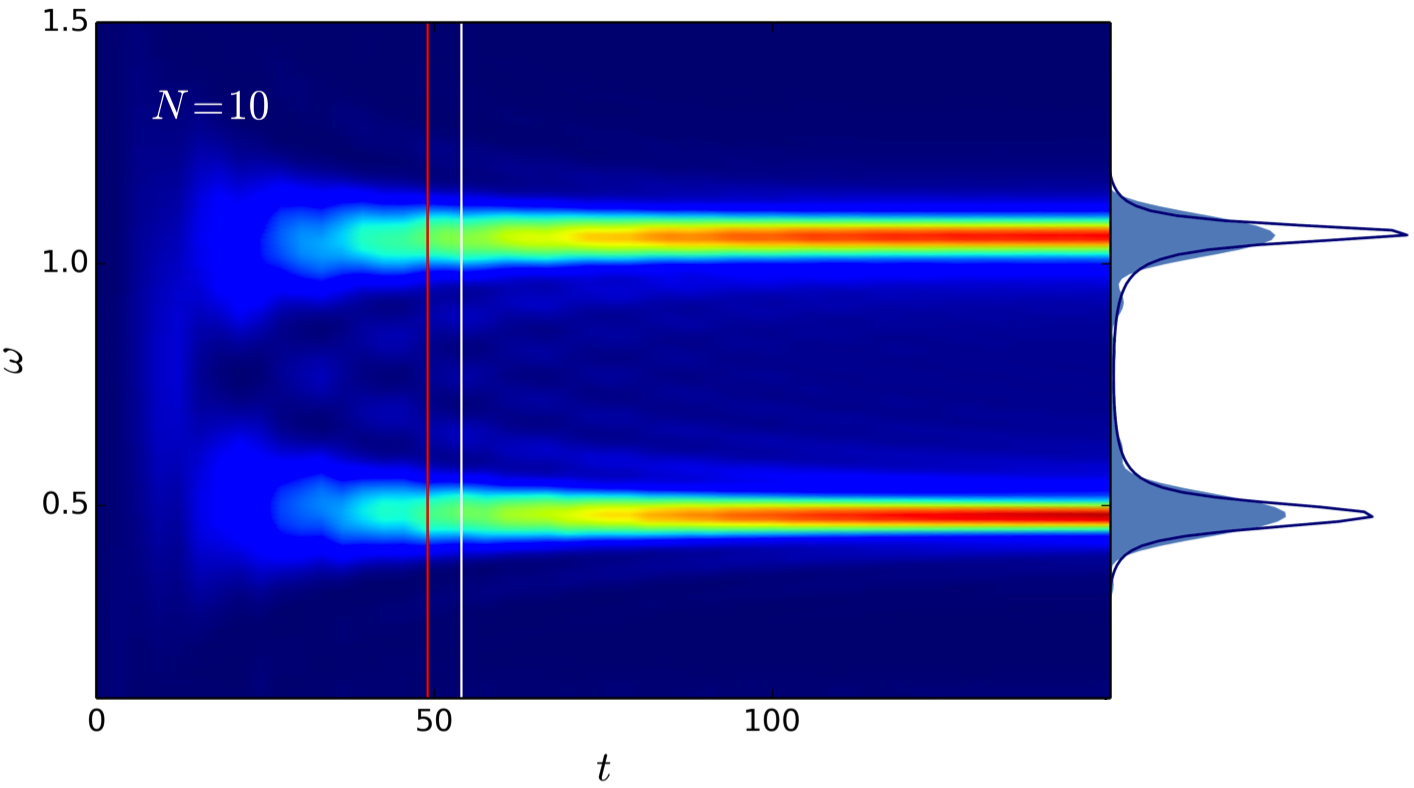}};
  \node at (4.5,-2.1) {\includegraphics[width=0.08\columnwidth]{colorbar}};
  \node[white] at (1.8, 3.8) {$(a)$};
  \node[white] at (1.8,-0.7) {$(b)$};
 \end{tikzpicture}
 \caption{Fluorescence spectra for an array with $N$ two-level systems, for $N = 2$ in panel $(a)$ and $N = 10$ in panel $(b)$. The parameters are given by $\omega_s = \omega_b = 1$, $\omega_a = 0.5$, and $|\alpha|^2 = 9$, and the bare couplings are taken as $g_a = 0.03$ and $g_b = 0.01$. The white (red) vertical lines indicate the time at which the height of the Rayleigh (SHG) peak reaches half of its maximum value. The spectral curves at such times and at the end of the simulation time are explicitly shown on the right side of the respective panels. The colormap is normalized to the maximum value of the Rayleigh peak at the final time. The units of energy and time are respectively given by $\omega_s$ and $\hbar/\omega_s$.}
 \label{fig:nlevel}
\end{figure}

In Fig.~\ref{fig:nlevel} we show the fluorescence spectra for an array of $N = 2$ and $N = 10$ two-level systems. We find that with increasing $N$, the time it takes for the peaks to develop is reduced, as indicated by the vertical lines in the figure. For $N = 2$ the resonant Rayleigh peak develops before the second harmonic peak, while for $N = 10$ the order is opposite. In addition, for $N = 10$ the SHG peak transiently exceeds the Rayleigh peak also in magnitude.

\begin{figure}
 \includegraphics[width=\columnwidth]{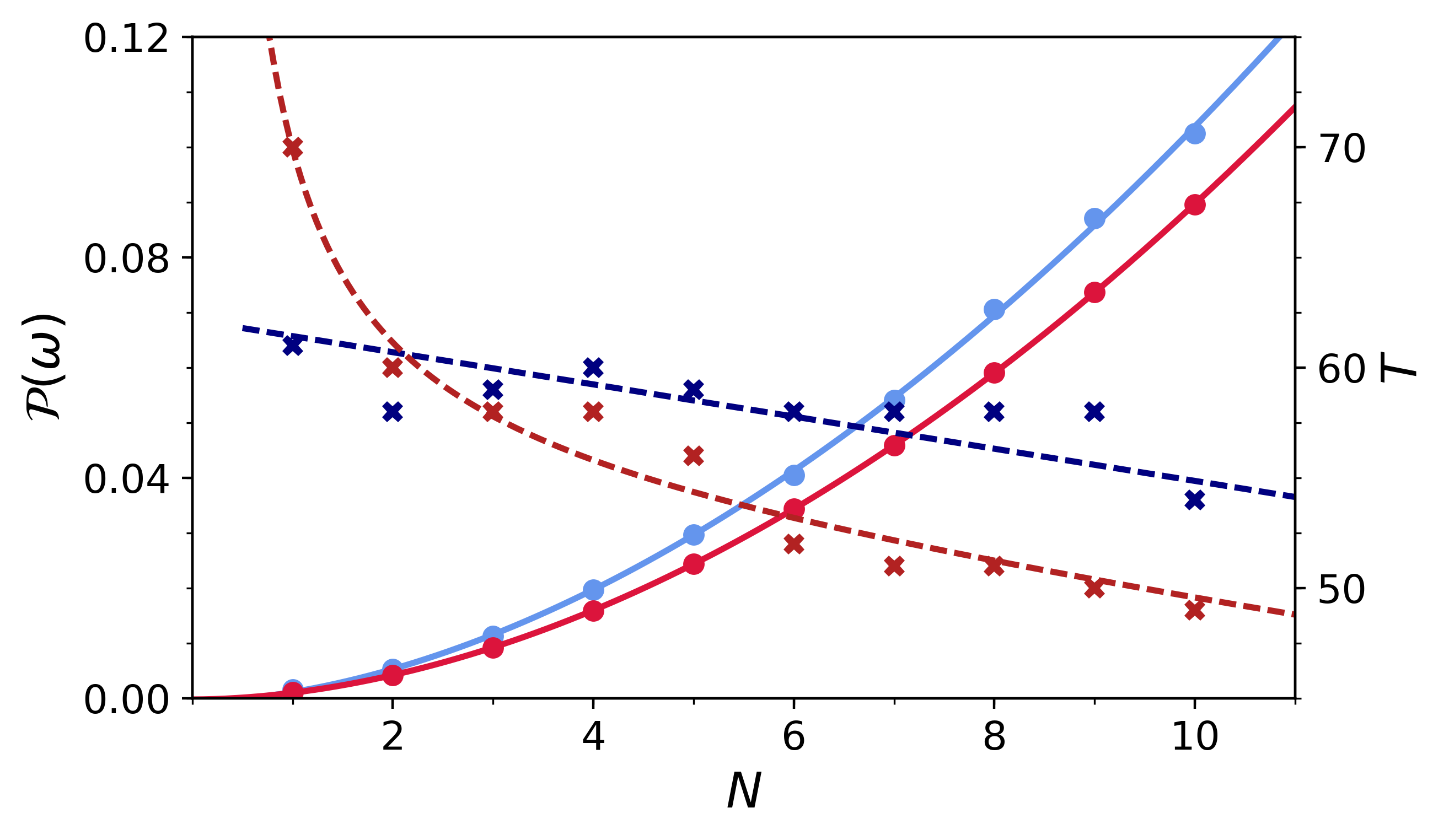}
 \caption{The rise time $T$ (dashed lines, crosses) and the number of emitted photons $\mathcal{P}(\omega)$ (solid lines, circles) of the Rayleigh (blue) and second harmonic (red) peaks as a function of the number of two-level systems $N$. The parameters are given by $\omega_s = \omega_b = 1$, $\omega_a = 0.5$, and $|\alpha|^2 = 9$, and the bare couplings are taken as $g_a = 0.03$ and $g_b = 0.01$.}
 \label{fig:scaling}
\end{figure}

To quantify these observations, we define $T$ as the time it takes a peak to reach half its maximum value. As shown in Fig.~\ref{fig:scaling}, we find that $T$ as a function of $N$ shows a crossover around $N = 2$, from a regime where the Rayleigh peak develops first to a regime where the SHG peak comes first. Further, we find for the Rayleigh peak that the dependence of $T$ on $N$ is approximately linear, while for the SHG peak is behaves as $1/N$. For the latter case the scaling is consistent with a superradiant behavior, where the duration of the superradiant burst decreases as $1/N$. In Fig.~\ref{fig:scaling} we also show the number of photons $\mathcal{P}(\omega_b = 1)$ in the fluorescent field as a function of $N$, and we find it increases as $N^2$ for both the Rayleigh and SHG peaks. Again this is consistent with a superradiant mechanism.

Taken together, the results presented here indicate that, compared to a single two-level system, the SHG signal can be enhanced by considering an array of $N$ of two-level systems. Furthermore, the dependence on $N$ of both the emission time and intensity of the field are consistent with a superradiance behavior.

%%%%%%%%%%%%%%%%%%%%%%%%%%%%%%%%%%%%%%%%%%%%%%%%%%%%%%%%%%%%
%%%%%%%%%%%%%%%%%%%%%%%%%%%%%%%%%%%%%%%%%%%%%%%%%%%%%%%%%%%%

\section{Motion of a two-level system in a cavity}\label{sec:atomic_motion}
The two-level systems considered in the previous sections were at fixed positions in space, 
like e.g. for a given pair of levels in a quantum dot.
However, if a two-level system is meant to model a pair of atomic orbitals, 
then the atomic center-of-mass motion can be an important, if not crucial, element to take into account. 

On the experimental side, compelling evidence comes for example from studies of quantum control, where atoms moving across an optical cavity provide information about cavity photons \cite{Haroche96}, or laser beams across an ion trap provide information about the internal state of the ions \cite{Vineland96}. 
On the theoretical side, the role of atom dynamics has been extensively considered \cite{Meystre89,Yakovlev,Demolition,Cirac93,Goldstein97, MeystreBook,Schleich,ScullyCapasso,CavityReview,BooKAdvances,Krumm,Mazer1,Mazer2,Mazer3},
often in terms of a generalized Jaynes-Cummings model where the standard two-level, one-mode Hamiltonian is augmented by a kinetic energy operator (for the center-of-mass motion). Furthermore, the light-matter coupling can become position dependent \cite{Schleich}, for example when the atom is moving inside/outside a optical cavity.

The solution of the generalized Jaynes-Cummings model has been approached in many different ways 
\cite{Meystre89,Yakovlev,Demolition,Cirac93,Goldstein97, MeystreBook,Schleich,ScullyCapasso,CavityReview,BooKAdvances,Krumm,Mazer1,Mazer2,Mazer3}. For example, with or without the RWA, with the center-of-mass motion described
classically or quantum mechanically, using density-matrix techniques
or resorting to a direct solution of the generalized Rabi equations in wavefunction space. 
Here, we again consider the exact numerical time-evolution of the full system wavefunction, 
thus avoiding the RWA. Since we are interested in how the atom dynamics affects fluorescence spectra, 
we consider a generalized Jaynes-Cummings model with two (pump and fluorescence) modes, and with a center-of-mass that moves longitudinally across a cavity of finite length. Transverse motion is not considered (i.e., space-wise, our system
is strictly one-dimensional). We treat the atom dynamics quantum mechanically, but for 
comparison we also consider a classical description via the Ehrenfest approximation.
Our approach includes all these element on equal-footing, and
in a single coherent description. This can offer an advantage: for example,
the cavity boundaries can have nontrivial effects on the spectra 
which depend also on the level of description.

Since we intend to look only in a preliminary and explorative
way at fluorescence spectra in this setup,  
we already here anticipate that in our calculations the ``atomic'' mass 
value is taken rather small, but not so small that it is necessary to take
into account spatial dispersion effects in the radiation-matter interaction. 
The purpose of this choice is twofold: on the one hand, the atom moves ``faster'', 
which alleviates the costs of the numerical time evolution to reach the long-time limit. 
On the other hand the role of quantum effects in the nuclear motion is enhanced, since
on increasing the value of the atomic mass a classical description becomes increasingly appropriate.

It is worth to mention that
excitons in solid-state systems (e.g. heterostructures \cite{BalslevBook}) can also be used for 
two-level atom optics in quantized light fields, with the exciton dynamics manipulated by optical means. This option has the merit that the value of the exciton electron/hole effective masses
(and thus of the total mass) can be tailored by manipulating the band-edge curvatures. Furthermore, using a mass-scaling transformation as described in Appendix~\ref{App_A0scal}, the calculations and results to follow can qualitatively relate to microwave transitions of atoms with realistic masses.

Out of this discussion, the Hamiltonian to consider is
\begin{align}  \label{atomoving}
 \hat{H}(t) &= \frac{\hat{p}^2}{2M} + \epsilon_1 \hat{c}_1^\dagger \hat{c}_1 + \epsilon_2 \hat{c}_2^\dagger \hat{c}_2 + \omega_a \hat{a}^\dagger \hat{a} + \omega_b \hat{b}^\dagger \hat{b} \\
 &+ \left[g_a(\hat{x},t)(\hat{a}^\dagger + \hat{a}) + g_b(\hat{x},t)(\hat{b}^\dagger + \hat{b})\right](\hat{c}_1^\dagger \hat{c}_2 + \hat{c}_2^\dagger \hat{c}_1) \nonumber
\end{align}
where $\hat{p}$ and $\hat{x}$ are the momentum and position operators of the atomic center of mass, $M$ is the atomic mass. As mentioned above, Eq.~(\ref{atomoving}) satisfies a useful scaling property (see Appendix~\ref{App_A0scal}.)
For the definition of the other quantities, we refer to Eqs.~(\ref{Heq1}-\ref{eq:ham_i}). 
As before, we assume that the atom is occupied by a single spinless electron, that the cavity field is of frequency $\omega_a$ and in a coherent state defined by $\hat{a}|\alpha\rangle = \alpha|\alpha\rangle$. Further, the fluorescent field is of frequency $\omega_b$ and initially in the vacuum state $\hat{b}|0\rangle = 0$. 

The interaction between the light and the atom is given by the couplings $g_a(\hat{x},t)$ and $g_b(\hat{x},t)$ respectively, with the spatial dependence of the coupling $g_a$ coming from the spatial profile of the cavity mode. We assume that the atomic motion happens only along the cavity axis, and restrict the length of the coordinate axis to the set $X = [0,L]$ of length $L$. We further divide the coordinate axis into two parts $X_{in}$ and $X_{out}$, corresponding respectively to inside and outside the cavity, where the set $X_{in} = [x_1,x_2]$ is of length $l = x_2 - x_1$ and $X_{out} = X\setminus X_{in}$ is of length $L - l$. For consistency we need to take $0 < x_1 < x_2 < L$. The cavity electric field $E$ is assumed to be in the lowest mode, with a spatial profile given by $E(x) = \sin((x-x_1)\pi/l)$ for $x\in X_{in}$ and $E(x) = 0$ otherwise. Using the characteristic function $\chi_I$, which is unity on the interval $I$ and zero otherwise, the light-matter couplings are then given by 
\begin{align}
g_a(\hat{x},t)  &= g\sin\left(\frac{\pi(\hat{x}-x_1)}{l}\right)\chi_{_{_{X_{in}}}}(\hat{x}) \label{barrier1}\\
g_b(\hat{x},t) &= g_1e^{-\Gamma_1 t}\chi_{_{_{X_{in}}}}(\hat{x}) + g_2e^{-\Gamma_2 t}\chi_{_{_{X_{out}}}}(\hat{x}).  \label{barrier2}
\end{align}
For the coupling to the fluorescent field we have assumed a constant coupling $g_1$ ($g_2$) inside (outside) the cavity, with a phenomenological decay $\Gamma_1$ ($\Gamma_2$) taking into account collision effects and an effective coupling to additional radiation modes in the continuum.

\subsection{Time dependent fluorescence spectrum}
Taking into account the quantum motion of the center of mass significantly increases the numerical effort necessary to obtain the fluorescence spectrum. To simplify the calculations we therefore consider the fluorescence response in the one-photon limit, where the coupling between the fluorescent field and the atom only acts once during the time evolution. For the initial state, it is assumed that i) the atom is prepared with 
a nuclear wavefunction the form of which is 
\begin{align}\label{eq:atom_wf}
\phi(x) = e^{-(x-x_0)^2/\sigma^2} e^{ixp_0},
\end{align}
and with the electron in level 1 (with energy $\epsilon_1$) ii) 
the cavity field is in a coherent state $\alpha$, and there are zero photons of the fluorescent field.
Thus the system's initial state is denoted by $|1,\phi,\alpha \rangle$ (the label for fluorescent state
being omitted, because of the zero-photon assumption). 

The spectrum is defined as the probability that at time $t$ there is one photon in the fluorescent field:
\begin{align}
P(t,\omega) = \sum_{ni} \int dx\,|\langle i,x,n|\hat{b}e^{-i\hat{H}t}|1,\phi,\alpha \rangle|^2. \label{equazio1}
\end{align}
Here, the state $|i,x,n\rangle$ also contains zero photons of the fluorescent field (note the $\hat{b}$ operator
immediately to the right of $\langle i,x,n |$). In the final state, $i$ denotes the electronic level (1 or 2), $x$ the atomic position, $n$ the number of photons in the cavity field, and $\sum\int$ the trace over $i,x,n$. The perturbative limit is obtained by assuming that the time-evolution operator can be written as
\begin{align}
 e^{-i\hat{H}t} \approx \int_0^t dt'\, e^{-i\hat{H}_0(t-t')}\hat{H}'(t')e^{-i\hat{H}_0t'} \label{equazio2}
\end{align}
where $\hat{H}'(t) = g_b(\hat{x},t)(\hat{c}_1^\dagger \hat{c}_2 + \hat{c}_2^\dagger \hat{c}_1)(\hat{b}^\dagger + \hat{b})$ and $\hat{H}_0 =\hat{H}- \hat{H}'$. 
In Appendix~\ref{App_B} we show that in this limit the probability is given by
\begin{align}\label{eq:spectrum}
P(t,\omega) &= \sum_{\lambda} \left|\sum_{\lambda'} \left(\frac{e^{-i(\epsilon_{\lambda}+\omega)t} -e^{-i\epsilon_{\lambda'}t-\Gamma_1 t}}{\omega +\epsilon_{\lambda}-\epsilon_{\lambda'}+i\Gamma_1} S^1_{\lambda\lambda'}\right.\right. \\
&+\left.\left. \frac{e^{-i(\epsilon_{\lambda}+\omega)t} -e^{-i\epsilon_{\lambda'}t-\Gamma_2 t}}{\omega +\epsilon_{\lambda}-\epsilon_{\lambda'}+i\Gamma_2} S^2_{\lambda\lambda'}\right)\langle\lambda'|1,\phi,\alpha\rangle\right|^2. \nonumber
\end{align}
where $\epsilon_\lambda$ are the eigenenergies of $\hat{H}_0$.
The coefficients $S_{\lambda\lambda'}^{1/2}$ are given by 
\begin{align}
 S_{\lambda\lambda'}^{1/2} = \sum_{ni}\int_{X_{in/out}} dx\, \langle\lambda|\hat{H}'|i,x,n\rangle\langle i,x,n|\lambda'\rangle,
\end{align}
where $\lambda,\lambda'$ label complete sets of states of $\hat{H}_0$ (but again with zero photons in the fluorescent field),
and $X_{in}$ and $X_{out}$ have been defined earlier.
This expression, used in the next section to calculate the fluorescence spectrum, is valid for all times $t$ and has the practical advantage of requiring only a single diagonalization of the Hamiltonian $\hat{H}_0$. However, in 
actual calculations the numerical grid for the nuclear coordinate $x$ is confined to the interval ($0\le x\le L$),
and, for a given $L$, the maximum useful $t$ is limited by the need to avoid wavepacket reflection at
the interval boundaries. 

\subsection{Classical versus quantum nuclear motion}
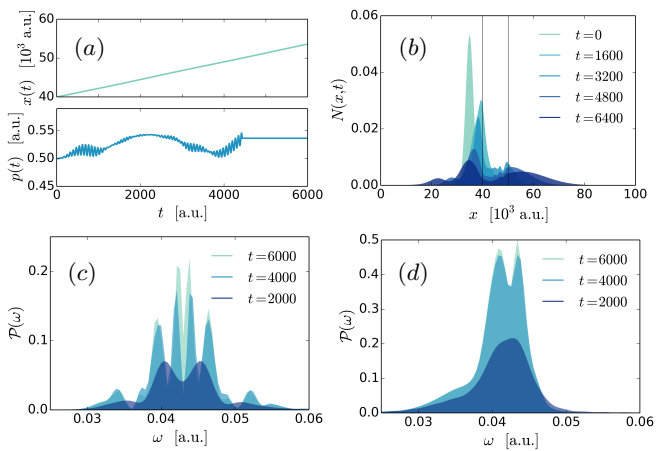
\begin{figure}
 \begin{tikzpicture}
  \node at (0.0, 1.5) {\includegraphics[width=0.5\columnwidth]{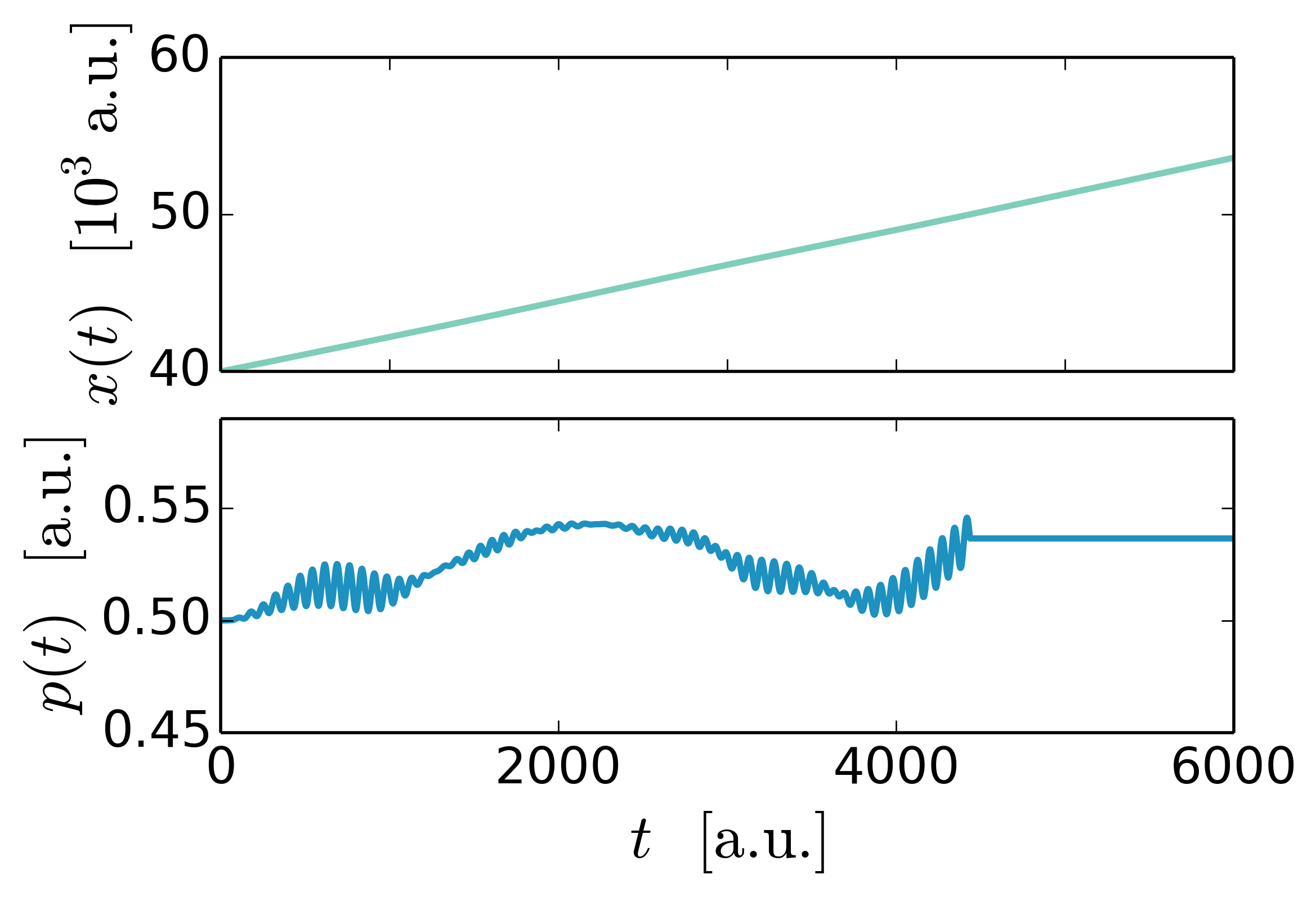}};
  \node at (4.3, 1.5) {\includegraphics[width=0.5\columnwidth]{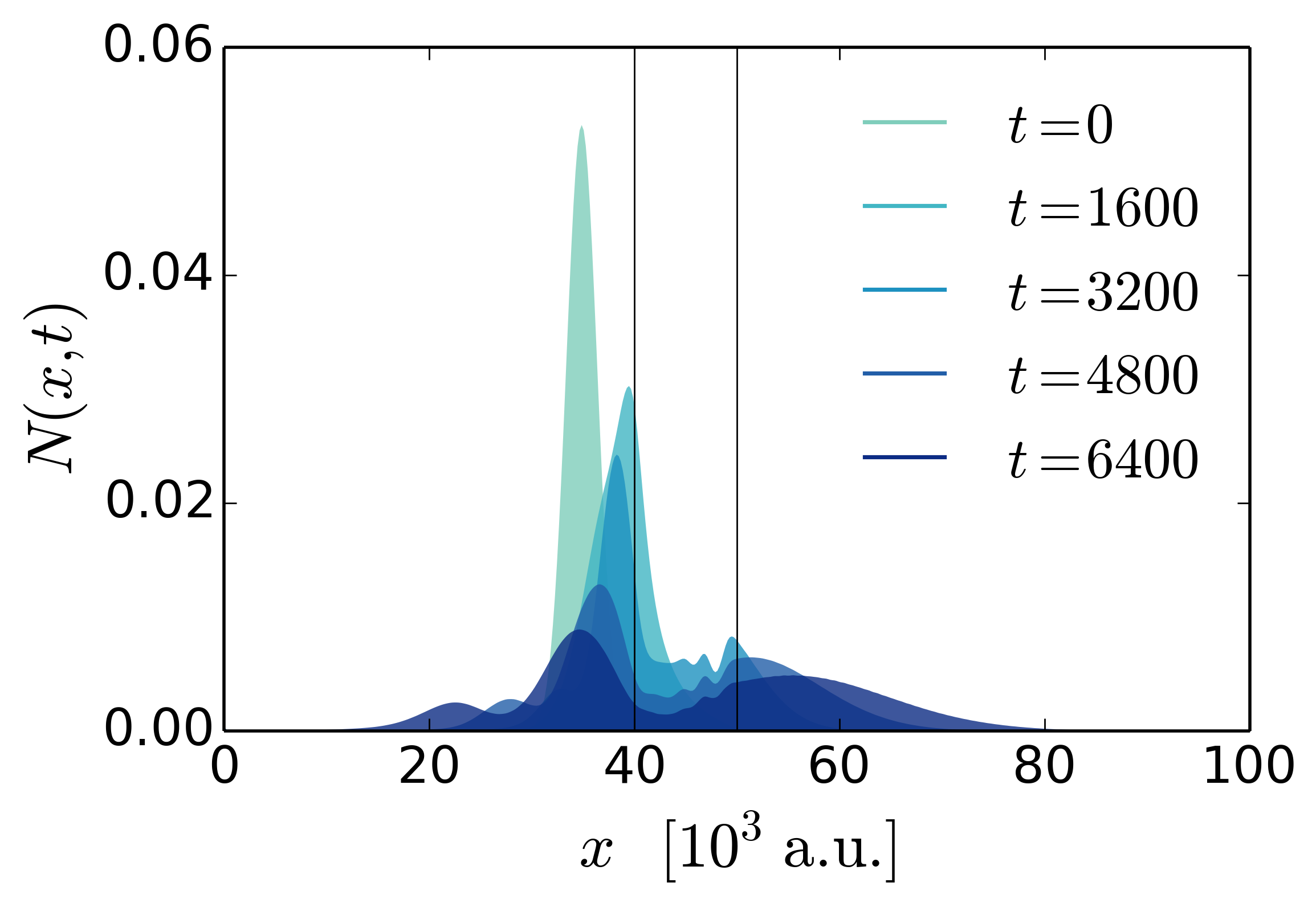}};
  \node at (0.0,-1.5) {\includegraphics[width=0.5\columnwidth]{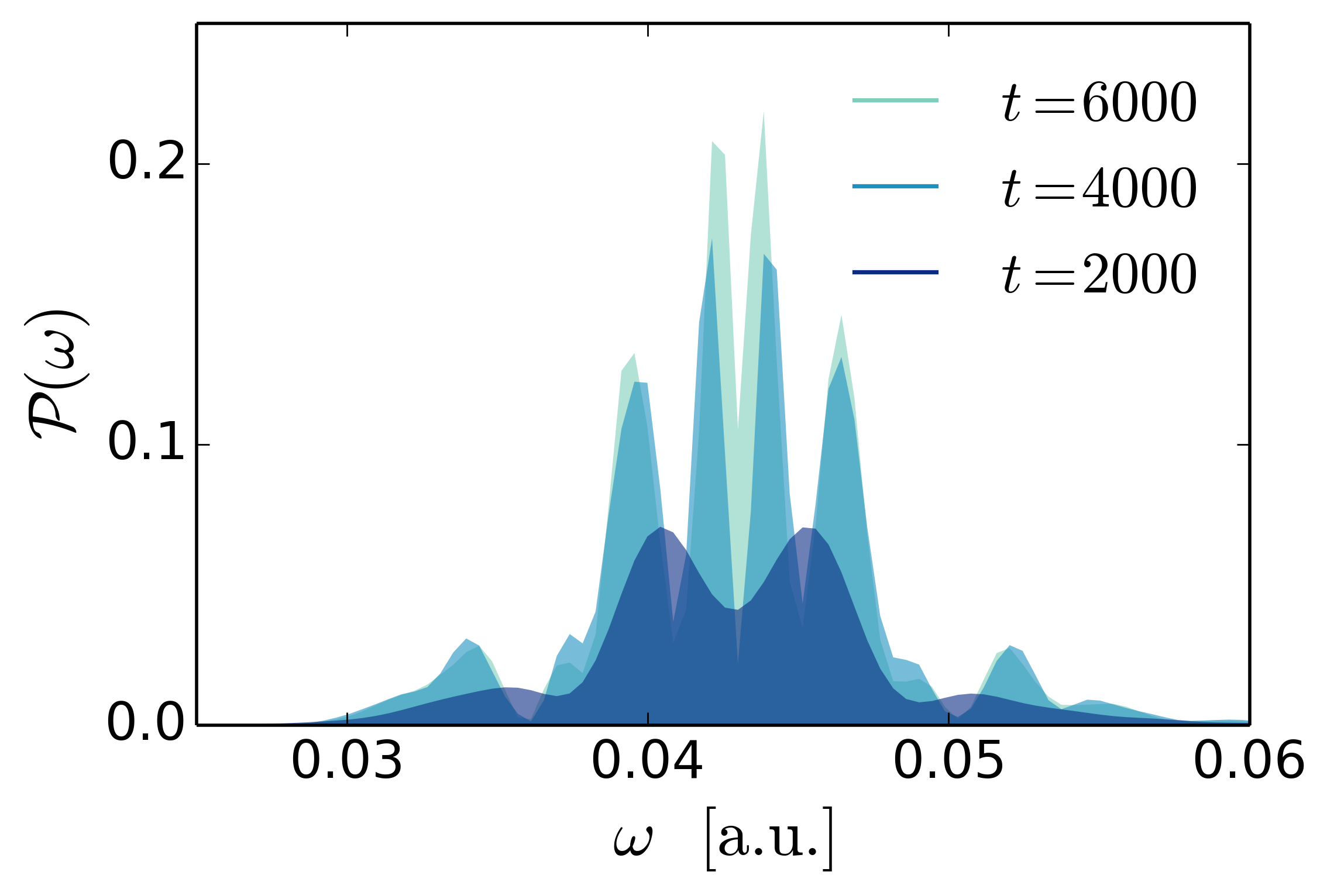}};
  \node at (4.4,-1.5) {\includegraphics[width=0.5\columnwidth]{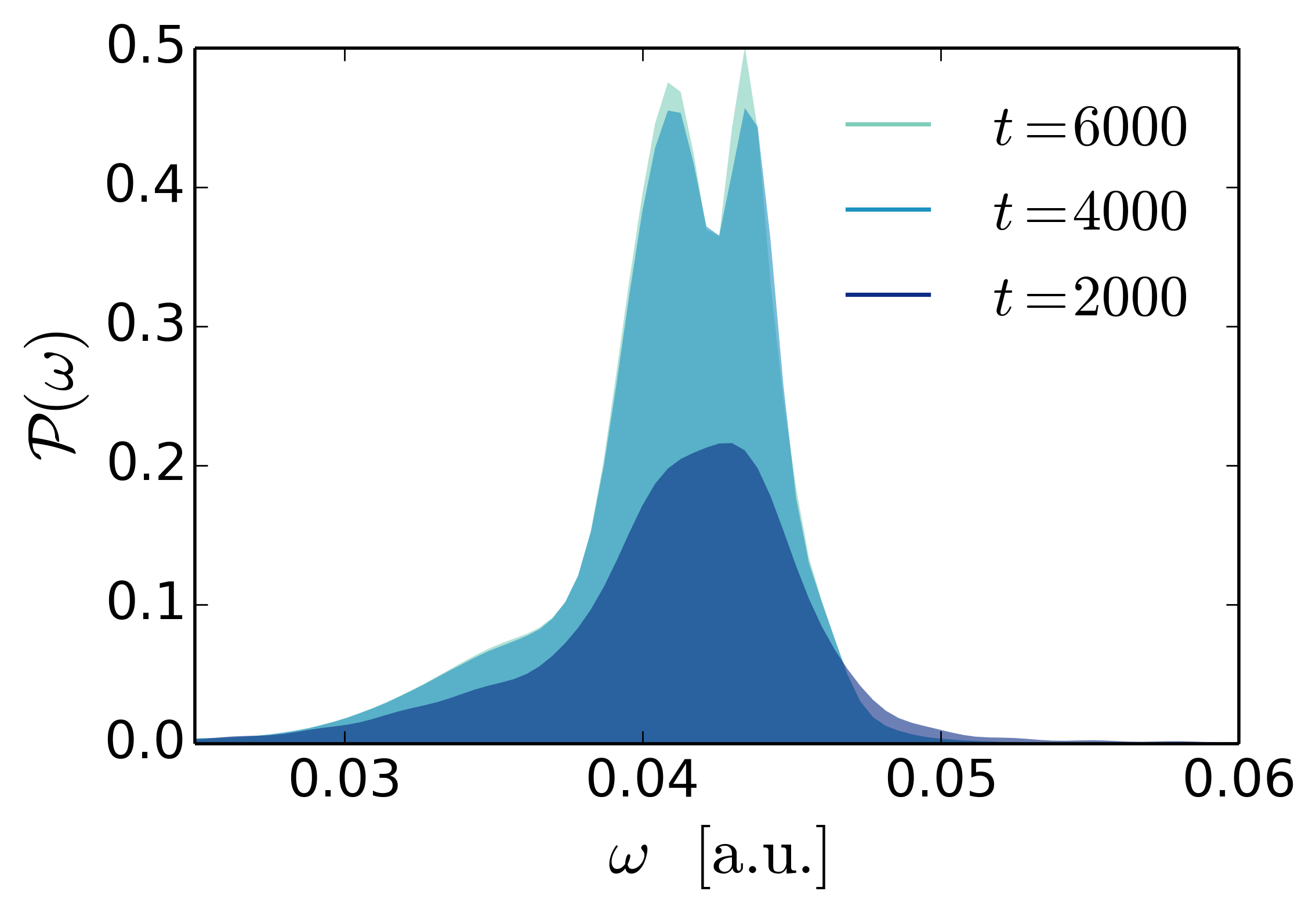}};
  \node[black] at (-1.0, 2.4) {$(a)$};
  \node[black] at ( 3.3, 2.4) {$(b)$};
  \node[black] at (-1.1,-0.6) {$(c)$};
  \node[black] at ( 3.3,-0.6) {$(d)$};
 \end{tikzpicture}
 \caption{Fluorescence spectra for classical and quantum atomic motion. Panel $(a)$: $x(t)$ and $p(t)$ for the classical atomic dynamics within the Ehrenfest approximation. Panel $(b)$: nuclear probability density $N(x,t)$ from the quantum dynamics. Panels $(c)$ and $(d)$ show snapshots of the corresponding fluorescence spectra $\mathcal{P}(\omega)$ at different times for classical and quantum atomic dynamics respectively. Taking $l_c = 10^4$ a.u., the initial conditions are given by $p_0 = 0.5$ a.u., $x_0 = 4l_c$, $x_1 = 4l_c$ and $x_2 = 5l_c$ in the classical case, and by $p_0 = 0.5$ a.u., $\sigma = 3l_c$, $x_0 = 3.5l_c$, $x_1 = 4l_c$, $x_2 = 5l_c$ and $L = 10l_c$ in the quantum case. The remaining parameters are $M = 10$ a.u., $\epsilon = \epsilon_2 - \epsilon_1 = 0.043$ a.u., $\omega_a = \epsilon$, $\alpha = 1$, $g_a = 0.1\epsilon$, $g_1 = 0.1\epsilon$, $g_2 = 0.01\epsilon$ and $\Gamma = 0.02\epsilon$.}
 \label{fig:ehrenfest}
\end{figure}

\begin{figure*}
 \begin{tikzpicture}
  \node at ( 0.0, 2.0) {\includegraphics[width=0.66\columnwidth]{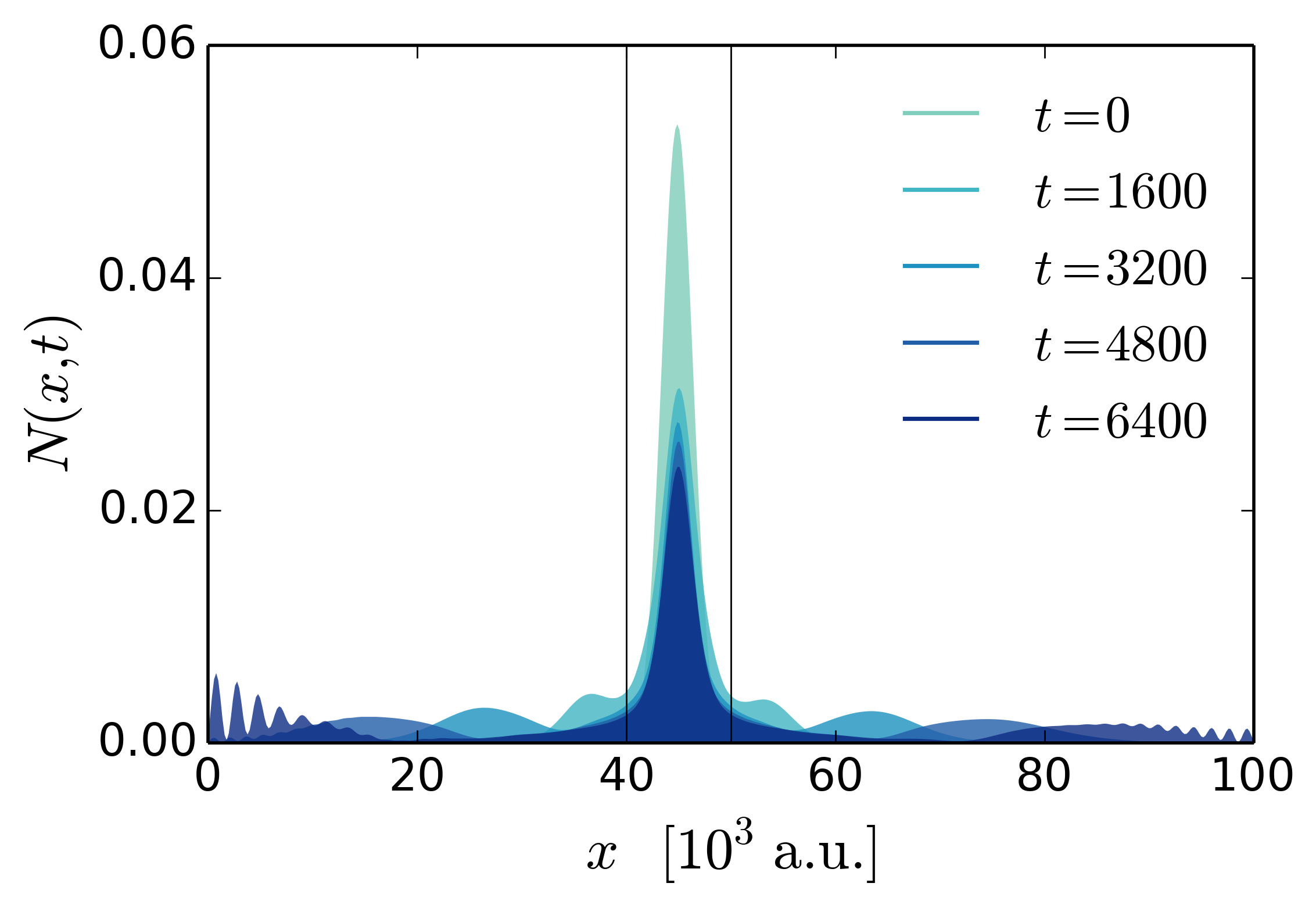}};
  \node at ( 5.8, 2.0) {\includegraphics[width=0.66\columnwidth]{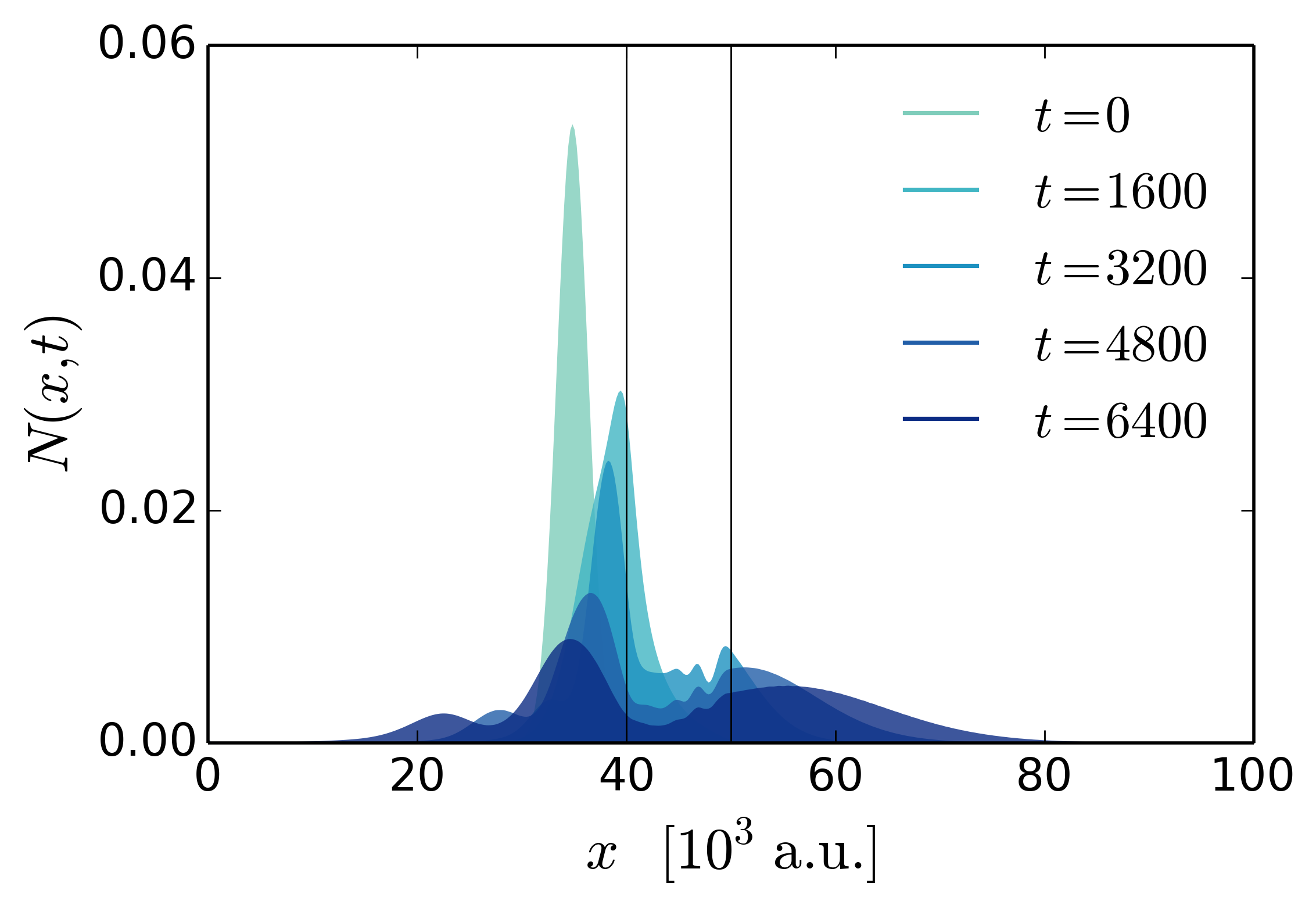}};
  \node at (11.6, 2.0) {\includegraphics[width=0.66\columnwidth]{nuclear_mollow_p05.png}};
  \node at ( 0.2,-2.0) {\includegraphics[width=0.64\columnwidth]{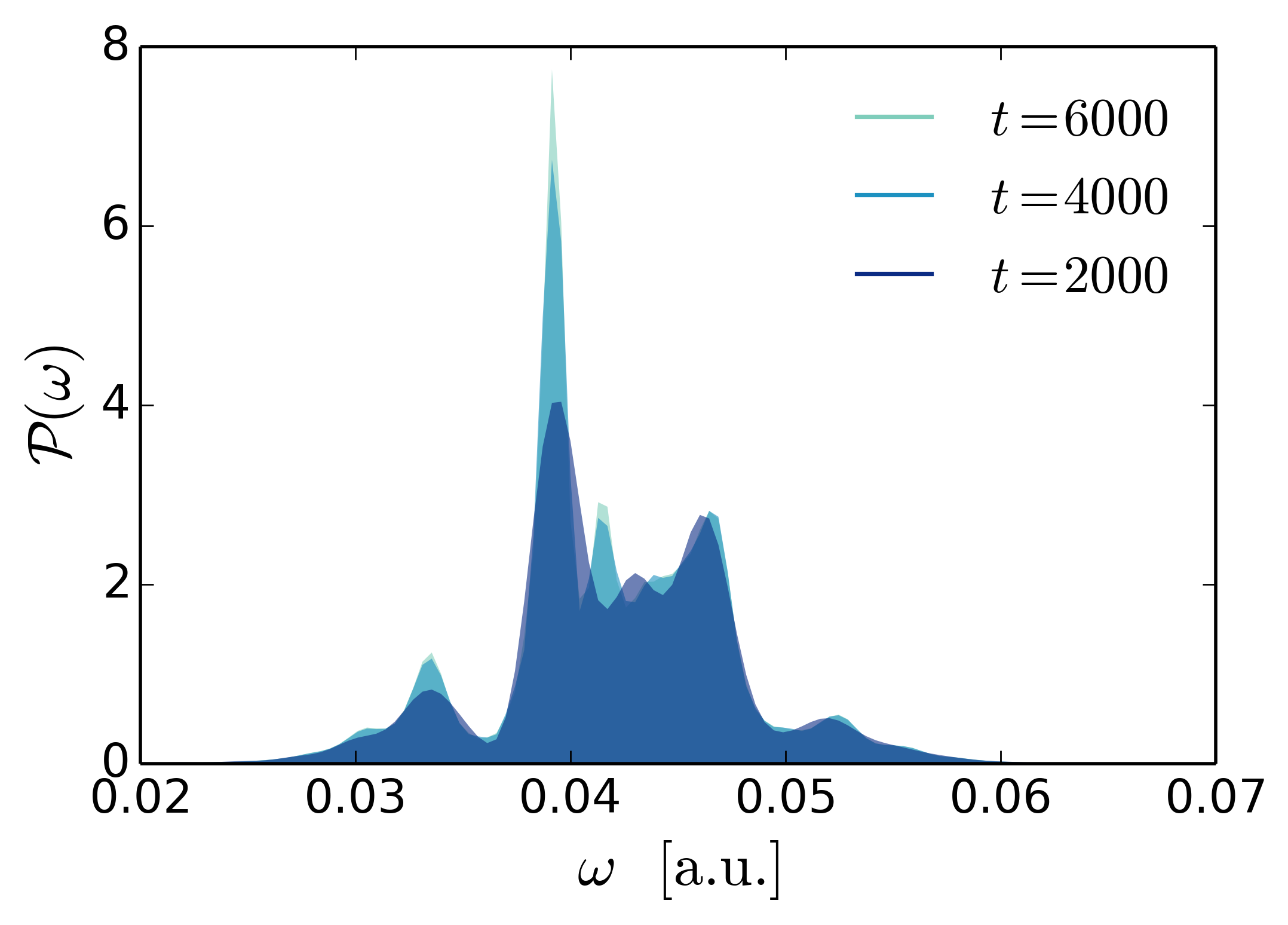}};
  \node at ( 5.9,-2.0) {\includegraphics[width=0.66\columnwidth]{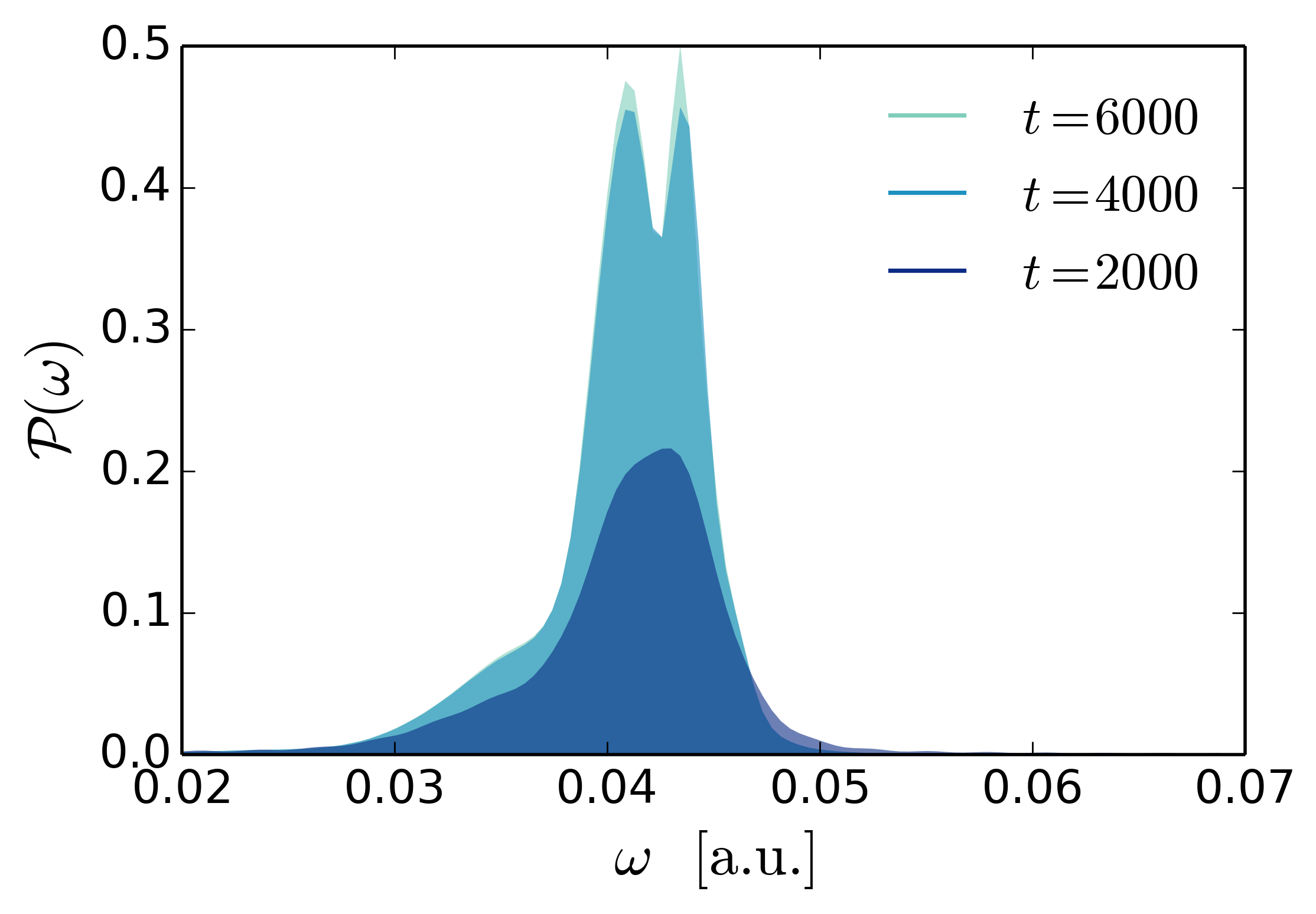}};
  \node at (11.7,-2.0) {\includegraphics[width=0.66\columnwidth]{emission_mollow_p05.png}};
  \node[black] at (-1.5, 3.2) {$(a)$};
  \node[black] at ( 4.3, 3.2) {$(b)$};
  \node[black] at (10.1, 3.2) {$(c)$};
  \node[black] at (-1.5,-0.8) {$(d)$};
  \node[black] at ( 4.3,-0.8) {$(e)$};
  \node[black] at (10.1,-0.8) {$(f)$};
 \end{tikzpicture}
 \caption{Nuclear probability densities (top row) and atomic fluorescence spectra (bottom row) for an atom moving through a cavity of frequency $\omega_a = \epsilon_2 - \epsilon_1 = 0.043$ a.u. The vertical lines denote the boundaries of the cavity. The initial atomic state is given by Eq.~(\ref{eq:atom_wf}), and we take $x_0 = 3.5l_c$ (except for panels $(a)$ and $(d)$ where $x_0 = 4.5l_c$) and $\sigma = 3l_c$ with $l_c = 10^4$ a.u. Panels $(a)$ and $(d)$ corresponds to $p_0 = 0$, panels $(b)$ and $(e)$ to $p_0 = 0.5$ a.u., and panels $(c)$ and $(f)$ to $p_0 = 2$ a.u. The remaining parameters are given by $M = 10$ a.u., $\epsilon = \epsilon_2 - \epsilon_1 = 0.043$ a.u., $\alpha = 1$, $g_a = 0.1\epsilon$, $g_1 = 0.1\epsilon$, $g_2 = 0.01\epsilon$, $\Gamma_1 = 0.01\epsilon$ and $\Gamma_2 = 0.02\epsilon$.}
 \label{fig:mollow_cavity}
\end{figure*}

Before studying the full dynamics of the system, we consider the classical limit of the nuclear dynamics as given by the Ehrenfest approximation. Assuming that $\hat{x}$ and $\hat{p}$ in Eq.~(\ref{atomoving})
are replaced by classical variables $x$ and $p$ evolving under the force $F(t) = -\langle\partial_x \hat{H}_t)\rangle$, the spatial dependence of $g_a$ turns into a time-dependence through $g_a(t) = g_a(x(t)) = g_a\sin((x(t)-x_1)\pi/L)$. For the coupling to the fluorescent field we assume (temporarily) that there is no decoherence in the cavity, so that 
\begin{align}
g_b(t) = g_1\chi_{_{_{X_{in}}}}(x(t)) + g_2e^{-\Gamma (t-t_0)} \chi_{_{_{X_{out}}}}(x(t))\label{barrier3}
\end{align}
with $t_0$ the time at which the atom has passed through the cavity. The quantum evolution is obtained by solving the time-dependent Schr\"odinger equation on a grid $x_n$, as described in more detail later on. In the following we work in atomic units (a.u.) and take $M = 10$ a.u., $\epsilon = \epsilon_2 - \epsilon_1 = 0.043$ a.u., $\omega_a = \epsilon$, $\alpha = 1$, $g_a = 0.1\epsilon$, $g_1 = 0.1\epsilon$, $g_2 = 0.01\epsilon$ and $\Gamma = 0.02\epsilon$. Except for the coupling $g_1$ of the fluorescent field to the atom inside the cavity, these values as the same as earlier in the paper  with the identification $\epsilon = 0.043$ a.u.. However, the value of $g_1$ was increased to enhance the emission into the fluorescent field, with $\langle \hat{b}^\dagger \hat{b}\rangle \ll 1$ still applying.

A physical notion of the chosen parameters can be gathered by noting that for a cavity of length $l_c = \lambda/2$ and $\epsilon = \hbar \omega_a/\kappa$ (for example, $\kappa = 0.5$ for SHG), we have $\epsilon = (\hbar c \pi)/(\kappa l_c)$. Choosing $l_c = 10^4$ a.u., one obtains  $\epsilon \approx 0.043/\kappa$ a.u., 
which can be a reasonable value for example for excitons, and fairly consistent with the value of $M=10$ a.u. introduced above \cite{caveat}.

Within the given units, the spatial simulation interval (cavity and outside) of our calculations is $L = 10l_c$ (i.e. about $5 \mu m$), and the cavity boundaries are set at $x_1 = 4l_c$ and $x_2 = 5l_c$. As initial conditions we take $p_0 = 0.5$ a.u. and $x_0 = 4l_c$ for the classical simulations, while in the quantum simulations the initial wave packet is given by the expression in Eq.~(\ref{eq:atom_wf}) with $p_0 = 0.5$ a.u., $x_0 = 3.5l_c$ and $\sigma = 3l_c$. This momentum corresponds to a velocity $v \approx  10^4$ cm/s. 

In Fig.~\ref{fig:ehrenfest} we compare the results obtained with the Ehrenfest approximation with the results of the full quantum evolution. To characterize the atomic motion, we look in the classical case at the functions $x(t)$ and $p(t)$, and in the quantum case at the nuclear probability density $N(x,t) = \sum_{in} |\langle i,x,n|\Psi(t)\rangle|^2$. 

We see in Fig.~\ref{fig:ehrenfest} that in the classical case the atom moves through the cavity with little resistance, and note that the effects oscillations of $p$ is not visible in $x$ due to the scale of the figure. In contrast, the quantum results show a splitting of the atomic wave packet. This is precisely the regime where the Ehrenfest approximation fails~\cite{Bostrom}, since in a classical description the atom must be either reflected or transmitted. However, at $t = 6400$ a.u. the quantum particle has both a (larger) reflected and (smaller) transmitted contribution, while the classical particle is out of the cavity already at $t = 3000$ a.u. Thus the atom has a reduced amplitude in the barrier 
and a reduced coupling to the field. Compared to the classical case, treating the atomic motion at the quantum level also has a large impact on the fluorescence spectrum. This is addressed in the bottom panels for different time snapshots (spectra at different times are rather similar to each other, and only the one at the latest time is fully visible). The Ehrenfest result resembles at great extent that of a stationary atom (cf. Fig.~\ref{fig:mollow}), while the spectrum corresponding to the quantum motion looks qualitatively different: It contains two main peaks instead of four, and is asymmetric with respect to the central frequency. 

To understand this dissimilarity in behavior, we note that   
in the classical approach
the nuclear wavepacket is perfectly localized both in position and momentum. By contrast, the quantum amplitude gets smaller in the repulsive barrier region. Further, the classical atom sees only a single resonant frequency (Doppler shifted due to the motion) and coupling to the cavity field at each given time of its travel through the cavity. Since the field is strongest in the center of the cavity, where it takes the same value as in the stationary case discussed above ($\alpha = 1$ and $g_a = 0.1\epsilon$), the main contribution to the fluorescence signal comes from when the atom is in this region. However, compared to the stationary case there is an enhancement of the spectrum for frequencies $\omega_b \approx \epsilon$, which most likely comes from fluorescent photons emitted in the regions where $g_a < 0.1 \epsilon$ and the fact that the classical atom couples to the photons more strongly.

In contrast, the quantum atom simultaneously experiences a range of resonance frequencies and field-atom couplings,
there is a lot of structure in the corresponding spectrum, and  its wave function gets low within the barrier region. From the shape of the nuclear wavepacket we expect the dominant contribution to the fluorescence signal to come from when the atom is in the initial and final part of the cavity, the probability distribution being mainly localized to these regions (see Fig.~\ref{fig:ehrenfest}). Consequently the spectrum is closer to what could be expected for a stationary atom weakly interacting with a light field ($g_a < 0.1 \epsilon$), leading to a smaller splitting between the Mollow peaks. However, a detailed explanation of the asymmetric form of the spectrum is difficult to give, but plausibly related to the varying Doppler shifts associated with the different parts of the atomic wavepacket.

\subsection{Fluorescence and quantum motion}
We now consider the fluorescence spectra resulting from a quantum evolution of the coupled atom-photon system. We take $L = 10l_c$, and, as before, the cavity is placed again between $x_1 = 4l_c$ and $x_2 = 5l_c$. To solve the Schr\"odinger equation we consider a grid $x_n$ for the atomic position, with $500$ points in the interval $[0,L]$. The fluorescence spectrum is computed from Eq.~(\ref{eq:spectrum}), and to get the atomic probability density $N(x,t) = \sum_{in} |\langle i,x,n|\Psi(t)\rangle|^2$ we solved the Schr\"odinger equation without the fluorescent field. We have verified that the atomic dynamics is highly insensitive to the presence of the fluorescent field, by explicitly solving the Schr\"odinger equation with the complete Hamiltonian for a number of values of the fluorescence frequency. This insensitivity is due to the weak atom-field coupling and the absence of the spatial dispersion effects. The weak coupling also guarantees that the first order fluorescence spectrum is a good approximation of the exact one. In the following we let $M = 10$ a.u. and $\omega_a = 0.043$ a.u. be fixed, and take $\epsilon = \omega_a$ or $\epsilon = 2\omega_a$ for the Mollow or SHG regimes respectively. The remaining parameters are $\alpha = 1$, $g_a = 0.1\epsilon$, $g_1 = 0.1\epsilon$, $g_2 = 0.05\epsilon$, $\Gamma_1 = 0.01\epsilon$ and $\Gamma_2 = 0.02\epsilon$.

\begin{figure*}
 \begin{tikzpicture}
  \node at ( 0.0, 2.0) {\includegraphics[width=0.66\columnwidth]{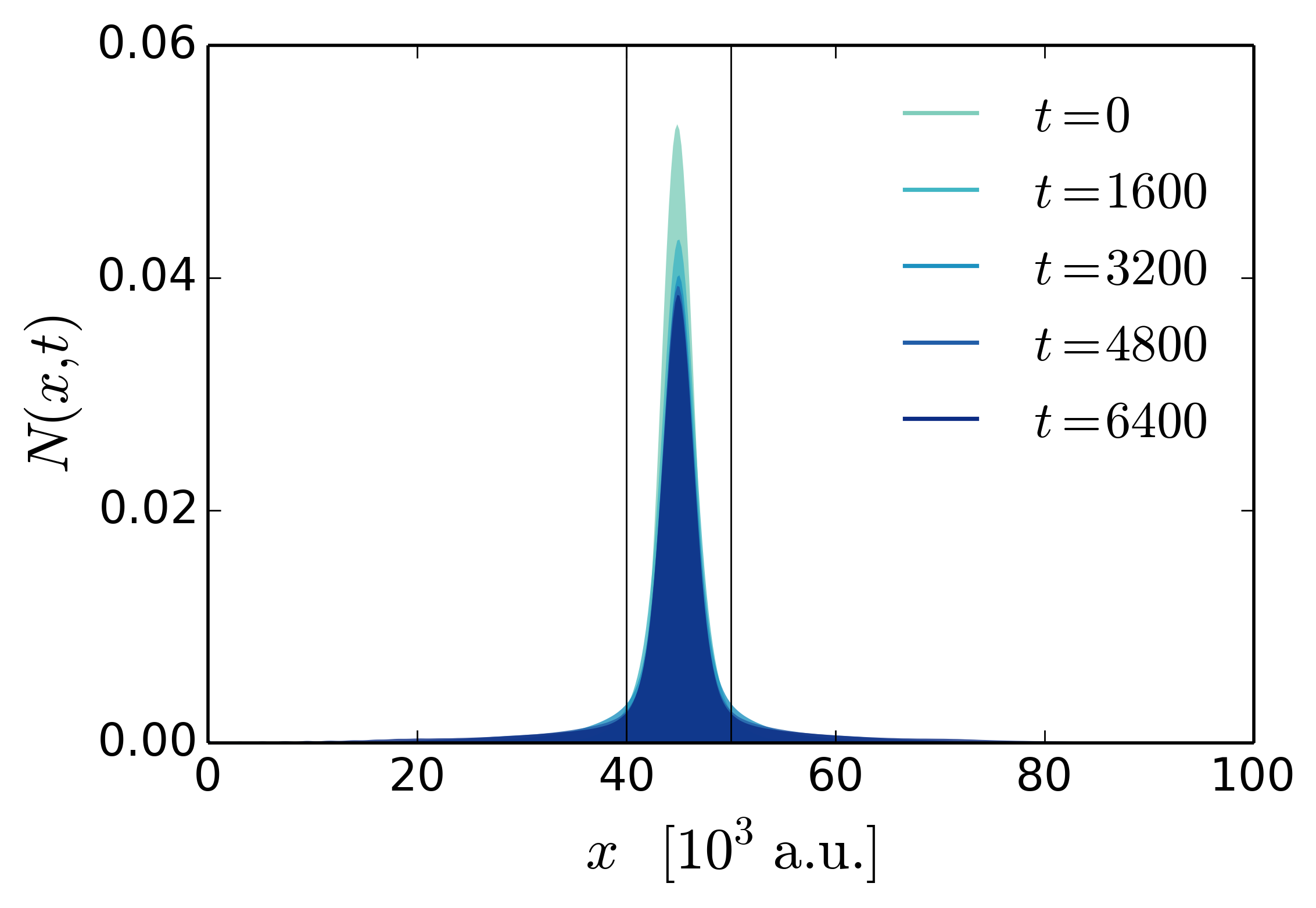}};
  \node at ( 5.8, 2.0) {\includegraphics[width=0.66\columnwidth]{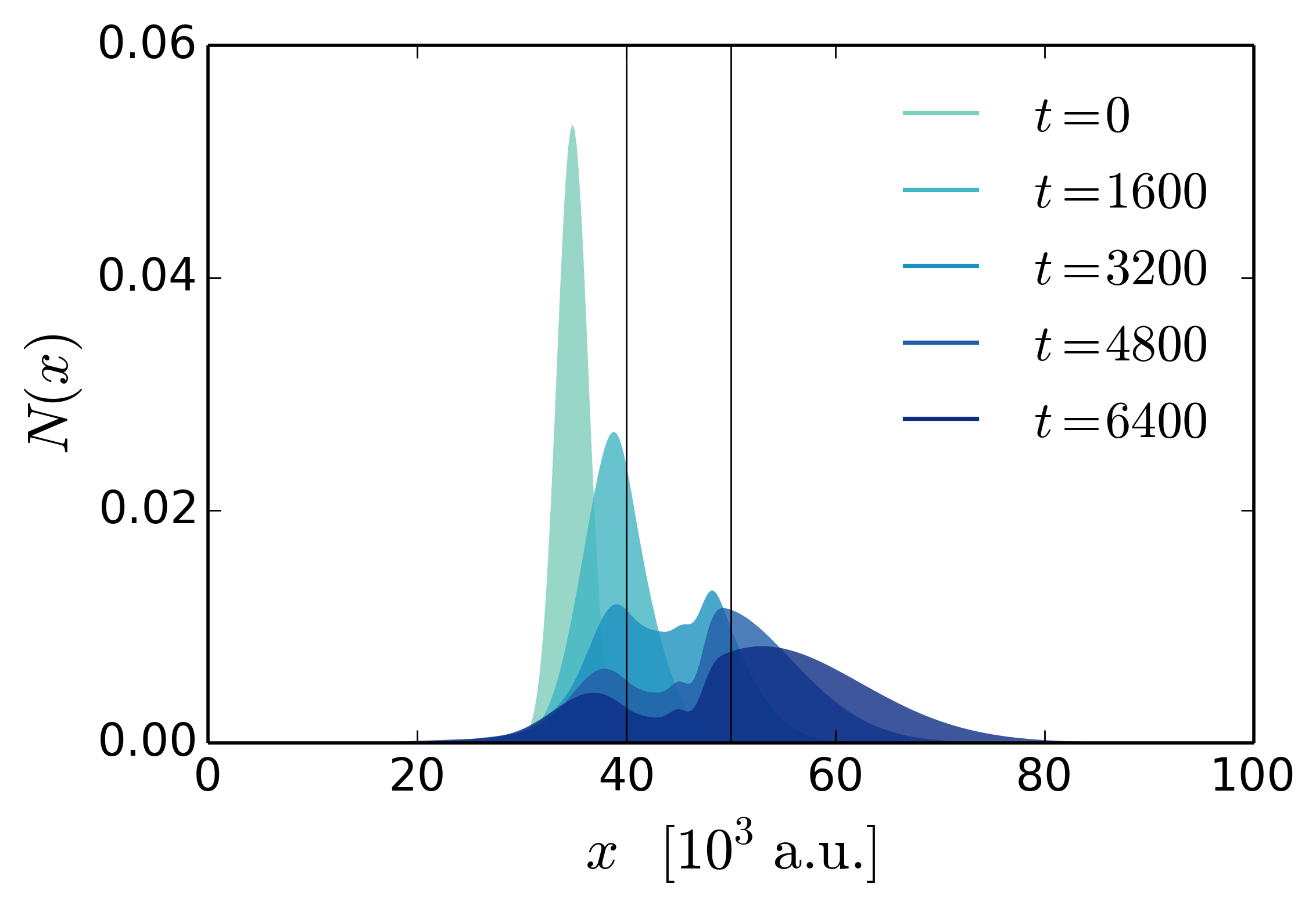}};
  \node at (11.6, 2.0) {\includegraphics[width=0.66\columnwidth]{nuclear_shg_p05.png}};
  \node at ( 0.1,-2.0) {\includegraphics[width=0.63\columnwidth]{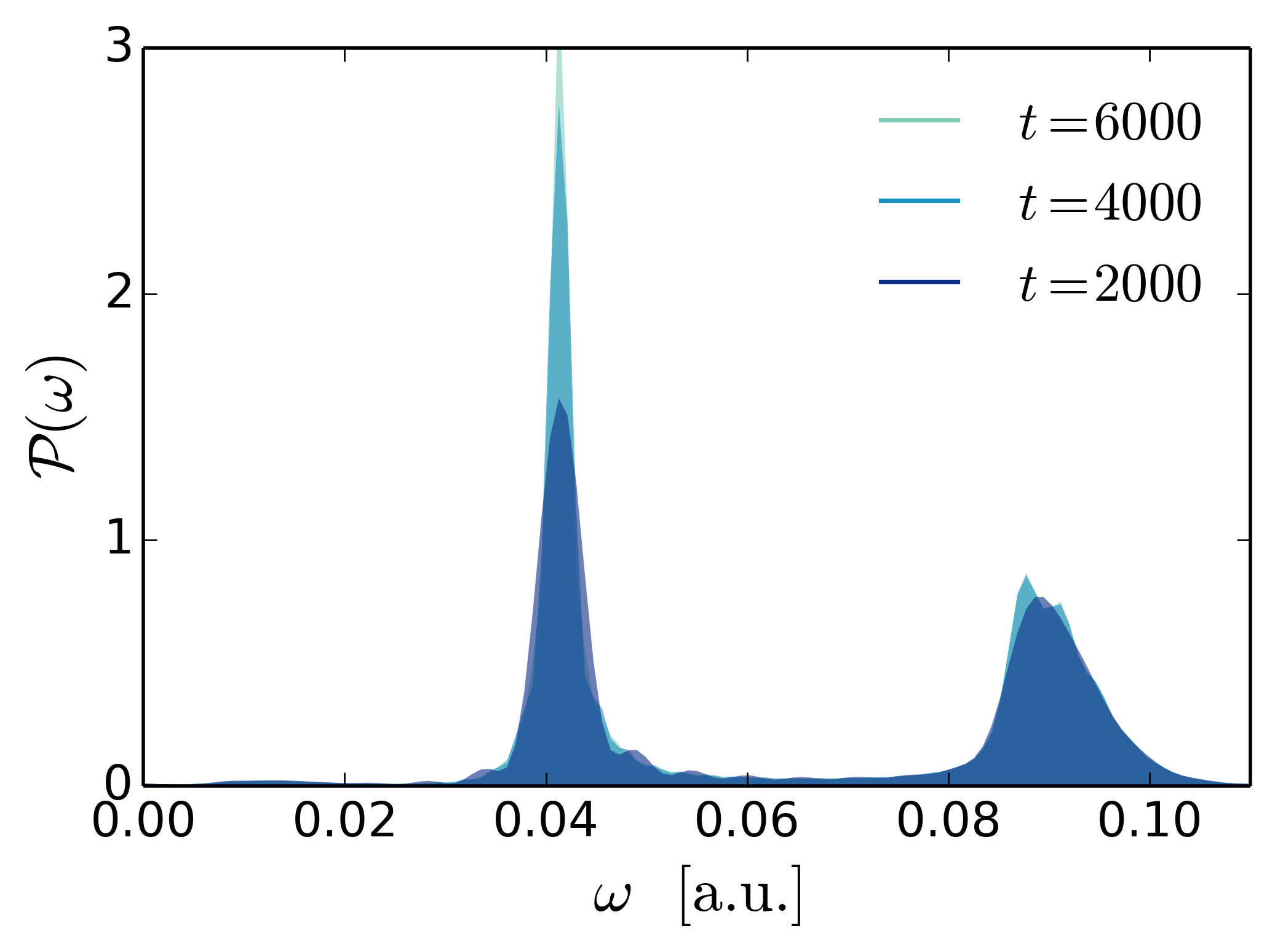}};
  \node at ( 5.7,-2.0) {\includegraphics[width=0.66\columnwidth]{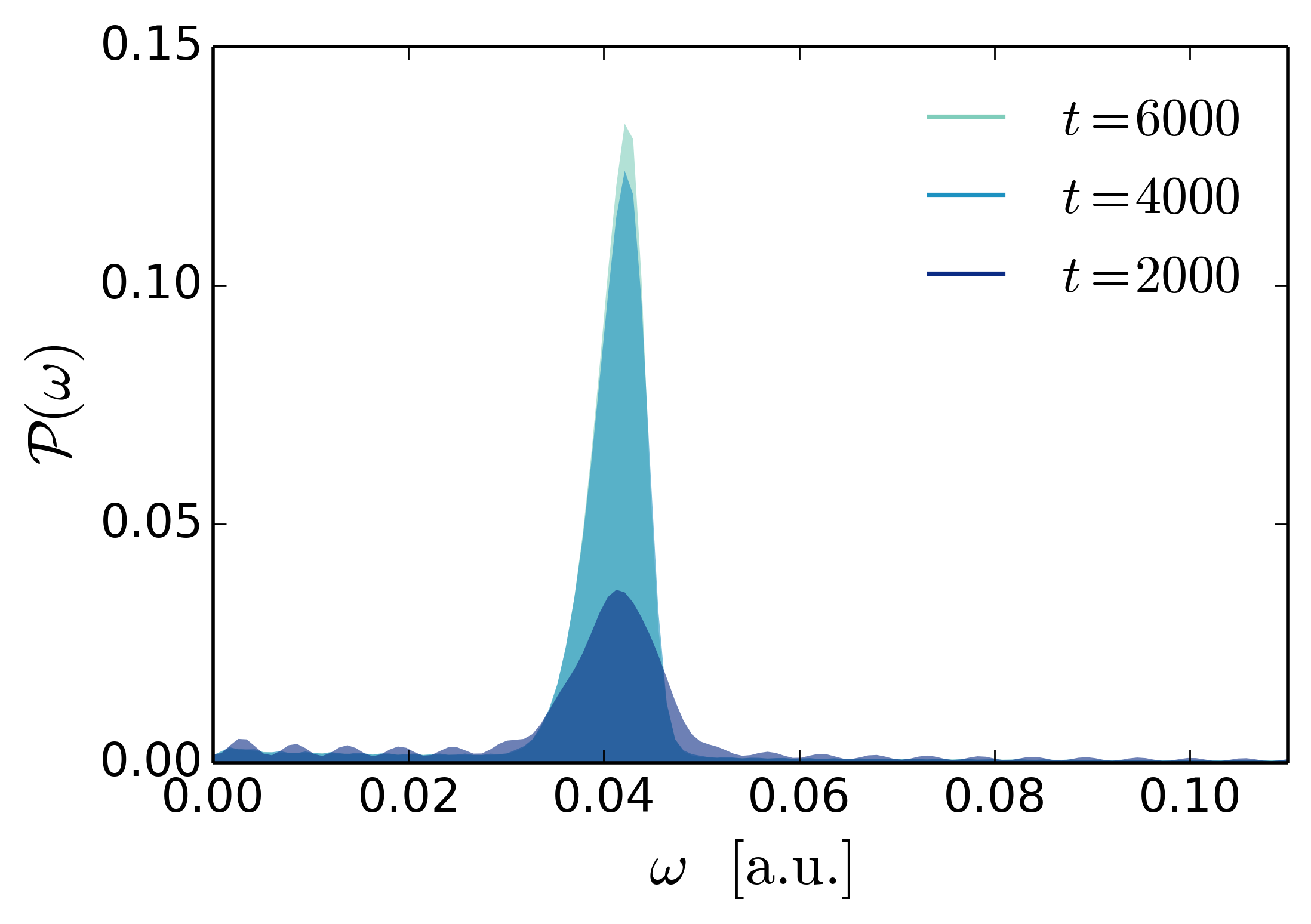}};
  \node at (11.5,-2.0) {\includegraphics[width=0.66\columnwidth]{emission_shg_p05.png}};
  \node[black] at (-1.5, 3.2) {$(a)$};
  \node[black] at ( 4.3, 3.2) {$(b)$};
  \node[black] at (10.1, 3.2) {$(c)$};
  \node[black] at (-1.5,-0.8) {$(d)$};
  \node[black] at ( 4.3,-0.8) {$(e)$};
  \node[black] at (10.1,-0.8) {$(f)$};
 \end{tikzpicture}
 \caption{Nuclear probability densities (top row) and atomic fluorescence spectra (bottom row) for an atom moving through a cavity of frequency $\omega_a = (\epsilon_2 - \epsilon_1)/2 = 0.043$ a.u. The vertical lines denote the boundaries of the cavity. The initial atomic state is given by Eq.~(\ref{eq:atom_wf}), and we take $x_0 = 3.5l_c$ (except for panels $(a)$ and $(d)$ where $x_0 = 4.5l_c$) and $\sigma = 3l_c$ with $l_c = 10^4$ a.u. Panels $(a)$ and $(d)$ corresponds to $p_0 = 0$, panels $(b)$ and $(e)$ to $p_0 = 0.5$ a.u., and panels $(c)$ and $(f)$ to $p_0 = 2$ a.u. The remaining parameters are given by $M = 10$ a.u., $\epsilon = \epsilon_2 - \epsilon_1 = 0.086$ a.u., $\alpha = 1$, $g_a = 0.1\epsilon$, $g_1 = 0.1\epsilon$, $g_2 = 0.01\epsilon$, $\Gamma_1 = 0.01\epsilon$ and $\Gamma_2 = 0.02\epsilon$.}
 \label{fig:shg_cavity}
\end{figure*}

In Fig.~\ref{fig:mollow_cavity} we show $N(x,t)$ and $P(\omega,t)$ for $\epsilon = \omega_a$. We see that for a stationary atom placed in the center of the cavity,
$x_0 = 4.5l_c$ and $p_0 = 0$, the spectrum resembles the Mollow spectrum in Fig.~\ref{fig:mollow}. We also find that although the atomic wave packet is initially contained in the cavity, parts of the probability distribution are ejected as time progresses. For higher initial momentum $p_0$, we see that an atom initially outside the cavity (at $x_0 = 3.5l_c$) is either split (for $p_0 = 0.5$ a.u.) or travels through the cavity (for $p_0 = 2$ a.u.). This is in agreement with the expectation based on an atom moving in the presence of a dipole force~\cite{Grynberg10}, where the force on the particle is proportional to the negative of the detuning and the gradient of the light intensity $F \sim -[\omega_a - (\epsilon_2 - \epsilon_1)]\partial_x I(x)$. For a field on resonance, an atomic motion in the positive direction leads to a positive detuning via a Doppler shift, so that atom is expelled from regions of higher intensity. This is why a minimal non-zero momentum is needed to pass through the cavity.

In Fig.~\ref{fig:shg_cavity} we show $N(x,t)$ and $P(\omega,t)$ for $\epsilon = 2\omega_a$.  We note that the results for $N(x,t)$
are rather similar to those in Fig.~\ref{fig:mollow_cavity}, presumably because the coupling to the radiation is too weak to make a larger difference. As for the Mollow regime above, we find that for an atom at rest with l $x_0 = 4.5l_c$ and $p_0 = 0$, the spectrum resembles the stationary SHG spectrum in Fig.~\ref{fig:shg}. For this initial state the atomic probability distribution is trapped in the cavity, consistent with motion under a dipole force as discussed above. For non-zero initial momentum $p_0$, we find that the SHG signal is strongly suppressed, and that the elastic scattering peak is broadened. Following considerations similar to those for Fig.~\ref{fig:mollow_cavity}, this effect is likely ascribable to the finite extent of the atomic wavepacket. Interestingly, and as for Fig.~\ref{fig:mollow_cavity}, on increasing $p_0$ the intensity of the SHG spectrum exhibits a non monotonic behavior. Further, due to the emission of a fluorescent photon being delayed with respect to the atomic excitation, the resonance frequency of the atom has time to change slightly between the two events. In fact, being a second order process, SHG is expected to be more sensitive to this type of detuning than the resonant scattering. Thus the combination of detuning and the decoherence induced by emission from different parts of the atomic wavepacket is likely the cause of the suppression of the SHG signal.

\section{Conclusions}\label{conclude}

In optics and in photonics, the two-level system plays the role of a 
Rosetta-stone for light-matter interactions, at the interface of quantum with 
classical and linear with nonlinear behavior. In this work, we used this paradigmatic 
system to address basic aspects of multi-photon fluorescence in the time-dependent 
and stationary regimes. The fluorescence response was considered in three cases 
of increasing complexity,  namely in a single two-level system interacting with an 
incident coherent field, as well as in an array of two-level systems, and a finally 
for a two-level atom moving across a cavity. 

By solving the Schr\"odinger equation through exact numerical time-propagation,
we showed that, depending on the system and field parameters, the time-dependent fluorescence 
spectrum develops distinct features that in some cases are not captured by 
a semi-classical treatment of the incident field. Some of these features offer direct 
evidence that the usual selection rules of perturbative optics, which consider photons 
and electrons separately, do not apply in the strong coupling regime.

As clear-cut example, we showed that a second harmonic signal 
(SHG) can occur in a two-level system. This result was analyzed 
in terms of the parity of the coupled electron-photon states, and we argued that this 
nonlinear process is allowed even for parity invariant Hamiltonians. We also studied 
the SHG process in a three-level system, and showed that in an appropriate limit 
the three-level system reproduces the results of the two-level system.  

The SHG signal gets enhanced in a setup with $N$ two-level 
systems (compared to the $N=1$ case), with trends suggestive 
of superradiant behavior. This conclusion was gathered by looking analytically at possible
signatures of a phase transition in the large $N$ limit, and by analyzing  the onset
of the SHG response in exact numerical calculations for $N\lesssim 10$.
However, compared to the case of one atom at rest,
both Mollow and SHG signals are suppressed by atomic motion.
This effect is especially strong in the quantum case:
For quantum atomic motion, the spread in space of the traveling nuclear wavepacket
is greatly increased by the presence of the barrier represented by the optical cavity, giving rise to
a position dependent Doppler shift. This in turn results in a large frequency dispersion  
in the fluorescence signal. It is important to specify here that the noted quantum effects
were enhanced by choosing artificially small nuclear masses. However, 
although we focussed here on model atomic systems, it is a fair assumption that
many of the results obtained should carry over to solid state two-level systems, 
where suitable (excitonic) masses could be engineered.

The aforementioned effects are weak, and
they manifest at low intensities; this is confirmed 
by the fact that, for the chosen parameters, a linear and an all-orders  
treatment of the fluorescence field provide an identical scenario.
This also means that our situation does not correspond to standard 
heterodyne setups, and no intensity renormalization is needed. 
Even so, said effects should be of some conceptual (if not practical)
interest. It can in fact be argued that the found superimposed Mollow-like structure to the SHG 
signal is specifically distinctive of the genuine ``two-level'' character of the material system, 
as also gathered by looking at three-levels results. We add that a similar (albeit weaker) superimposed structure
also occurs for higher-order harmonics.

Concerning dissipation effects, we expect on purely speculative grounds that the peculiar four-peak structure of the Mollow
spectrum could be dimmed, while the SHG signal would considerably change but still survive for not too strong dissipation.

In conclusion, we have addressed general features of multi-photon fluorescence, but
only in very simple model systems: The inclusion of additional radiation modes,
more general time-dependent couplings, a careful inclusion of bath effects,
and more realistic atomic and solid-state setups
are possible directions for future work, and to confirm in a broader sense the robustness of present results.
Ultimately, true validation comes from experiment, and we hope that 
our work will stimulate investigations in that direction.

\acknowledgments
We wish to thank W. P. Schleich for useful discussions.
\noindent E. V. B was supported by Crafoordska Stiftelsen. C. V. was supported by the Swedish Research Council.\\

%%%%%%%%%%%%%%%%%%%%%%%%%%%%%%%%%%%%%%%%%%%%%%%%%%%%%%%%%%%%
%%%%%%%%%%%%%%%%%%%%%%%%%%%%%%%%%%%%%%%%%%%%%%%%%%%%%%%%%%%%
\appendix

\section{}\label{App_A0}
We provide here some additional motivation and detail about the model and the method of solution.

\subsection{About dissipative effects}\label{App_A0_Lind}

 For the first two typologies of
systems considered, i.e. one or many two-level system(s) at rest,
we assume that (via e.g. a cavity-geometry or an
high-optical quality sample in ultra-high vacuum 
and at helium temperature), the inhomogeneous
broadening has been made as negligible as possible. In this way,
the focus is solely on the homogeneous broadening. This is due
to both radiative and non-radiative components, 
that in our treatment are accounted for by a total phenomenological 
Lorentzian damping.

For the the third typology of system (i.e. the ``atom'' in motion, 
and where the dissipative environment
can be  considerably different) we have followed a common practice in the literature, 
considering only the moving material system and the relevant modes of interest.
Ultimately, this choice is also dictated by computational convenience, since
the full quantum treatment of nuclear motion adds considerable
complexity to the numerics.

Looking ahead, a possible way to include dissipative 
effects is via Lindblad-type master equations 
or, alternatively, via non-equilibrium Green's functions (NEGF).
NEGF permit to include memory effects and 
the dispersive contribution of the environment in a very direct and
systematic way, and it would 
be rather interesting to perform a comparison between NEGF and master equation
results. These calculations and comparisons
are deferred to future work. 

\subsection{Computational details}\label{App_A0_Lanc}
The short iterated Lanczos method is an efficient algorithm to approximate the time-evolution operator $U$. This is done by constructing $U$ in a small optimized subspace (the Krylov space), which allows to maintain unitarity of $U$ (in contrast to a straightforward Taylor expansion) while being numerically efficient. We used this algorithm to propagate the many-particle Schr\"odinger equation, and additional details can be found in Ref.~\cite{ParkLight}.

Regarding the choice of basis, we use two basis states $|1\rangle$ and $|2\rangle$ two describe the ``atomic'' electron states, and the number states $|n_a\rangle$ and $|n_b\rangle$ for the coherent and fluorescent fields respectively. In the last part of the manuscript, where we study the motion through a cavity, we use the position basis $|x_n\rangle$ on an equidistant grid to describe the ``atomic'' center of mass. 

Finally, the choice of cut-off number(s) for the radiation modes is determined by the convergence (i.e. by increasing the number of states until the results are converged within machine accuracy). For the fluorescent field we found it was sufficient in all cases considered to use a maximum $n_b \approx 10$. For the coherent field the numerical cut-off depends on $\alpha$, since the coherent field follows a Poissonian distribution when written in terms of number states. For $\alpha = 1$ and $\alpha = 5$ respectively, we found that a cut-off at $n_a \approx 30$ and $n_a \approx 150$ is enough.

\subsection{A scaling property}\label{App_A0scal}
The Hamiltonian in Eq. (\ref{atomoving}) satisfies a scaling property
relating the full quantum dynamics of systems with different masses. 
Specifically, we
start by considering a scaling parameter $Z$, and the Schr\"odinger equation $i\partial_t \psi(t)=\hat{H}(t)\psi(t)$. Dividing by $Z$, and 
setting $t'=Z t$, we get $i\partial_{t'} \phi(t')=Z^{-1}\hat{H}(t'/Z) \phi(t')$, where $\phi(t')= \psi(t'/Z)$. By
relabeling the time variable, $t'\rightarrow t$, we then have $i\partial_t \phi(t)=\tilde{H}(t)\phi(t)$, where $\tilde{H}(t)=Z^{-1}\hat{H}(t/Z)$.
According to this scaling prescription, a given numerical calculation represents in fact a entire one-parameter set of numerical simulations, where the integration interval, the time dependence in $\hat{H}$ and the fermion-boson interactions
are changed.

\section{}\label{App_A}
We consider here the adiabatic elimination approximation (AEA) for the three-level Hamiltonian given by the sum of Eqs.~(\ref{Hs3}),(\ref{Hi_a}) and the free field part. 
Most often, the AEA is done in connection with the rotating wave approximation (RWA) (see e.g. \cite{Alsing87,Gou89,Brion07,Fewell05}), and choosing the order in which AEA and RWA are performed can be important \cite{Fewell05}. Since our considerations here aim to be qualitative and general in character, we use for simplicity the more common protocol where the RWA is introduced before the AEA ~\cite{Gou89}, and before the pump field undergoes a transformation to a coherent photon picture. Proceeding in this way, we obtain
\begin{align}
 \hat{H}_{RWA} &= f(t)[\hat{c}^\dagger_3 \hat{c}_2 + \hat{c}^\dagger_2 \hat{c}_1]\hat{a} + g_b(t)\hat{c}^\dagger_3 \hat{c}_1 \hat{b}+\text{H.c.} \nonumber \\
 &+ \epsilon_1 \hat{c}^\dagger_1 \hat{c}_1+\epsilon_2 \hat{c}^\dagger_2 \hat{c}_2+\epsilon_3 \hat{c}^\dagger_3 \hat{c}_3 + \omega_a \hat{a}^\dagger \hat{a} + \omega_b \hat{b}^\dagger \hat{b} 
\end{align}
After an AEA of the intermediate level $|2\rangle$ (therefore, the dynamical Stark effect due to the intermediate level is neglected) we get 
\begin{align}\label{eq:H_AEA}
 \hat{H}^{AEA}_{RWA} &= \hat{c}^\dagger_3 \hat{c}_1\big[ f(t) \hat{a}^2 + g_b(t) \hat{b}\big] +\text{H.c.} \nonumber \\
 &+ \frac{\epsilon_3-\epsilon_1}{2}(\hat{c}^\dagger_3 \hat{c}_3 -\hat{c}^\dagger_1 \hat{c}_1)+  \omega_a \hat{a}^\dagger \hat{a} + \omega_b \hat{b}^\dagger \hat{b}.
\end{align}
At this point the coherent state picture could be introduced. However, already at this stage, the AEA two-level model of Eq.~(\ref{eq:H_AEA}) seems rather different from the original two-level model of Eqs.~(\ref{Heq1})-(\ref{eq:ham_i}). Therefore, the results of the main text for the SHG and Mollow spectra in a two-level system should not be ascribed to an adiabatic suppression of the virtual level.\\

\section{}\label{App_B}
We want to calculate the probability defined in Eq.~(\ref{equazio1}) and repeated here for convenience:
\begin{align}
P(t,\omega) = \sum_{ni} \int dx\,|\langle i,x,n|\hat{b}e^{-i\hat{H}t}|1,\phi,\alpha \rangle|^2. \label{equazio0}
\end{align}
In the perturbative limit, the time-evolution operator becomes:
\begin{align}
 e^{-i\hat{H}t} \approx \int_0^t dt'\, e^{-i\hat{H}_0(t-t')}\hat{H}'(t')e^{-i\hat{H}_0t'}, 
\end{align}
and to further simplify the analysis, we define the probability amplitude
\begin{align}
A_{in}^x(t,\omega) = -i&\theta(t)\int_0^t dt'\,\langle i,x,n|\hat{b}e^{-i\hat{H}_0(t-t')}\hat{H}' \\ \nonumber
                      &\times e^{-i\hat{H}_0t'}|1,\phi,\alpha\rangle. 
\end{align} \label{equazio4}
Since $\hat{H}_0$ is independent of time, the probability amplitude can be found through a straightforward expansion in the eigenstates of $\hat{H}_0$. In the expression for the probability $P$ above, we trace over a
complete set of final 
\begin{widetext}
\noindent states $|i,x,n\rangle$, but since any complete set is allowed we can instead choose to trace over the eigenstates of $\hat{H}_0$. In the following we therefore consider the probability amplitude $A_\lambda(t,\omega)$, and by inserting a set of complete states we find
\begin{align}
 \langle\lambda|\hat{b}e^{-i\hat{H}_0(t-t')}\hat{H}'(t')e^{-i\hat{H}_0t'}|1,\phi,\alpha\rangle &= \sum_{\lambda''\lambda'} \langle \lambda|\hat{b}|\lambda''\rangle e^{-i(\epsilon_{\lambda''}+\omega)(t-t')} H'_{\lambda''\lambda'}(t')e^{-i\epsilon_{\lambda'}t'}\langle\lambda'|1,\phi,\alpha\rangle \nonumber \\
 &= \sum_{\lambda'} e^{-i(\epsilon_{\lambda}+\omega)(t-t')-i\epsilon_{\lambda'}t'}H'_{\lambda\lambda'}(t')\langle\lambda'|1,\phi,\alpha\rangle.
\end{align}

\noindent Now the matrix elements $H'_{\lambda\lambda'}(t')$ can be broken into two parts according to
\begin{align}
 H'_{\lambda\lambda'}(t') &= \sum_{ni}\int dx\, \langle\lambda|\hat{H}'(t')|i,x,n\rangle\langle i,x,n|\lambda'\rangle \\
 &= \sum_{ni}\int_{X_{in}} dx\, g_1e^{-\Gamma_1 t'}\langle\lambda|\hat{H}'|i,x,n\rangle\langle i,x,n|\lambda'\rangle + \sum_{ni}\int_{X_{out}} dx\, g_2e^{-\Gamma_2 t'}\langle\lambda|\hat{H}'|i,x,n\rangle\langle i,x,n|\lambda'\rangle \nonumber \\
 &= g_1e^{-\Gamma_1 t'}S^1_{\lambda\lambda'} + g_2e^{-\Gamma_2 t'}S^2_{\lambda\lambda'} \nonumber
\end{align}
where the coefficients $S_{\lambda\lambda'}^{1}$ and $S_{\lambda\lambda'}^{2}$ are given by 
\begin{align}
S_{\lambda\lambda'}^{1} = \sum_{ni}\int_{X_{in}} dx\, \langle\lambda|\hat{H}'|i,x,n\rangle\langle i,x,n|\lambda'\rangle \nonumber \\
S_{\lambda\lambda'}^{2} = \sum_{ni}\int_{X_{out}} dx\, \langle\lambda|\hat{H}'|i,x,n\rangle\langle i,x,n|\lambda'\rangle.
\end{align}
Integrating over $t'$ we find the probability amplitude to be
\begin{equation}
A_\lambda(t,\omega) = \sum_{\lambda'} \left(\frac{e^{-i(\epsilon_{\lambda}+\omega)t} -e^{-i\epsilon_{\lambda'}t-\Gamma_1 t}}{\omega 
+\epsilon_{\lambda}-\epsilon_{\lambda'}+i\Gamma_1} S^1_{\lambda\lambda'} + \frac{e^{-i(\epsilon_{\lambda}+\omega)t} -e^{-i\epsilon_{\lambda'}t-\Gamma_2 t}}{\omega +\epsilon_{\lambda}-\epsilon_{\lambda'}+i\Gamma_2} S^2_{\lambda\lambda'}\right)\langle\lambda'|1,\phi,\alpha\rangle, 
\end{equation}
and inserting this into the expression for the probability we find
\begin{equation}\label{eq:spectra}
P(t,\omega) = \sum_{\lambda} \left|\sum_{\lambda'} \left(\frac{e^{-i(\epsilon_{\lambda}+\omega)t} -e^{-i\epsilon_{\lambda'}t-\Gamma_1 t}}{\omega +\epsilon_{\lambda}-\epsilon_{\lambda'}+i\Gamma_1} S^1_{\lambda\lambda'} + \frac{e^{-i(\epsilon_{\lambda}+\omega)t} -e^{-i\epsilon_{\lambda'}t-\Gamma_2 t}}{\omega +\epsilon_{\lambda}-\epsilon_{\lambda'}+i\Gamma_2} S^2_{\lambda\lambda'}\right)\langle\lambda'|1,\phi,\alpha\rangle\right|^2. 
\end{equation}
If necessary, it is possible to further simplify Eq.~(\ref{eq:spectra}) by going at long times
(i.e. where the exponentials $e^{-\Gamma_k t}$ tend to zero) provided that $L$ is 
correspondingly taken large enough to avoid atom reflection at the ends of the 
$x$-coordinate domain. Arguing that the cross terms vanish for $t \to \infty$, the asymptotic limit becomes 
\begin{align}
P(\omega) &= \sum_{\lambda} \left|\sum_{\lambda'}\left(\frac{1}{\omega +\epsilon_{\lambda}-\epsilon_{\lambda'}+i\Gamma_1} S^1_{\lambda\lambda'}\right.\right. +\left.\left. \frac{1}{\omega +\epsilon_{\lambda}-\epsilon_{\lambda'}+i\Gamma_2} S^2_{\lambda\lambda'}\right)\langle\lambda'|1,\phi,\alpha\rangle\right|^2. \nonumber
\end{align}
This latter result makes contact with the long time limit of the static-atom case of Sec. \ref{sec:single_system}. However, 
to calculate the time-dependent fluorescence spectrum for a moving atom we will go back to the full expression for $P(t,\omega)$ in Eq.~(\ref{eq:spectra}) here, or Eq.~(\ref{eq:spectrum}) in the main text.

\end{widetext}


\begin{thebibliography}{999}
\bibitem{PARAMS} With the chosen units, the parameters values employed in our model calculation, while generic in character, fall well within the the physical range of validity.
\bibitem{Since} 
It might be worth to add that for exciton-polaritonics in 3D crystals, 
another class of template systems where our results might be of potential relevance, 
the spontaneous decay occurs because of the fluctuations in the polaritonic vacuum.
Furthermore, being described by coupled
Schr\"odinger and Maxwell equations, 
the polaritonic eigenfunction does not spontaneously decay in the bulk, but rather
at the solid/vacuum interface.
\bibitem{confuse} The system in Eqs.~(\ref{Heq1}-\ref{eq:ham_i})
should not be confused with the effective
two-level one obtained from a three level system via adiabatic elimination, as described in appendix \ref{App_A}.
\bibitem{caveat} Caution should be exerted in associating too closely our parameter values
to the case of excitons: inside the cavity, it is more appropriate to consider a polaritonic picture.
Thus, the reference to excitons rather aims to showing that the parameters
of our model study are plausible in the physical range, rather than to connecting
to a concrete and specific case.
\end{thebibliography}
\end{document}